\documentclass[12pt]{article}
\usepackage[height=8.85in,width=6.45in]{geometry}
\usepackage{amsmath,amssymb}
\usepackage[svgnames]{xcolor}
\usepackage[debug,pageanchor=false]{hyperref}
\definecolor{link}{rgb}{.8,.15,.1}
\hypersetup{colorlinks=true,linkcolor=link,citecolor=link,urlcolor=link,linktocpage}
\usepackage{cite}
\usepackage[export]{adjustbox}

\def\tr{\mathop{\mathrm{tr}}\nolimits}

\usepackage{times}
\usepackage{courier}
\usepackage{mathtools}
\numberwithin{equation}{section}

\let\bar\overline

\def\Nequals#1{$\mathcal{N}{=}\,#1$}

\newcommand{\bC}{\mathbb{C}}

\newcommand{\bZ}{\mathbb{Z}}

\newcommand{\sB}{\mathsf{B}}
\newcommand{\sH}{\mathsf{H}}

\newcommand{\SL}{SL}
\newcommand{\Spin}{Spin}
\newcommand{\SU}{SU}
\newcommand{\OO}{O}
\newcommand{\SO}{SO}

\newcommand{\fso}{\mathfrak{so}}
\newcommand{\fsu}{\mathfrak{su}}
\newcommand{\fusp}{\mathfrak{usp}}
\newcommand{\fg}{\mathfrak{g}}
\newcommand{\fh}{\mathfrak{h}}
\newcommand{\fe}{\mathfrak{e}}
\newcommand{\fu}{\mathfrak{u}}

\def\rank{\mathop{\mathrm{rank}}}

\newcommand{\be}{\begin{equation}}
\newcommand{\ee}{\end{equation}}
\newcommand{\bea}{\begin{eqnarray}}
\newcommand{\eea}{\end{eqnarray}}

\DeclareMathOperator{\Tr}{Tr}

\DeclareMathOperator{\CS}{CS}
\usepackage{dsfont}

\begin{document}

\begin{titlepage}

\begin{flushright}
IPMU-18-0192
\end{flushright}

\vfill

\begin{center}

{\Large\bfseries Compactifications of $6d$ $\mathcal{N} = (1,0)$ SCFTs\\[1em]
 with  non-trivial Stiefel-Whitney classes}

\vskip 1cm
Kantaro Ohmori${}^{a}$, Yuji Tachikawa${}^{b}$ and Gabi Zafrir${}^{b}$
\vskip 1cm

\begin{tabular}{ll}
 & ${}^a$ School of Natural Sciences, Institute for Advanced Study, \\
 & 1 Einstein Drive Princeton, NJ 08540, USA \\
 & ${}^b$ Kavli Institute for the Physics and Mathematics of the Universe (WPI), \\
 & University of Tokyo,  Kashiwa, Chiba 277-8583, Japan\\
\end{tabular}

\vskip 1cm

\end{center}

\noindent

We consider compactifications of very Higgsable $6d$ $\mathcal{N}{=}(1,0)$ SCFTs on $T^2$
with non-trivial Stiefel-Whitney classes (or equivalently 't Hooft magnetic fluxes) introduced for their flavor symmetry groups.
These systems can also be studied as twisted  $S^1$ compactifications of the corresponding 5d theories.
We deduce various properties of the resulting 4d $\mathcal{N}{=}2$ SCFTs by combining these two viewpoints.
In particular, we find that all 4d rank-1 $\mathcal{N}{=}2$ SCFTs with a dimension-6 Coulomb branch operator
with flavor symmetry $\fe_8$, $\fusp(10)$, $\fsu(4)$ and $\fsu(3)$ 
can be uniformly obtained by starting from a single-tensor theory in $6d$.

We also have a mostly independent appendix where we propose a rule to determine the Coulomb branch dimensions of $4d$ $\mathcal{N}{=}{2}$ theories obtained by $T^2$ compactifications of $6d$ very Higgsable theories with and without Stiefel-Whitney twist.

\vfill
\vfill

\end{titlepage}

\tableofcontents

\section{Introduction}

Compactification of $6d$ supersymmetric field theories on two-dimensional surfaces continues to shed new light on the dynamics and the dualities of $4d$ supersymmetric field theories.
The classic example is to start from $6d$ \Nequals{(2,0)} theories on Riemann surfaces to obtain 4d \Nequals2 theories \cite{Gaiotto:2009we,Gaiotto:2009hg};
the resulting theories are called the class S theories and have been studied in extreme detail.

More recently, we started to understand the more general $6d$ \Nequals{(1,0)} theories much better than we did previously, thanks to the developments initiated by \cite{Heckman:2013pva,DelZotto:2014hpa} and reviewed in \cite{Heckman:2018jxk}.
Compactification of these $6d$ \Nequals{(1,0)} theories on general Riemann surfaces leads to 4d \Nequals1 theories;
some cases starting from a few parent $6d$ theories have been studied in great detail,
showing intricate webs of duality of $4d$ \Nequals1 theories, see e.g.~\cite{Gaiotto:2015usa,Razamat:2016dpl,Bah:2017gph,Kim:2017toz,Kim:2018bpg,Kim:2018lfo,Razamat:2018gro}.
Along the way, we gradually came to realize the importance of identifying the flavor symmetry as a group,
not simply as a Lie algebra, since it affects which flavor background flux can be introduced 
on the Riemann surface used in the compactification.
The global structural of the global symmetry group is generally defined by demanding it to faithfully act on the local operators.

Our aim in this paper is to study the effect of the flavor symmetry \emph{group} on a slightly simpler setup of $6d$ \Nequals{(1,0)} theories on a flat torus $T^2$, preserving $4d$ \Nequals2 supersymmetry.
There are  several previous studies of the compactification of $6d$ \Nequals{(1,0)} theories on $T^2$ preserving 4d \Nequals2 supersymmetry e.g.~in \cite{Ganor:1996pc,Ohmori:2015pua,DelZotto:2015rca,Ohmori:2015pia,Zafrir:2015rga,Ohmori:2015tka,Hayashi:2015zka,Mekareeya:2017jgc,Mekareeya:2017sqh},
where in most of the papers no flavor symmetry background was introduced,
with the sole exception of the early papers \cite{Ganor:1996pc,Eguchi:2002fc,Eguchi:2002nx} on the compactification of the E-string theory, where arbitrary flat backgrounds of $E_8$ were considered.

Let us now suppose that the flavor symmetry group is $\SU(3)/\bZ_3$.
We can then consider not only the standard compactification without any flavor background,
but also a more intricate one where the background flavor holonomies along two directions of the torus are \begin{equation}
P=\begin{pmatrix}
1 &0&0\\
0 & e^{2\pi i/3} &0 \\
0&0&e^{4\pi i/3} 
\end{pmatrix},\qquad
Q=\begin{pmatrix}
0& 1 & 0\\
0 & 0& 1\\
1 & 0 & 0
\end{pmatrix}
\label{eq:SU3holonomy}
\end{equation}
which \emph{do not} commute within $\SU(3)$ but \emph{do} commute within $\SU(3)/\bZ_3$.
Such a flavor background  is said to have a nontrivial (generalized) Stiefel-Whitney class;
or equivalently, we are using the 't Hooft magnetic flux \cite{tHooft:1979rtg} in the flavor background.

In this paper, we restrict our choice of the parent $6d$ theories to the so-called \emph{very Higgsable theories} introduced in \cite{Ohmori:2015pia},
i.e.~those theories which, if realized in the F-theory language, allows a sequence of blow-downs of ($-1$)-curves to a single point.
This is because their $T^2$ compactifications are easier to analyze than more general theories, as can be seen by comparing the analysis of $T^2$ compactifications of very Higgsable theories in \cite{Ohmori:2015pua} and that of more general theories in \cite{DelZotto:2015rca,Ohmori:2015pia}.

One of the highlights of our analysis is that our methods allow us to construct uniformly
all the $4d$ \Nequals2 rank-1 theories with non-Abelian flavor symmetry
whose Coulomb branch operator $u$ has dimension $6$.
Let us recall that Argyres, Lotito, L\"u, and Martone have classified in a series of impressive works \cite{Argyres:2015ffa,Argyres:2015gha,Argyres:2016xmc,Argyres:2016xua} all the possible 4d \Nequals2 rank-1 theories,
by a meticulous study of the consistency conditions imposed on the Coulomb branch.
They found that the possible non-Abelian flavor symmetries are $\mathfrak{e}_8$,
$\mathfrak{usp}_{10}$, $\mathfrak{su}_4$ or $\mathfrak{su}_3$,
and determined $a$, $c$ and $k$, as tabulated in Table~\ref{table:rank1data}.

\begin{table}
\[
\begin{array}{c||c|cc|cc|c||c}
 \text{flavor sym.} & k & 12c& 24a &  n_v & n_h & (n_h)_\text{gen.Coulomb} & \text{prev. realizations} \\
 \hline
 \mathfrak{e}_8 &  12 & 62& 95 &  11 & 40 &  0  & \text{\cite{Minahan:1996cj,Ganor:1996xd,Ganor:1996pc}} \\
 \mathfrak{usp}({10}) &  7 & 49& 82 &  11 & 27 &  5 & \text{\cite{Argyres:2007tq,Chacaltana:2013oka}}\\
 \mathfrak{su}({4}) &  14 & 49& 75 &  11 & 20 &  4 &\text{\cite{Chacaltana:2016shw}}\\
 \mathfrak{su}({3}) &  14 & 38& 71 &  11 & 16 &  3 &\text{none}
\end{array}
\]
\caption{List of $4d$ \Nequals2 rank-1 theories with non-Abelian flavor symmetry
whose Coulomb branch operator $u$ has dimension $6$. Here, $k$, $c$, and $a$ are flavor and conformal central charges in the standard normalization; the effective numbers $n_v$, $n_h$ of vector multiplets and hypermultiplets are defined as always by the formula $24a=5n_v+n_h$ and $12c=2n_v+n_h$. $(n_h)_\text{gen.Coulomb}$ is the Higgs branch dimension at a generic point on the Coulomb branch.
\label{table:rank1data}}

\[
\begin{array}{c||c|c|c}
 \text{$4d$ flavor}  & \text{$6d$ parent} &\text{$6d$ flavor} & \text{twist} \\
 \hline
 \mathfrak{e}_8 &  \text{E-string}   & E_8&\bZ_1\\
 \mathfrak{usp}({10})  & \text{$\fsu(2)+ 10F$ } &\Spin(20)/\bZ_2 & \bZ_2 \\
 \mathfrak{su}({4}) &  \text{$\fsu(3)+12F$ } & \SU(3)/\bZ_3 & \bZ_3 \\
 \mathfrak{su}({3}) &   \text{$\fsu(4)+12F+AS$} & (\SU(2)\times\SU(12))/\bZ_4 & \bZ_4 
\end{array}
\]
\caption{$6d$ realizations of $4d$ \Nequals2 rank-1 theories listed in Table~\ref{table:rank1data}.
Here, $\fg + nF$ means a $\fg$ gauge theory with $n$ hypermultiplets in the fundamental representation; $+AS$ means that there is also a hypermultiplet in the second rank antisymmetric tensor.
\label{table:6dreal}}
\end{table}

We find that by starting from a $6d$ single-tensor theory with a suitable gauge group and hypermultiplet matter content and putting it on $T^2$ with a Stiefel-Whitney class,
we can realize all these four cases uniformly.
The $6d$ theories are listed in Table~\ref{table:6dreal}.
Among others, ours is the first realization of the 4d rank-1 theory with $\mathfrak{su}_3$ symmetry;
the work by Argyres et al.~only provides a necessary condition for any rank-1 theory, 
whereas ours provide an F-theory realization.

We will study the properties of such $T^2$ compactifications by utilizing various techniques.
One is to go directly from $6d$ to $4d$; this allows us to carry over the knowledge of the anomaly polynomial in $6d$ to that in $4d$ and to determine the central charges $a$, $c$ and $k$ of the $4d$ theories, by generalizing the method of \cite{Ohmori:2015pua}.
Another is to first identify the 5d theory after compactifying on an $S^1$ with the holonomy $P$.
This is a mass deformation of the 5d theory obtained by compactifying on an $S^1$ with a trivial holonomy,
whose brane-web description has been studied e.g.~in \cite{Hayashi:2015fsa,Zafrir:2015rga,Ohmori:2015tka,Hayashi:2015zka}.
Then a compactification on a further $S^1$ with the holonomy $Q$ is realized as a compactification by a $\bZ_k$ twist, which has been studied in \cite{Zafrir:2016wkk}.

The rest of the paper is organized as follows. 
In Sec.~\ref{sec:rank1}, we study compactifications with Stiefel-Whitney twists leading to rank-one theories in $4d$.
We first study in Sec.~\ref{sec:2.1}, \ref{sec:2.2}, \ref{sec:2.3} the symmetries and the spectra on generic points on the $u$-plane of the compacitifications of $6d$ $\fsu(2)$, $\fsu(3)$ and $\fsu(4)$ theories in turn, 
and see that they match  the data of  the $4d$ theories with $\fusp(10)$, $\fsu(4)$ and $\fsu(3)$ symmetries, respectively. 
In Sec.~\ref{sec:rank1central}, we compute the central charges $a$, $c$, $k$ of the compactified theory from the data of the 6d theory, and show that they indeed reproduce those of the $4d$ theories.
In Sec.~\ref{subsec:rank1-5d}, we carry out another way to analyze the compactification, first by identifying the 5d theory after the compactification on $S^1$ with a nontrivial holonomy,
and then further compactifying the 5d theory on $S^1$ with a further twist.
This is done by utilizing the brane web description.
Up to this point, we only use a single choice of holonomies on $T^2$ with a given Stiefel-Whitney class.
In Sec.~\ref{sec:hol}, we study the effect of changing the holonomies within the same Stiefel-Whitney class, using the case of $6d$ $\fsu(2)+10F$ theory as the example.

In Sec.~\ref{sec:general}, we extend the analysis in Sec.~\ref{sec:rank1} in several ways.
In Sec.~\ref{sec:conformal-matter}, we analyze  compactifications with Stiefel-Whitney twists of minimal $(\fg,\fg)$ conformal matters.
We identify the resulting $4d$ theories with a twisted class S construction.
In Sec.~\ref{sec:quivers}, we consider $6d$ theories whose F-theory description consists of a chain of $-1$, $-2$, \ldots, $-2$ curves decorated with $\fusp$ and $\fsu$ gauge algebras. The cases discussed in this section can be studied by first reducing to $5d$.  
In Sec.~\ref{sec:Z5}, we consider cases, with the same F-theory description as before, but for which the reduction to $5d$ does not appear to be as useful. Specifically, the global symmetry in these cases is of the form $\frac{G}{\bZ_5}$, so the Stiefel-Whitney twists are valued in $\bZ_5$.

We  have an appendix~\ref{app:A} where we carefully study the periodicity of the phase of the $4d$ scalar $u$ which comes from the $6d$ tensor multiplet, in the presence of the Stiefel-Whitney twist. 
The result uncovered there plays an important role when computing the $4d$ central charges from the $6d$ data.

We also have an appendix~\ref{app:B} which can be read independently from the rest of the paper. 
There,  we discuss how we can compute the Coulomb branch operator dimensions of $4d$ theories obtained by $T^2$ compactifications of $6d$ very Higgsable theories, with and without Stiefel-Whitney twists.
Our derivation of the rule is not foolproof; instead we provide many examples.

\section{Illustrative examples: rank-one theories in $4d$}
\label{sec:rank1}

We shall first start with some simple examples where the resulting $4d$ theories can be identified with rank-one SCFTs classified in \cite{Argyres:2015ffa,Argyres:2015gha,Argyres:2016xmc,Argyres:2016xua}. Many of these rank-one SCFTs can be constructed through a $6d$ compactification by compactifying the $6d$ $(2,0)$ theory on a three punctured Riemann sphere; for example, the rank-1 theory with $\fsu(4)$ flavor symmetry was originally found in  \cite{Chacaltana:2016shw} using this technique. 
For the case of the standard $E_8$ theory of Minahan and Nemeschansky \cite{Minahan:1996cj}, we can also realize it by a torus compactification of the rank-one E-string theory \cite{Ganor:1996pc}. The constructions we consider in this section can be thought of as a generalization of this latter technique for more involved cases. 

Particularly, the $6d$ SCFTs we consider here are generalizations of the rank-one E-string theory by the addition of vector multiplets and hypermultiplets. 
These 6d theories can be engineered in string theory in various ways: as orbifolds of the E-string SCFT\cite{DelZotto:2014hpa,Mekareeya:2017jgc}, as the low-energy theory on D$6$-branes partitioned by an NS$5$-brane in the presence of an O$8^-$ plane\cite{Brunner:1997gf,Hanany:1997gh}, or as F-theory compactifications with a $-1$ curve as the base\cite{Heckman:2013pva}. Unlike the E-string theory though, they can also be realized as UV completions of $6d$ gauge theories, and we shall mostly use this description.

\subsection{4d $\fusp(10)$ theory from 6d $\fsu(2)+10F$}
\label{sec:2.1}
In the first case we consider the $6d$ SCFT having an $\fsu(2)$ gauge theory with ten doublet hypermultiplets as its low-energy description on the tensor branch. This SCFT has an $\fso(20)$ global symmetry which is manifested in the gauge theory as the symmetry rotating the hypermultiplets. In what follows we shall need to determine also the global structure of the symmetry group. For this we need to consider the operator spectrum of the SCFT. First, we have the moment map operators which are in the adjoint representation of $\fso(20)$. There is one additional Higgs branch chiral ring generator that is not visible from the gauge theory. This operator is in a chiral spinor of $\fso(20)$ and has as its lowest component a scalar in the $\bold{4}$ of $\fsu(2)_R$\cite{Hanany:2018uhm,Kim:2018bpg}. This structure is consistent with the group being $\Spin(20)/\bZ_2$ where the $\bZ_2$ is the center element acting non-trivially on the vector and the other chiral spinor.

We next consider compactifications of this theory on a torus to $4d$. This should reduce to $4d$ \Nequals2  theories. The direct torus compactification of this theory was considered in \cite{Ohmori:2015pua}, and leads to a strongly coupled SCFT which can also be described in class S. However, the global symmetry of this theory appears to be $\Spin(20)/\bZ_2$ which is not simply connected. Thus, we can consider compactifications with a non-trivial $\Spin(20)/\bZ_2$ bundles on the torus. 
Such compactifications were considered in \cite{Witten:1997bs} in the case of Type I string with its $\Spin(32)/\bZ_2$ symmetry, and called compactifications without vector structure.
Equivalently, this can be described as turning on two almost commuting holonomies. More specifically, when compactifying theories we are free to turn on holonomies along the two cycles of the torus. These, however, are forced to commute due to the homotopy properties of the torus, since we want the background to be flat. Yet, as the group here is $\Spin(20)/\bZ_2$, we  can also support two holonomies that commute up to the $\bZ_2$ center element, as it acts trivially on everything. Such a flat background is not connected to the trivial background (in the moduli of flat connections), and therefore this has a profound effect on the compactified theory.

In the case at hand, we can construct such a compactification as follows. We consider the subalgebra\footnote{In this paper we abuse the symbol $\times$ to denote the direct sum of Lie algebras.}  $\fsu(2) \times \fusp(10)$ of $\fso(20)$, where the vector of $\fso(20)$ decomposes as $\bold{20} \rightarrow (\bold{2},\bold{10})$. We can then consider the following two holonomies, where $J_{10 \times 10}$ is the antisymmetric matrix preserved by $\fusp(10)$:
\be
P=
\begin{pmatrix}
  J_{10 \times 10} & 0 \\
  0 &  -J_{10 \times 10}
\end{pmatrix}
,\qquad  Q=
\begin{pmatrix}
  0 &  J_{10 \times 10} \\
  - J_{10 \times 10} & 0
\end{pmatrix} .
\ee   
These commute up to the center of $\SO(20)$. Furthermore, these break the $\fsu(2)$ part of the $\fsu(2) \times \fusp(10)$, and thus should lead to a $4d$ theory with $\fusp(10)$ global symmetry. We conjecture that this theory is in fact the rank-one $\fusp(10)$ SCFT first proposed in \cite{Argyres:2007tq}. We next give a few supporting tests in favor of this conjecture. More shall be performed in the upcoming sections.

The first striking feature here is the Coulomb branch. The $6d$ theory on a circle is expected to have a two dimensional Coulomb branch: one from the $\fsu(2)$ vector on the circle and one from the tensor. However, as we shall now argue, the contribution from the vector on the circle is projected out by the holonomies. First note that while there are no gauge invariant operators in the vector of $\fso(20)$, there is a gauge variant state, the quarks. These are also in the doublet of the gauge $\fsu(2)$ so the action of the center of $\SO(20)$ can be canceled by the center of the gauge $\SU(2)$. As a result for the consistency of the whole configuration, we must also incorporate two holonomies for the $\SU(2)$ gauge group that commute up to its $\bZ_2$ center. These break the $\fsu(2)$ gauge symmetry completely and so the vectors are projected out in this reduction, and we indeed expect a rank-one 4d theory.

We can extract yet more information from this analysis. As we mentioned the quarks can be thought of as transforming in the $(\bold{2},\bold{2},\bold{10})$ of $\fsu(2)_\text{gauge} \times \fsu(2) \times \fusp(10)$ where we have decomposed $\fso(20)$ to its $\fsu(2) \times \fusp(10)$ subgroup. In this presentation we have almost commuting holonomies for both $\fsu(2)_\text{gauge}$ and $\fsu(2)\subset \fsu(2)\times \fusp(10) \subset \fso(20)$ which can be thought of as residing in the diagonal $\fsu(2)$. Under the decomposition of $\fsu(2)_\text{gauge} \times \fsu(2)$ to the diagonal group, we have $(\bold{2},\bold{2})\rightarrow \bold{3}\oplus \bold{1}$. The $\bold{3}$ should be projected out in this reduction, however, the $\bold{1}$ should survive. This implies that on a generic point on the Coulomb branch of the resulting $4d$ theory we expect to have ten half-hypermultiplets. This indeed agrees with the properties of the rank-one $\fusp(10)$ SCFT\cite{Argyres:2016xua}.

Another supporting test is matching the spectrum of gauge invariant operators. The rank-one $\fusp(10)$ SCFT has, as low-lying $\fusp(10)$ charged operators, the moment map operators, in the adjoint of $\fusp(10)$ and in the $\bold{3}$ of $\fsu(2)_R$, and a Higgs branch chiral ring operator in the $\bold{132}$ of $\fusp(10)$ and in the $\bold{4}$ of $\fsu(2)_R$. These can be seen from the superconformal index, which we can evaluate from one of its class S realization, like the ones in \cite{Chacaltana:2011ze}, using the Hall-Littlewood formula in \cite{Lemos:2012ph}. Alternatively, it can be argued from the dualities in \cite{Argyres:2007tq} involving this SCFT.   
These should come from the basic $6d$ multiplets we mentioned. Under its $\fsu(2)\times \fusp(10)$ subalgebra, the $\fso(20)$ adjoint and chiral spinor\footnote{The group $\fso(20)$ has two chiral spinors which decompose differently. Here we are interested in the representation appearing in the $6d$ SCFT which is singled out as the one invariant under the $\bZ_2$ center of the $\SU(2)$.} decompose as:
\begin{align}
\bold{190} &\rightarrow (\bold{3},\bold{1}) \oplus (\bold{3},\bold{44}) \oplus (\bold{1},\bold{55}) , \\
\bold{512} &\rightarrow (\bold{5},\bold{10}) \oplus (\bold{3},\bold{110}) \oplus (\bold{1},\bold{132}) .
\end{align} 
We indeed see that projecting to the $\fsu(2)$ invariant sector we obtain results consistent with  the known operator spectrum of the rank-one $\fusp(10)$ theory.

In Sec.~\ref{sec:rank1central} and \ref{subsec:rank1-5d} we will preform additional consistency checks on this and related proposals. First, we will compute the central charge and match it against the $4d$ result, under some mild assumptions regarding the structure of the $4d$ Coulomb branch. We shall further study the reduction by first going to $5d$, where we shall relate this to a twisted reduction of a $5d$ SCFT of the type studied in \cite{Zafrir:2016wkk}. From the $5d$ picture one can perform additional tests that complement those performed here. But first we would like to consider the other cases related to rank-one $4d$ SCFT.

\subsection{4d $\fsu(4)$ theory from 6d $\fsu(3)+12F$}
\label{sec:2.2}
For our next case we shall consider the $6d$ SCFT with a low-energy description on the tensor branch as an $\fsu(3)$ gauge theory with twelve hypermultiplets in the fundamental representation. This $6d$ SCFT has an $\fsu(12)$ algebra as its local global symmetry, while the baryonic $\fu(1)$ is anomalous. We can again wonder what is the global structure of the group. Perturbatively, we have two basic states charged under the global symmetry. The first ones are the moment map operators in the adjoint of $\fsu(12)$ and in the $\bold{3}$ of $\fsu(2)_R$. The second ones are the baryons which are in the $\bold{220}$ of $\fsu(12)$ and in the $\bold{4}$ of $\fsu(2)_R$.

We also expect non-perturbative states in the $6d$ SCFT. One way to handle these is to compactify to lower dimensions on circles, and read these from the resulting theory. For the case at hand, we can consider torus compactifications to $4d$, where it is known that this theory reduces to a class S theory associated with an $A_8$ $(2,0)$ theory on a three punctured sphere with punctures $[1^9]$, $[3^3]$ and $[5,4]$ \cite{Mekareeya:2017jgc}. We can now use the Hall-Littlewood formula describing its Higgs branch chiral ring \cite{Gadde:2011uv,Gaiotto:2012uq,Gaiotto:2012xa} to discover that there is also an operator in the $\bold{924'}$ of $\fsu(12)$ and in the $\bold{5}$ of $\fsu(2)_R$, that is not visible in the $6d$ gauge theory description of the SCFT. Note that as this operator is part of the Higgs branch chiral ring, and as torus compactifications preserve the Higgs branch, the analogous operator must also appear in $6d$.     

From the spectrum of states, we see that the global symmetry appears to be $\SU(12)/\bZ_3$. Thus, we can again consider a $4d$ torus reduction with two holonomies that commute up to the $\bZ_3$ action. It is straightforward to find matrices possessing such commutation relations\cite{tHooft:1979rtg}: 
\be
P=
\begin{pmatrix}
  I_{4 \times 4} & 0 & 0 \\
  0 & \omega^4 I_{4 \times 4} & 0  \\
	0 & 0 & \omega^{8} I_{4 \times 4}  
\end{pmatrix}
,\qquad  Q=
\begin{pmatrix}
  0 & I_{4 \times 4} & 0 \\
	0 & 0 & I_{4 \times 4} \\
	I_{4 \times 4} & 0 & 0 
\end{pmatrix} ,
\label{eq:SU12holonomy}
\ee
where we have used $\omega$ for the generator of $\bZ_{12}$. Here we have used the $\fsu(3)\times \fsu(4)$ subalgebra of $\fsu(12)$ and embedded the matrices $P,Q$ of \eqref{eq:SU3holonomy} in the $\fsu(3)$ part. Due to the holonomies, the $\fsu(3)$ part will be broken and we expect a $4d$ theory with $\fsu(4)$ global symmetry.

We propose that the theory generated in this way is the rank-one $\fsu(4)$ SCFT. We can motivate this conjecture in a similar manner to the previous case. First we note that consistency also requires us to incorporate two almost commuting holonomies for the gauge $\fsu(3)$. This implies that the $\fsu(3)$ gauge symmetry is projected out in the reduction and we indeed expect a rank-one theory. We can further use the $6d$ gauge theory description to compute the matter spectrum at a generic point on the Coulomb branch. Here, the quarks are in the $(\bold{3},\bold{3},\bold{4})$, of $\fsu(3)$ gauge and $\fsu(3)\times \fsu(4)\subset \fsu(12)$. We again embed the holonomies in the diagonal $\SU(3)$ where the $\bold{3}$ of $\fsu(3)$ gauge is mapped to the $\bar{\bold{3}}$ of $\fsu(3)$ global. This is dictated as the holonomies in these two groups must cancel one another when acting on the quarks. Under this subgroup, the $(\bold{3},\bold{3})$ decomposes to $\bold{8} \oplus \bold{1}$, and particularly, we have an $\fsu(3)$ singlet. This implies that on a generic point on the Coulomb branch we expect $4$ hypermultiplets, transforming in the fundamental of $\fsu(4)$, in addition to a single vector multiplet. This agrees with the results of \cite{Argyres:2016xua} for the rank-one $\fsu(4)$ SCFT.

We can also compare the spectrum of operators expected from the $6d$ SCFT. The low-lying Higgs branch chiral ring operators in the rank-one $\fsu(4)$ SCFT were computed by \cite{Chacaltana:2016shw} who found them to be: the moment map operators in the $\bold{15}$ of $\fsu(4)$ and the $\bold{3}$ of $\fsu(2)_R$, an operator in the $\bold{20''}$ of $\fsu(4)$ and the $\bold{4}$ of $\fsu(2)_R$ and an operator in the $\bold{50}$ of ${\fsu(4)}$ and the $\bold{5}$ of $\fsu(2)_R$.

From our previous discussion, we identified some of the representations expected to exist in the $6d$ SCFT. From group theory, these are known to decompose from $\fsu(12)$ to $\fsu(3)\times \fsu(4)$ as:
\begin{align}
\bold{143} &\rightarrow (\bold{15},\bold{1})  \oplus (\bold{15},\bold{8})  \oplus (\bold{1},\bold{8}) ,\\
\bold{220} &\rightarrow (\bold{20''},\bold{1})  \oplus (\bold{20},\bold{8})  \oplus (\bar{\bold{4}},\bold{10}) ,\\
\bold{924'} & \rightarrow  (\bold{50},\bold{1})  \oplus (\bold{64},\bold{8})  \oplus (\bold{10},\bold{10})  
 \oplus  (\bar{\bold{10}},\bar{\bold{10}})  \oplus (\bold{6},\bold{27}) .
\end{align}
We now see that projecting to the $\fsu(3)$ invariant sector indeed reproduces the expected Higgs branch chiral ring generators of the rank-one $\fsu(4)$ SCFT.
We can again perform several additional tests that we shall go more in-depth on in  Sec.~\ref{sec:rank1central} and \ref{subsec:rank1-5d}.

\subsection{4d $\fsu(3)$ theory from 6d $\fsu(4)+12F+1AS$}
\label{sec:2.3}
As our last rank-one example, we consider the $6d$ SCFT with a low-energy description on the tensor branch as an $\fsu(4)$ gauge theory with an antisymmetric hyper and twelve hypermultiplets in the fundamental representation. This SCFT should have an $\fsu(2)\times \fsu(12)$ global symmetry. Note that the $\fsu(2)$ flavor symmetry acts on the antisymmetric of $\fsu(4)$, since it is a strictly real representation. 
We can again consider the global structure of the symmetry group, for which we consider the charged operator spectrum of the theory. Besides the moment map operators, we expect there to be an operator in the $(\bold{2},\bold{66})$ of $\fsu(2)\times \fsu(12)$ in the $\bold{4}$ of $\fsu(2)_R$, and one in the $(\bold{1},\bold{495})$ of $\fsu(2)\times \fsu(12)$ in the $\bold{5}$ of $\fsu(2)_R$. 

There may also be non-perturbative operators that are only visible in the SCFT. Again these can be conveniently inferred from the class S theory resulting from the torus compactification of the $6d$ SCFT using the known formula for the Hall-Littlewood index of such theories. In this case, the corresponding class S theory is of type $A_9$ on a three punctured sphere with punctures $[1^{10}]$, $[4,3^2]$ and $[5^2]$ \cite{Mekareeya:2017jgc}. From this we see that there should also be an operator in the $(\bold{2},\bold{924'})$ of $\fsu(2)\times \fsu(12)$ and in the $\bold{6}$ of $\fsu(2)_R$.

The above spectrum is consistent with the global symmetry being $(\SU(2)\times \SU(12))/\bZ_4$, where the generator of the $\bZ_4$ acts as $-1$ of the $\SU(2)$ and the element $\omega^3$ for the $\SU(12)$, where $\omega$ is the generator of the center of $\SU(12)$.
We can then again consider a torus compactification with almost commuting holonomies. Here we can use for the $\fsu(12)$ the holonomies:
\be
P=
\begin{pmatrix}
  I_{3 \times 3} & 0 & 0 & 0 \\
  0 & \omega^3 I_{3 \times 3} & 0 & 0  \\
	0 & 0 & \omega^{6} I_{3 \times 3} & 0 \\
	0 & 0 & 0 & \omega^{9} I_{3 \times 3}
\end{pmatrix}
,\qquad  Q=
\begin{pmatrix}
  0 & I_{3 \times 3} & 0 & 0 \\
	0 & 0 & I_{3 \times 3} & 0 \\
	0 & 0 & & I_{3 \times 3} \\
	I_{3 \times 3} & 0 & 0 & 0 
\end{pmatrix} ,
\ee
and
\be
P=
\begin{pmatrix}
  i & 0 \\
  0 & -i
\end{pmatrix}
,\qquad  Q=
\begin{pmatrix}
  0 & i \\
	i & 0 
\end{pmatrix} , \label{HolSU2}
\ee
for the $\SU(2)$. Here we have again used the $\fsu(3)\times \fsu(4)$ subgroup of $\fsu(12)$, but have now embedded the holonomies in the $\fsu(4)$ part. These are expected to completely break the $\fsu(2)$ and the $\fsu(4)$ subalgebra of $\fsu(12)$. Thus, we expect a $4d$ theory with $\fsu(3)$ global symmetry.

We conjecture that this theory is the rank-one $\fsu(3)$ SCFT speculated to exist in \cite{Argyres:2016xua}. As there is no previously known construction of this SCFT, to our knowledge, aside from the Seiberg-Witten geometry, we know less about this SCFT, and so there are less things we can check. However, this does suggest that this theory exist, since  we have proposed a method to realize it by a dimensional reduction.

One test that we can perform is to consider the effect of the compactification on the gauge symmetry. Similarly to the previous case we need to accompany the holonomies for the flavor symmetry with ones for the gauge symmetry, when we consider reductions on a generic point on the tensor branch. The analysis is similar to the previous cases, and we find that the gauge symmetry is completely projected out, and that on a generic point, we have three hypermultiplets transforming as the fundamental of $\fsu(3)$. This agrees with the results of \cite{Argyres:2016xua}.

We can also consider the operator spectrum expected based on the $6d$ picture. Here we have nothing to compare against so this is not a check, but rather a prediction of the construction. Besides the moment map operators, which just lead to the moment map operators of the $\fsu(3)$ global symmetry, the only other operator that can contribute is the one in the $(\bold{1},\bold{495})$ of $\fsu(2)\times \fsu(12)$. Under the decomposition of $\fsu(12)\rightarrow \fsu(3)\times \fsu(4)$ we have:
\be
\bold{495}\rightarrow (\bold{45},\bold{3})  \oplus (\bold{20'},\bold{6})  \oplus (\bold{15},\bold{15})  \oplus (\bold{1},\bold{15'}) .
\ee 
From this we see that we expect the rank-one $\fsu(3)$ SCFT to have a Higgs branch chiral ring generator in the $\bold{15'}$ of $\fsu(3)$ and in the $\bold{5}$ of $\fsu(2)_R$. 

We next discuss other consistency checks that we can perform on our proposal. 

\subsection{Calculating the $4d$ central charges}
\label{sec:rank1central}

We can compute the central charges expected from our construction. The strategy of the computation is to use the structure of the Coulomb branch as well as the results in \cite{Shapere:2008zf} to compute the central charges. Such a strategy was previously used to compute the central charges for $4d$ theories resulting from the direct compactification of this type of $6d$ SCFTs in \cite{Ohmori:2015pua}, and our discussion here can be thought of as a generalization of this to the case with a non-trivial Stiefel-Whitney class.

\subsubsection{Structure of the $u$-plane}
We consider the Coulomb branch of the $6d$ theory on the torus with non-trivial Stiefel-Whitney class. This should be a one dimensional complex space, whose spanning parameter should be related to the $6d$ tensor multiplet on the circle. It should contain various singularities. Physically, we expect one at the origin, associated with the $4d$ theory we expect to be the result of the compactification. Additionally we expect at least two more, associated with points where the Kaluza-Klein modes associated with the two circles become massless. Finally, the behavior at large Coulomb branch vevs should be describable by the $6d$ low-energy gauge theory on the torus. Now, holomorphy relates the behavior at infinity to that around the other singularities. Thus, if we understand the singularity structure we can extract the behavior at the singularity associated with the $4d$ theory point from which we can extract the central charges of the $4d$ theory.

To determine the singularity structure of the $4d$ theory we shall rely on a self-consistency argument. Particularly, we shall assume that the singularity structure is that of the $4d$ theories we proposed, and use this to derive the central charges. If our proposal is correct then we must get the central charges expected from these theories. For the case at hand, all three theories we discussed are forced to have the Coulomb branch structure of the Minahan-Nemeschansky $E_8$ theory at the conformal point.

We denote by $u$ the Coulomb branch parameter. 
The Coulomb branch can be called the $u$-plane as is customary. 
As a theory obtained by compactifying a $6d$ theory on $T^2$, the asymptotic infinity of the $u$-plane has the metric of a semi-infinite cylinder.
We assume that there is a Kodaira singularity of type $II^*$ at the origin $u=0$.
This means that the Coulomb branch can be more properly called the $u$-cigar; we will use the common terminology $u$-plane interchangeably.

The placement of Kodaira singularities on the $u$-plane determines the metric.
For the asymptotic infinity to be cylindrical, we need two more singularities of Kodaira type $I$.
The local physics at each of these singularities is that of a free hypermultiplet becoming massless\cite{Ganor:1996pc},
signaling that a Kaluza-Klein mode is becoming massless.

\subsubsection{From functions on the $u$-planes to central charges}
To determine the central charges we use the results of \cite{Shapere:2008zf}, that relates them to properties of the topologically twisted theory. Particularly, consider the \Nequals2 theory on a curved manifold coupled to a non-trivial flavor symmetry background field, and twist it by turning on also a non-trivial $\fsu(2)_R$ connection that is locked to be the same as that of one of the $\fsu(2)$ groups in the spacetime $\fsu(2)\times \fsu(2) = \fso(4)$ rotation symmetry.
The path integral of the twisted theory then has a schematic form given by:
\be
Z = \int [du][dq] A(u)^{\chi} B(u)^{\sigma} C(u)^{n} e^{-S} ,
\ee       
where we use $\chi$ and $\sigma$ for the Euler characteristic and signature of the manifold respectively and $n$ for the instanton number associated with the flavor background. 
We also use $u$ for a gauge invariant coordinate on the Coulomb branch, and denote by $q$ other fields that may exist in the system. 

In superconformal theories, the central charges can be related to the R-charges of $A$, $B$ and $C$ as found in \cite{Shapere:2008zf}:
\begin{align}
a_\text{SCFT} &= \frac{1}{4} R[A] + \frac{1}{6} R[B] + a_\text{generic},\label{Rela}\\
c_\text{SCFT} &= \frac{1}{3} R[B] + c_\text{generic}, \label{Relc}\\
k_\text{SCFT} &= R[C] + k_\text{generic}, \label{Relk}
\end{align}
where $a_\text{SCFT}$, $c_\text{SCFT}$, $k_\text{SCFT}$ are the central charges of the SCFT at the origin,
$R[A]$, $R[B]$, $R[C]$ are the R-charges of the objects enclosed in $[\cdots]$,
and $a_\text{generic}$, $c_\text{generic}$ and $k_\text{generic}$ are the contributions of the multiplets that exist on a generic point on the Coulomb branch.

In order to determine the central charges then we need to determine $R[A]$, $R[B]$ and $R[C]$ for the $4d$ interacting superconformal point. To do this we first consider the behavior at large $u$. As we previously mentioned, this region should be well captured by the $6d$ gauge theory. It is known that the latter contains a Green-Schwarz coupling 
which couples the self-dual tensor $\sB$ to  some combination $I_\text{GS}$ of characteristic classes of $4$-manifold. 
Therefore, when dimensionally reduced in the topologically twisted theory, this term should give the desired coupling between $u$, which is associated with the scalars coming from the tensor multiplet on a torus, and the characteristic classes $\chi$, $\sigma$ and $n$.

In our case the Green-Schwarz term can be determined from the gauge theory description, and is known to be of the form:
\be
I_\text{GS} = d\, c_2 (R) + \frac{1}{4} p_1 (T) + \frac{1}{4} \Tr(F^2_F) - \frac{1}{4} \Tr(F^2_G).\label{GS}
\ee 
Here we use $c_2 (R)$ for the second Chern class of the R-symmetry bundle and $p_1 (T)$ for the first Pontryagin class of the tangent bundle. 
The coefficient $d$ depends on the gauge theory description corresponding to the $-1$ curve in question.
If it only has $\fg$ gauge multiplets and standard hypermultiplets without any strongly-coupled SCFT charged under $\fg$, $d$ is known to be given simply by $d=-h^\vee(\fg)$.
More details and examples can be found in \cite{Ohmori:2014kda}.

In the topologically twisted theory the relation above translates to
\be
I_\text{GS} = -\frac{d}{2} \chi + \frac{3}{4} (1-d) \sigma +  n_F -  n_G ,
\ee 
where we have used $\sigma = \frac{1}{3} p_1 (T)$, $n = \frac{1}{4} \Tr(F^2)$ and $c_2 (R) = -\frac{1}{2}\chi - \frac{1}{4} p_1 (T)$ due to the twisting.
From this the dependence of $A$, $B$ and $C$ around infinity was found in \cite{Ohmori:2014kda} to be
\be
A^{\chi} B^{\sigma} C^{n} \sim (u_\text{naive})^{-\frac{d}{2}\chi} (u_\text{naive})^{\frac{3}{4}(1-d)\sigma} (u_\text{naive})^{n_F} 
\ee
where $u_\text{naive}$ is related to the scalar $a$ in the tensor multiplet and the tensor field $\sB$ via \begin{equation}
u_\text{naive}=\exp\left(a + 2\pi i \int_{T^2} \sB \right)
\end{equation} in the semi-classical region, in the standard normalization that $\sB$ has period 1.

A crucial difference in our case is the following. 
Due to the Green-Schwarz coupling \eqref{GS}, under a $G$ gauge transformation by $g$, we have
\begin{equation}
\delta_g \int B= \delta_g \int  CS(F_G) \label{OOO}
\end{equation}
where $CS(F_G)=\frac14\Tr(A_G dA_G + (2/3)A_G^3) $ is the Chern-Simons term
normalized so that it is defined modulo 1 for simply-connected $G$.
In the previous analysis in \cite{Ohmori:2014kda},
the background holonomy on $T^2$ was trivial, 
and therefore the right hand side of \eqref{OOO} was also trivial.

On $T^2$,  we now  have nontrivial holonomies $P$, $Q$ satisfying $PQ=QP \omega$,
where $t$ specifies the order of the Stiefel-Whitney class.
A large gauge transformation by $\omega$ is known to shift $\int \frac14 CS(F_G)$ by $2\pi/t$ \cite{Borel:1999bx}.\footnote{%
This is due to the fact that the Chern-Simons invariant of the $\SU(t)/\bZ_t$ bundle on $T^3$ with holonomies $P$, $Q$, $\omega$ along the three 1-cycles is $1/t$.
This is also related to the fact that the minimal instanton number for $\SU(t)/\bZ_t$ gauge configuration is $1/t$.
For more details, see Appendix~\ref{app:A}.
}
This means that the periods of $\sB$ given by $\int_{T^2} \sB=0$ and $=2\pi/t$ have to be identified,
and therefore the correct coordinate $u$ on the Coulomb branch is given by $u=(u_\text{naive})^t$.
With this, we obtain
\be
A^{\chi} B^{\sigma} C^{n} \sim (u^{\frac{d}{2t}})^{\chi} (u^{\frac{3}{4 t}(d-1)})^{\sigma} (u^{-\frac{1}{t}})^{n_F} .
\ee 

We can now proceed with the calculation of the central charges. Under the phase rotation $u \rightarrow e^{2 \pi i} u$ we have that: 
\be
A^{\chi} B^{\sigma} C^{n} \rightarrow e^{2 \pi i (\frac{d \chi}{2 t} + \frac{3}{4 t}(d-1)\sigma -\frac{n_F}{t})} A^{\chi} B^{\sigma} C^{n} .
\ee
 Around each singularity we have a superconformal theory and can represent this phase shift as a $\frac{2\pi}{R[u]}$ in $\fu(1)_R$, where $R[u]$ is the R-charge of $u$. We then have:
\be
A^{\chi} B^{\sigma} C^{n} \rightarrow e^{2 \pi i (\frac{R[A]}{R[u]}\chi + \frac{R[B]}{R[u]}\sigma + \frac{R[C]}{R[u]} n_F} ) A^{\chi} B^{\sigma} C^{n} .
\ee

We know the behavior of $A$, $B$, $C$ on two out of three singularities on the $u$-plane. We also know their asymptotic behavior at $u\to \infty$.
This allows us to determine the behavior of $A$, $B$, $C$ at the singularity which supports a nontrivial SCFT.
Let us carry this out.

\subsubsection{The formulas}
For the flavor central charge $k_\text{SCFT}$, only the SCFT point contribute.
Using (\ref{Relk}), we find:
\begin{equation}
k_\text{SCFT}-k_\text{generic}=\frac{12 I}{t},
\label{rec-k}
\end{equation}
where $I$ is the embedding index of the $4d$ global symmetry into the $6d$ global symmetry, and we have used the fact that $R[u] = 12$ for the Minahan-Nemeschansky $E_8$ theory. 

 We can apply the same procedure to $B$ in order to determine $c_\text{SCFT}$. Now, we need to take the contribution from all three points, and using (\ref{Relc}), we find:
\begin{equation}
c_\text{SCFT} - c_\text{generic}=\frac{3(1-d)}t -1.
\label{rec-c}
\end{equation}

The computation of $a$ is reduced to the computation of $R[A]$ via the relation $(2a-c)_\text{SCFT}-(2a-c)_\text{generic} = R[A]/2$ which follows from \eqref{Rela} and \eqref{Relc}.
The function $A(u)$ is not a simple rational function, but $A(u)/A_\text{E-string}(u)$ is.
Using the known behavior of $A_\text{E-string}(u)$, we conclude that \begin{equation}
(2a-c)_\text{SCFT}-(2a-c)_\text{generic} = -\frac{3d}{t}-\frac12.
\label{rec-nv}
\end{equation}

\subsubsection{Examples}
We can now apply the formalism for the calculation of the central charges
in the three examples we discussed so far.
We note that the crucial numbers $(d,t,I)$ appearing in the formulas above are given as follows:\begin{equation}
\begin{array}{c||c|c|c|c}
\text{4d flavor} & \text{6d gauge $\fg$} & -d=h^\vee(\fg) & t & I \\
\hline
\fusp(10) & \fsu(2) & 2 & 2 & 1 \\
\fsu(4) & \fsu(3) & 3 & 3 & 3 \\
\fsu(3) & \fsu(4) & 4 & 4 & 4
\end{array}
\end{equation}
Below, each of the three examples are referred to by the 4d flavor symmetry.

Let us first study the flavor central charge $k$.
From our previous arguments, we know that there are charged matter multiplets on a generic point on the Coulomb branch, specifically, a fundamental half-hyper for $\fusp(10)$ case and a fundamental hyper for the $\fsu(4)$ and $\fsu(3)$ cases. This gives $k^{\fusp(10)}_\text{generic} = 1$ and $k^{\fsu(4)}_\text{generic} = k^{\fsu(3)}_\text{generic} = 2$. 
Combining all of these and plugging in to \eqref{rec-k}, we find that $k_{\fusp(10)} =7$ and $k_{\fsu(4)} = k_{\fsu(3)} = 14$. This agrees with the results of \cite{Argyres:2016xua}.
This computation also shows that the $\fusp(10)$ flavor symmetry has Witten's anomaly,
since the half-hyper on a generic point on the Coulomb branch has.
Again this is consistent with the known fact about the $\fusp(10)$ theory.

Next, let us study the conformal central charge $c$.
At a generic point we expect there to be a free vector as well as $5$, $4$ and $3$ free hypers for the $\fusp(10)$, $\fsu(4)$ and $\fsu(3)$ cases respectively. From all this, we determine via \eqref{rec-c} that $c_{\fusp(10)} = \frac{49}{12}$, $c_{\fsu(4)} = \frac{42}{12}$ and $c_{\fsu(3)} = \frac{38}{12}$. Again this is in agreement with the results of \cite{Argyres:2016xua}.

In the rank-1 cases, once $c_\text{SCFT}$ is known, the $a_\text{SCFT}$ central charge is essentially guaranteed to work due to our working assumptions. The reason is that $a = \frac{c}{2} + \frac{n_v}{8}$, where $n_v$ is the effective number of vector multiplets which in turn is determined by the dimension of the Coulomb branch operator by the standard formula $n_v = R[u]-1 = 11$ by assumption.  
Of course we can use the general formula \eqref{rec-nv}.
Either way, we conclude $a_{\fusp(10)} = \frac{41}{12}$, $a_{\fsu(4)} = \frac{25}{8}$ and $a_{\fsu(3)} = \frac{71}{24}$,
again reproducing the results of  \cite{Argyres:2016xua}.

\subsection{Relation to $5d$ compactifications}
\label{subsec:rank1-5d}
We can attempt to identify the 4d theory directly by reducing the $6d$ theory first to $5d$ and then proceed to $4d$. In the first step, we reduce the $6d$ SCFT along a circle with a holonomy which we will choose to be $P$ in the pair of almost commuting holonomies. This should lead us to some $5d$ theory. This type of relations for the $6d$ SCFTs we are considering here was studied in \cite{Hayashi:2015fsa,Zafrir:2015rga,Ohmori:2015tka,Hayashi:2015zka}, and we can use their results here. The important feature here is that this $5d$ theory can be described via a brane web system, which we will use to represent the resulting $5d$ theory\footnote{In some cases, these $5d$ theories can also be described by $5d$ gauge theories. This requires special choices of the holonomies that may differ from the choice we make here. In any case, for our purposes the brane web description suffices.}.

\begin{figure}
\center
\includegraphics[width=1\textwidth]{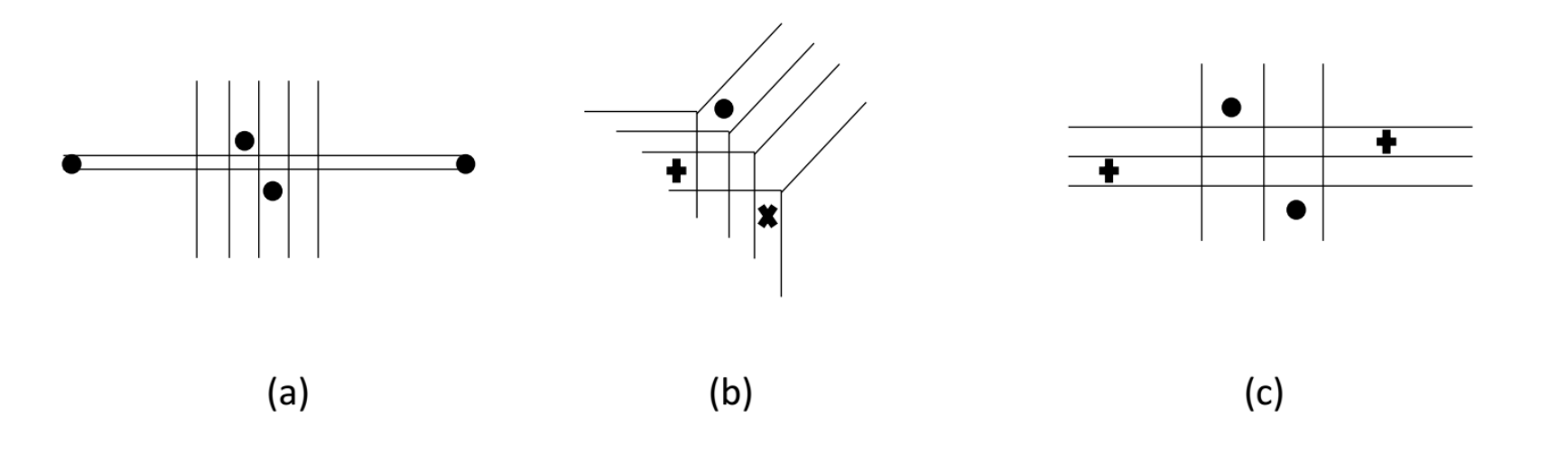} 
\caption{The brane webs describing the effective $5d$ theory that results from the compactification of the previously discussed $6d$ SCFTs. Specifically, (a) is for the $\fsu(2)+10F$ SCFT, (b) for the $\fsu(3)+12F$ SCFT and (c) for the $\fsu(4)+1AS+12F$ SCFT. Here we use black dots for D$7$-branes, black $+$ for $(0,1)$ $7$-branes and black $X$ for $(1,1)$ $7$-branes. The configuration are drawn such that they show the discrete symmetry permuting the flavor symmetry, here manifested by a rotation in the plane of the web accompanied with the appropriate $\SL(2,\bZ)$ transformation.}
\label{Webs}
\end{figure}

We have drawn the webs describing these theories in figure \ref{Webs}. Here we used a specific representation involving $7$-branes without any $5$-brane ending on them. The reason for this choice is twofold. First, it makes the web easier to draw, as pulling out all the $7$-branes tend to lead to infinitely spiraling webs\cite{Kim:2015jba}. Another reason is that this choice makes the discrete symmetry that will be modded out manifest. This will be important to us later, when we reduce to $4d$, as the second holonomy essentially acts on this system via the action of this discrete symmetry.   

Next we take the zero radius limit. Without the holonomies, this is known to amount to taking one of the free $7$-branes in figure \ref{Webs} to infinity, which leads to a $5d$ SCFT\cite{Zafrir:2015rga}. However, here we have non-trivial holonomies whose effect we must take into account. The first holonomy is expected to break the global symmetry to one whose non-abelian sector is $\fsu(10)$, $\fsu(4)^3$ and $\fsu(3)^4$ for $\bZ_2$, $\bZ_3$ and $\bZ_4$ respectively. Also there must be the analogous discrete symmetry acting on the configuration. Taking both of these into account, it is strongly likely that the correct theory to consider is the one where we take to infinity not only one $7$-brane, but all those related to it by the discrete action. This leads to the webs in figure \ref{Webs} but with all the free $7$-branes removed, which now describe $5d$ SCFTs possessing the desired non-abelian symmetry, as well as the desired discrete symmetry acting on it. We can now consider compactifying these SCFTs with the presence of the second holonomy. Due to its action of permuting factors of the global symmetry, it is most natural to interpret this as a compactification of the SCFT with a twist under the relevant discrete symmetry element.     

Such type of reductions were considered in \cite{Zafrir:2016wkk}, which studied them in detail for the $\bZ_2$ and $\bZ_3$ cases. The results of \cite{Zafrir:2016wkk} are consistent with ours. Particularly, theories \ref{Webs} (a) and (b) without the free $7$-branes, were indeed identified with the rank-one $\fusp(10)$ and $\fsu(4)$ theories, respectively. This identification was motivated by various consistency conditions. Some, like analyzing the operator spectrum, mimic the tests we preformed in $6d$. However, some of the tests, particularly studying mass deformations and dualities, are different, and can be used to lend further support to our identification.

One test one can perform is to study dualities. Particularly, we consider gauging part of the global symmetry of the $5d$ SCFT, in a manner preserving the discrete symmetry. We then consider compactifying the resulting theory with a twist, which should lead to some $4d$ theory. We can easily analyze the resulting theory in the limit of infinitely weak gauging, where this should just describe the analogous gauging in $4d$ of the theory resulting from the twisted compactification. Now, if the gauging is conformal in $4d$ then we get a $4d$ SCFT. In these cases we can consider moving to the $5d$ SCFT point, that is infinitely strong gauging, and then continue towards negative gauge coupling\footnote{The gauge coupling can be identified as a vev to the scalar in a background vector multiplet coupled to the topological $\fu(1)$ symmetry that exist for every non-abelian gauge factor\cite{Seiberg:1996bd}. Such scalars are known to be real and so can be both positive and negative.}. When doing this the original description in $5d$ breaks down, but taking the limit of negative but weak  coupling, a new description usually emerges, where the same $5d$ SCFT, in that region of parameter space, is described as a potentially different weak gauging of some $5d$ SCFT. We can next consider reducing this limit to $4d$, where again we expect to find the analogous gauging in $4d$ of the theory resulting from the twisted compactification.

In the cases we are considering, the varied gauge coupling is mapped to an exactly marginal deformation in $4d$, and we expect these two different descriptions to be two weakly coupled limits of the same underlying SCFT in different corners of the conformal manifold, that is they should be dual to each other. Indeed in all cases studied in \cite{Zafrir:2016wkk}, when one side is conformal in $4d$, so is the other side. Furthermore the dualities derived in this way can sometimes be mapped to known dualities. 

\begin{figure}
\center
\includegraphics[width=1\textwidth]{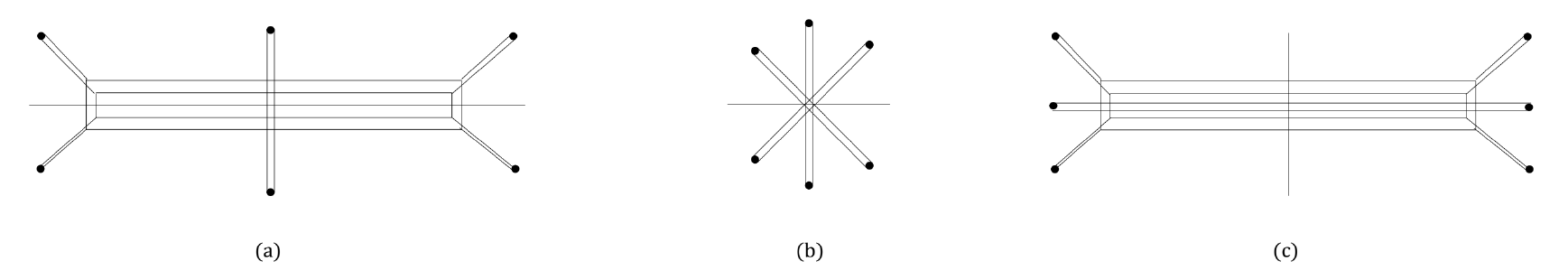} 
\caption{The brane web one gets by gauging part of the global symmetry of the $5d$ $\fsu(10)$ SCFT we get through the compactification of the $6d$ $\fsu(2)+ 10F$ theory with the appropriate holonomy. Specifically, we gauge both of the $\fsu(5)$ subgroups inside the $\fsu(10)$ global symmetry group, with an hypermultiplet in the antisymmetric for both $\fsu(5)$ groups. In (a), we show the resulting brane web in the limit where both the $\fsu(5)$ gauge groups are weakly coupled. In (b), we show the resulting brane web at the SCFT point. In (c) we show the web, after an S-duality, in the limit of negative but weak (in absolute value) coupling. At this point the system is described by a dual $\fsu(6)\times \fsu(6)$ gauge theory, with a bifundamental hypermultiplet, and a fundamental hyper plus a half-hyper in the $\bold{20}$ for each $\fsu(6)$ gauge group.
\label{fig:continuation}}
\end{figure}

For instance in the case of the rank-one $\fusp(10)$ theory, one can gauge the $5d$ SCFT with an $\fsu(5)$ gauge theory with an antisymmetric hypermultiplet, which leads to a conformal theory in $4d$. The resulting 5d brane web is shown in figure \ref{fig:continuation} (a). In the middle of the web we have the $5d$ $\fsu(10)$ SCFT, which we get from the webs in figure \ref{Webs} (a) by removing the free $7$-branes, where we have also performed an S-duality with respect to that figure. On the $5$ internal D5-branes connecting to it on each side live the two $\fsu(5)$ gauge groups. The configurations at the two ends of the web provide the antisymmetric hyper. This can be seen by realizing the antisymmeric hyper using an $O7^-$-plane with a stuck NS$5$-brane and then decomposing it to a pair of $7$-branes, see \cite{BZ}. 

The coupling constant of the $\fsu(5)$ gauge groups is inversely proportional to the lengh of the D$5$-branes on which they live, and so they become weakly coupled at large distances. Alternatively, when all the $5$-branes intersect at a point, we get the $5d$ SCFT associated with this system. This is shown in figure \ref{fig:continuation} (b).

We can next continue past infinite coupling by performing S-duality and separating the dual web. The resulting system is shown in figure \ref{fig:continuation} (c). It describes an $\fsu(6)\times \fsu(6)$ quiver theory with a bifundamental hypermultiplet, which is visible in the middle of the web. The configurations at the edges of the web provides matter for both $\fsu(6)$ groups, which turns out to be a fundamental hyper as well as a half-hyper in the $\bold{20}$\footnote{See \cite{Hayashi:2019}. Another way to argue this is to use the class S fixture in \cite{TT} associated with a collection of free hypers transforming under its $\fsu(6)$ global symmetry as a half-hyper in the $\bold{20}$ and three fundamentals. This can be lifted to a brane web configuration, using the results of \cite{Benini:2009aa}, which is precisely the one appearing at the edges of the web in figure \ref{fig:continuation} (c) with the addition of two D$7$-branes on top of the D$5$-branes. Removing these corresponds to removing two fundamentals under the $\fsu(6)$ leading to our identification.}. When compactified with a twist, the latter description gives a $4d$ $\fsu(6)$ gauge theory with a fundamental hyper, a symmetric hyper, and a half-hyper in the $\bold{20}$. This duality matches the duality found in \cite{Argyres:2007tq}, see the entry 6 of their Table 2.

One can also use this to find other dualities for the rank-one $\fusp(10)$ theory as well as dualities for the rank-one $\fsu(4)$ theory. However, these usually lead to  new dualities, and so, while that makes them quite interesting, they cannot be used for consistency checks. For the $\fsu(3)$ case, the situation is less favorable, as the $\fsu(3)$ here cannot be conformally gauged.   

Another test one can perform is to study mass deformations. First there are mass deformations leading to other SCFTs, like the one taking the rank-one $\fusp(10)$ SCFT to the rank-one $\fsu(2)\times \fusp(6)$ SCFT. 
These can be seen from the web as mass deformations, preserving the discrete symmetry, that sends the $5d$ SCFT to another $5d$ SCFT. This can then be compared against the expected flow patterns of these theories\cite{Argyres:2015gha,Argyres:2016xua}. One can then employ some of the previous consistency conditions also for the study of these cases, which then provide further consistency checks.

We can also study mass deformation of the $5d$ SCFT into a $5d$ gauge theory. The end point here is an IR free theory, which in this case can be analyzed explicitly, and this should map to flows of the $4d$ SCFT to IR free $4d$ gauge theories. For instance, consider the $5d$ SCFT whose twisted compactification we associate with the rank-one $\fusp(10)$ theory. It has two deformations sending it to $5d$ gauge theories. One leads to an $\fsu(3)$ gauge theory with $8$ fundamental hypermultiplets and Chern-Simons level $0$, while the other gives an $\fsu(2)\times \fsu(2)$ quiver gauge theory with a bifundamental hyper and $3$ doublet hypers for each group. In the first case, the discrete symmetry acts as charge conjugation, and the result of the reduction should be an $\fso(3)$ gauge theory with $4$ hypermultiplets in the vector representation. In the second case, the discrete symmetry acts as quiver reflection, and the result of the reduction should be an $\fsu(2)$ gauge theory with $3$ fundamental hypermultiplets and one symmetric hyper. Both of these deformations agree with those expected from the Seiberg-Witten curve\cite{Argyres:2015gha}. 

The same game can be also played for the rank-one $\fsu(4)$ theory, where here the mass deformation leads to a $4d$ $\fsu(2)$ gauge theory with $2$ fundamental hypermultiplets and one half-hyper in the $\bold{4}$, see \cite{Zafrir:2016wkk} for the details. These again agree with results derived from the Seiberg-Witten curve\cite{Argyres:2016xua}. For the rank-one $\fsu(3)$ theory the situation is more complicated, as here the mass deformations sending the $5d$ SCFT to a gauge theory are not consistent with the discrete symmetry, unless one also goes on the Coulomb branch.  

\subsection{Behavior under holonomy changes}
\label{sec:hol}

So far we considered compactification with two holonomies that commute up to a central element of the simply connected version of the global symmetry group that acts trivially on the matter content of the $6d$ SCFT. We have also written specific examples for such holonomies. However, these are not the only possible choices. In fact, it is known that there is a continuous space of possible holonomies, all having the same commutation relation, that can be generated by conjugating the pairs we have used by some otherwise commuting elements, see \cite{Borel:1999bx}. The specific choices we have used were such that the maximal symmetry possible was preserved. However, generic holonomy choices are expected to break that symmetry to its maximal torus. 

In this subsection, we study the effects of this on the $4d$ theories. At first sight, the interpretation of this is straightforward. Holonomies in circle compactification are related to mass deformations associated to that symmetry. So we expect there to be an SCFT at the maximal symmetry point, and changes in the holonomies are then interpreted as mass terms under the associated symmetry. Generic configurations indeed break the global symmetry to its maximal torus.

This interpretation goes through without problems when the global symmetry is of type $\fsu$. However, for more general global symmetries there are interesting phenomena that should be addressed. Particularly, in these cases it is known that depending on the choice of holonomies, different global symmetries can be preserved\cite{Borel:1999bx}. Stated differently,  in those cases there could be several maximal symmetry points each preserving a different symmetry. We can then inquire what happens if we compactify using this choice.

\subsubsection{The $\fso(10)$ point of 6d $\fsu(2)+10F$ theory}
The only case of relevance to us in this section is the $6d$ $\fsu(2)+10F$ SCFT. Here the global symmetry is $\Spin(20)/\bZ_2$, and such phenomena are relevant. 
In this case it is convenient to introduce a physical manifestation of this, originally introduced by \cite{Witten:1997bs}.
There,  torus compactification of type I string theory with almost commuting holonomies for its $\Spin(32)/\bZ_2$ symmetry was studied, and called  compactifications without vector structure. A useful result there is that this configuration is T-dual to a system of $8$ D$7$-branes in the presence of three O$7^-$-planes and one O$7^+$-plane. The choice of holonomies that we mentioned are here dual to the location of the D$7$-branes on this space. 

In our case, the group we consider is of the same type though somewhat smaller, so we can essentially use the same construction to visualize the space of the possible holonomy, but using only $5$ D$7$-branes. Specifically, We can choose to place all $5$ D$7$-branes on the O$7^+$-plane. The symmetry on the $7$-branes then is $\fusp(10)$, and this corresponds to the $\fusp(10)$ preserving holonomies that we have previously introduced. Under an infinitesimal change in the holonomies, corresponding to moving and separating the D$7$-branes from the O$7^+$-plane, the symmetry will be broken down to $\fu(1)^5$. The best one can preserve is $\fu(1)\times \fsu(5)$, where one moves all five D$7$-branes together away from the O$7^+$-plane. This corresponds to the appropriate mass term in the $4d$ SCFT.

The curious observation here is that we can next move all five D$7$-branes to one of the O$7^-$-planes. This then corresponds to a choice of holonomies preserving $\fso(10)$. Indeed, one can use the $(\OO(2)\times \OO(10))/\bZ_2$ subgroup of $\SO(20)$ to construct a pair of almost commuting holonomies by embedding them in $\OO(2)$\cite{Witten:1997bs}:
\be
P=
\begin{pmatrix}
  I_{10 \times 10} & 0 \\
  0 & - I_{10 \times 10}
\end{pmatrix}
,\qquad
 Q=
\begin{pmatrix}
  0 &  I_{10 \times 10} \\
  - I_{10 \times 10} & 0
\end{pmatrix} . \label{HolO2}
\ee  

 We can then ask what does this implies for the $4d$ theories. First, we note that the zero area limit of the torus, where we expect the $4d$ SCFT, corresponds to the infinite dual torus, where the distance between the orientifold planes becomes infinite. Thus, at this limit, we cannot move continuously to the other choice, which is encouraging as we do not expect such a phenomena in the $4d$ SCFT. However, we can still consider the same compactification but with the $\fso(10)$ preserving holonomies. We now expect a rank-one theory with $\fso(10)$ global symmetry. We next argue that this is just the IR free $\fsu(2)+5F$ $4d$ gauge theory. For this we shall use the previous study of the structure of the Coulomb branch.

The general idea is as follows. We first assume that the structure of the Coulomb branch for the torus compactified theory, is as previously stated, that is we have the SCFT point where the rank-one $\fusp(10)$ theory emerges, as well as two other singular points where there is just a free hyper. As previously stated, on a finite area torus we should be able to continuously vary the holonomy so as to reach the point where the $\fso(10)$ should emerge. From the brane picture, this transition must involve a mass deformation breaking the $\fusp(10)$ group to $\fu(1)\times \fsu(5)$. This splits the type $II^*$ singularity, associated with the rank-one $\fusp(10)$ theory, to three singularities of type $I_5$, $I_4$ and $I_1$. Here the singularity $I_5$ is associated with five massless free hypers with charge $1$, $I_4$ with one massless free hyper with charge $2$, and $I_1$ with one massless free hyper with charge $1$, and is the same as the two other singularities that exists beside the $\fusp(10)$ one. Here the charges are all under the $\fu(1)$ gauge group that remains on generic points on the Coulomb branch.

After splitting the singularity, we can consider moving the $I_5$ singularity and merging it with one of the $I_1$ singularities associated with a massless Kaluze-Klein mode. These can then form the type $I^*_1$ singularity associated with the $\fsu(2)+5F$ theory. The resulting picture then is that at the $\fso(10)$ preserving holonomy, the Coulomb branch should again have three singularities, one associated with the rank-one $\fso(10)$ theory and two where there is a massless free hyper, associated with the Kaluze-Klein mode, but here the two singularites are not the same rather differing by the charge of the massless hyper under the $\fu(1)$.

All that remains is to compute the central charges and check whether this indeed agrees with those of the $4d$ $\fsu(2)+5F$ theory. This can be done using the method we introduced in the previous subsection. First, we need to determine the matter content on a generic point on the coulomb branch. For this we again consider the $\fsu(2)\times \fso(10)$ half-hyper bifundamental. This feels the combination of the holonomies (\ref{HolSU2}) for $\SU(2)$ gauge and (\ref{HolO2}) for $\SO(20)$. As one contains $i$ while the other does not, it is clear that these act non-trivially on all the fields so there should be no additional fields on a generic point on the Coulomb branch\footnote{Recall that to get to this point we performed a mass deformation breaking $\fusp(10)$ to $\fu(1)\times \fsu(5)$. This has the effect of equal non-zero masses to the free hypers that existed on a generic point on the Coulomb branch, explaining their absence now.}.

We can now proceed with the calculation. The behavior at infinity is the same as before so the computation is only modified by the different generic contribution, the change in the R-charge of the singularity and the change to the other singularity. Repeating the calculation we now find:
\be
k_{SO(10)} = - R[u] \frac{I}{t} = 4,\qquad
c_{SO(10)} = R[u] (\frac{d-1}{4t}-\frac{1}{12}) + \frac{1}{6} = \frac{4}{3} .
\ee 
where we again assumed that $R[u]$ is the R-charge expected from the singularity type, here $R[u] = 4$.
This matches the central charges of the $4d$ $\fsu(2)+5F$ theory.

\subsubsection{Other choices for  6d $\fsu(2)+10F$ theory}
\label{sec:USp4}
There are other natural intermediate choices of holonomies which interpolate the two extremes $\fusp(10)$ and $\fso(10)$ we already studied.
They preserve $\fso(10-2n)\times \fusp(2n)$ symmetry, for $n=0,1,2,3,4,5$.
We already discussed the $n=5$ case as the main example,
and then the $n=0$ case just now. 
Let us examine the other cases $n=1,2,3,4$ in turn.
It is useful to break the symmetry further to $\fu(1)\times \fsu(5-n)\times \fusp(2n)$.

\paragraph{$\fu(1)\times \fsu(4)\times \fusp(2)$:}
To study the case $n=1$, we deform the $\fusp(10)$ theory by a particular mass term, so that the original $II^*$ singularity splits to an $I^*_0$ and an $I_4$.
This deformation was already discussed in \cite{Argyres:2016xua}: 
the $I_4$ is a point with four free hypers with $\fsu(4)$ symmetry, 
and the $I^*_0$ is for an $\fsu(2)$ gauge theory with a triplet hypermultiplet (also known as the \Nequals4 super Yang-Mills) with $\fusp(2)$ symmetry.
In total, we have two $I_1$, one $I_4$ and one $I^*_0$ singularity on the $u$-cigar.

We can now enhance $\fsu(4)$ to $\fso(8)$ by collapsing these two $I_1$s and the $I_4$ to form an $I^*_0$ singularity.
In the end, on the Coulomb branch, we simply have an $I^*_0$ singularity with $\fso(8)$ symmetry and another $I^*_0$ with $\fusp(2)$ symmetry.

\paragraph{$\fu(1)\times \fsu(3)\times \fusp(4)$:}
To generalize, we note that the $I^*_{n-1}$ singularity can support the $\fusp(2n)$ symmetry,
realizing $\SO(3)$ gauge theory with $n$ hypermultiplets in the triplet representation.
In the standard normalization for $\SU(2)$ gauge theory, the beta function from $n$ hypers in the triplet is equal to $4n$ hypers in the doublet, so this would usually be associated to the $I^*_{4(n-1)}$ singularity.
By changing $\SU(2)$ to $\SO(3)$, we rescale the normalization of the electric charge by $2$,
changing the monodromy by a factor of $2^2$, ending up with the $I^*_{n-1}$ singularity.

Then, we can have a $\fu(1)\times \fsu(5-n)\times \fusp(2n)$ point with four singularities on the Coulomb branch, two $I_1$'s, one $I_{5-n}$ and an $I^*_{n-1}$.
One $I_1$ can be pushed to either $I_{5-n}$ or $I^*_{n-1}$ to get a more interesting situation.

Let us take $n=2$, and push one $I_1$ to $I_3$.
We have $I_1$, $I_4$ and $I^*_1$ on the Coulomb branch.
As $I_4$ has  $\fsu(4)\simeq \fso(6)$ and $I^*_1$ has $\fusp(4)$, we can identify this configuration as the $\fso(6)\times \fusp(4)$ point.

If we instead push $I_1$ to $I^*_1$, we can obtain the Coulomb branch with  $I_1$, $I_3$ and $IV^*$. 
The last singularity can be identified with the rank-1 $\fusp(4)\times \fu(1)$ symmetry, whose Coulomb branch parameter $u$ has dimension 3.

\paragraph{$\fu(1)\times \fsu(2)\times \fusp(6)$: }
Our general discussion, applied to $n=3$, gives 
the Coulomb branch with two $I_1$, one $I_2$ and one $I^*_2$ singularity with $\fu(1)\times \fsu(2)\times \fusp(6)$ symmetry. 

We expect that there is an $\fso(4)\times \fusp(6)$ point.
Pushing $I_1$ and $I_2$ does not achieve this goal. 
Luckily, it is known that  the  $III^*$ singularity we get by pushing $I_1$ to $I^*_2$ carries the $\fsu(2)\times \fusp(6)$ symmetry, whose Coulomb branch parameter $u$ has dimension $4$ \cite{Argyres:2016xua}.
Together with the additional $I_2$ singularity we already had, we indeed realize a point with $\fsu(2)\times \fsu(2)\times \fusp(6)$ symmetry.

\paragraph{$\fu(1)\times \fusp(8)$: }
Finally, taking $n=4$ in our general discussion, we have a $u$-cylinder with three $I_1$'s and one $I^*_3$.
This setup has $\fu(1)\times \fusp(8)\simeq \fso(2)\times \fusp(8)$ symmetry.
Tuning the $\fu(1)$ mass parameter and collapsing $I_1$ to $I^*_3$, we have the $II^*$ singularity with $\fusp(10)$ symmetry, which was the starting point of our discussion.

\section{Generalizations}
\label{sec:general}

Having illustrated the main idea and methods for the simple case of rank-one theories, we next explore the implications of this construction to other cases. 
Concretely, we consider two sets of examples: 
in Sec.~\ref{sec:conformal-matter} we study $6d$ minimal conformal matter SCFTs,
and in Sec.~\ref{sec:quivers} and Sec.~\ref{sec:Z5} we study very Higgsable theories whose gauge theory description on a generic point on the tensor branch is a quiver gauge theory of $\fsu$ and $\fusp$ gauge algebras.

\subsection{Minimal Conformal matter} 
\label{sec:conformal-matter}

\subsubsection{General discussions}
In this section, we study the generalization to the minimal conformal matter family of $6d$ SCFTs\cite{DelZotto:2014hpa}. These are the SCFTs living on $1$ M$5$-brane probing a $\mathbb{C}^2/\Gamma_{\fg}$ singularity, where $\Gamma_{\fg}$ is a finite subgroup of $\SU(2)$ associated with an ADE algebra $\fg$. The latter are known to obey an ADE classification, and the global symmetry of the $6d$ SCFT is known to  be given locally by $\fg \times \fg$. The case of $\fg = A_{k-1}$ is just a collection of $k^2$ free hypers transforming as the bifundamental of $\fsu(k) \times \fsu(k)$. The case of $\fg = D_{k+4}$ can be described by a low-energy $\fusp(2k)$ gauge theory with $2k+8$ fundamental hypermultiplets at a generic point on its one dimensional tensor branch.

The cases of $E_6$, $E_7$ and $E_8$ are more involved, and these can no longer be described as gauge theories by going to the tensor branch. Nevertheless, on the tensor branch these can be described by gauge theories connected to E-string theories via gauging of part of their global symmetry. Specifically, the case of $\fg = E_{6}$ can be described as the UV completion of an $\fsu(3)$ gauge theory weakly gauging an $\fsu(3)$ subalgebra of $E_8$ belonging to two copies of the rank-one E-string theory. For $\fg = E_{7}$, the low-energy description now involves an $\fsu(2) \times \fso(7) \times \fsu(2)$ gauge theory with two half-hypers, one in the $(\bold{2}, \bold{8}, \bold{1})$ and one in the $(\bold{1}, \bold{8}, \bold{2})$, and two copies of the rank-one E-string theory. The latter are connected via weakly gauging an $\fsu(2)$ subalgebra of their $E_8$ global symmetry, each by one of the gauge $\fsu(2)$. We will not discuss the $E_8$ case here as it will not be relevant in what follows.

A natural question then is what is the actual global symmetry of these $6d$ SCFTs. One way to tackle this problem is to study the operator spectrum similar to the studies we performed in the previous check. Some preliminary checks of this kind were done in \cite{Kim:2018lfo}, for not necessarily minimal conformal matter theories\footnote{See also \cite{Bah:2017gph} for a related discussion, including cases with a non-trivial Stiefel-Whitney class, in the $\fg=A$ case.}. It was concluded there that the global symmetry appears to $\frac{G \times G}{Z_G}$, where $Z_G$ is the center of the group $G$. For the minimal case of $\fg = D_{k+4}$, the symmetry algebra enhances to $\fso(4k+16)$, and the corresponding group is thought to be $\Spin(4k+16)/\bZ_2$\cite{Hanany:2018uhm,Kim:2018bpg}, generalizing the $k=1$ case discussed in the previous section. This group indeed contains $(\Spin(2k+8)\times \Spin(2k+8))/Z_{\Spin(2k+8)}$ as a subgroup.

As a result the global symmetry group in these SCFTs appears to be non-simply-connected, and it is thus natural to consider torus compactifications with non-trivial Stiefel-Whitney class. We shall next propose an identification of the resulting $4d$ theories with certain class S theories, for a suitable choice of holonomies. To achieve this, it is convenient to first recall the analogous connection for the torus compactifications with trivial Stiefel-Whitney class. 

In \cite{Ohmori:2015pua}, the theory that the $(\fg,\fg)$ minimal conformal matter on $T^2$ with trivial flavor holonomy flows into was identified with class S theory of type $\fg$ with two full punctures and one simple puncture:
\begin{equation}
	\text{minimal $(\fg,\fg)$ conformal matter on $T^2$}
	\to
	T_{{\fg}}(\text{full},\text{full},\text{simple}).
	\label{without}
\end{equation}
Here,$T_\fg(A,B,C)$ stands for the class S theory of type $\fg$ on a sphere with three punctures $A$, $B$ and $C$.
We also recall here that  the full puncture is the puncture with full $\fg$ symmetry, and the simple puncture is obtained by partially closing the full puncture with the subregular orbit of $\fg$.

We would like to generalize the statement into the current case of a torus compactification with a Stiefel-Whitney twist.
First let us briefly review how to derive the relation \eqref{without} without a twist by a Stiefel-Whitney class.
For the case without a twist, the resulting class S construction can be read off from the duality
\begin{equation}
	\text{M-theory on $\mathbb{C}^2/\Gamma_\fg\times T^2$} \leftrightarrow \text{type IIB on $\mathbb{C}^2/\Gamma_\fg\times S^1$}.
	\label{eq:MIIBdual}
\end{equation}
This duality can be understood by first going down to IIA from M-theory and then take the T-dual along the remaining $S^1$ direction, or as an M-theory/F-theory T-duality with the trivial elliptic fiber.
The $6d$ theory can be realized by putting an M$5$-brane on the ALE space, and in the dual side we get a $D$3-brane. Therefore we get a $6d$ \Nequals{(2,0)} on a cylinder probed by a D$3$-brane. Regarding the two infinites of the cylinder as two full punctures of class S theory, we expect that the resulting $4d$ theory is the class S theory of a sphere with two full punctures and one some puncture $P$ realized as a D$3$-brane.
The puncture $P$ is identified with the simple puncture in \cite{Ohmori:2015pua} and 
this identification was consistent with anomaly calculations and other consistency checks done in the reference.

To generalize into the case with a twist with a Stiefel-Whitney class, 
we need to have a generalization of \eqref{eq:MIIBdual}.
Let us assume that the left hand side of \eqref{eq:MIIBdual} still has a IIB dual description with a compactification manifold $X$ even with a twist with a Stiefel-Whitney class. 
$X$ needs to be a fibration over $S^1$, and the fiber $Y$ should be hyperk\"ahler to preserve the supersymmetry. Further, in the limit where the $S^1$ shrinks, the manifold $X$ should degenerate into $\mathbb{C}^2/\Gamma'$ for some $\Gamma'$, since in the dual M-theory side the $T^2$ becomes infinitely large and the Stiefel-Whitney class should become irrelevant.
With these constraint, we propose that
\begin{equation}
	X = (\mathbb{C}^2/\Gamma_{\tilde{\fg}}\times S^1)/\mathbb{Z}_t,
\end{equation}
where $t$ is the order of the Stiefel-Whitney class, $\mathbb{Z}_t$ acts on $S^1$ as a $2\pi/t$ shift, and $\mathbb{C}^2/\Gamma_{\tilde{\fg}}$ is the other ALE space admitting a $\mathbb{Z}_t$ hyperk\"ahler isometry and satisfying
\begin{equation}
	(\mathbb{C}^2/\Gamma_{\tilde{\fg}})/\mathbb{Z}_t = \mathbb{C}^2/\Gamma_\fg.
	\label{eq:ALEquotient}
\end{equation}
Possible hyperk\"ahler isometries of an ALE space, which are equivalent to possible extensions of the corresponding finite group inside $\SU(2)$, was studied in \cite{reid1985young};
the final list is given in  Table~\ref{tab:tildeG}.\footnote{%
Note that, as mentioned in the footnote 15 of \cite{Witten:1997kz}, the original table in p.376 of \cite{reid1985young} has a typo, which is corrected in our table~\ref{tab:tildeG}.
}
From there, we can read off $\tilde{\fg}$ with given $\fg$ and $t$.

Summarizing, our proposal for the generalization of \eqref{eq:MIIBdual} is
\begin{equation}
	\text{M-theory on $\mathbb{C}^2/\Gamma_\fg\times T^2_{w_2}$} \leftrightarrow \text{type IIB on $(\mathbb{C}^2/\Gamma_{\tilde \fg}\times S^1)/\mathbb{Z}_t$},
	\label{eq:MIIBdualSW}
\end{equation}
where $T^2_{w_2}$ means the torus with the Stiefel-Whitney twist.

\begin{table}[]
	\centering
	$\begin{array}{c|c|c||c|c}
		\fg & t & \tilde{\fg} & \fh & o\\
		\hline
		A_{k-1} & t\mid k & A_{k/t-1} & A_{k/t-1} & 1\\
		D_k & 2 & A_{2k-5} & B_{k-2} & 2\\ 
		D_{2k} & 2 & D_{k+1} & C_{k}  & 2\\
		D_{2k+1} & 4 & A_{2k-2} & C_{k-1} & 2\\ 
		E_6 & 3 & D_4 & G_2 & 3\\
		E_7 & 2 & E_6 & F_4 & 2
	\end{array}$
	\caption{$\tilde{\fg}$ satisfying \eqref{eq:ALEquotient} for given $\fg$ and $t$.
	The second row corresponds to the Stiefel-Whitney class related to the element generating $\pi_1(\SO(2k))$, while the third row is for the non-trivial element in $\pi_1(\SO(4k)/\{1,-1\})$ that is not induced from $\pi_1(\SO(4k))$. The fourth row is for the element generating $\pi_1(\SO(4k+2)/\{1,-1\}) = \mathbb{Z}_4$.
	We also listed a maximal subalgebra $\fh$ which is preserved by the corresponding Stiefel-Whitney class of the adjoint type Lie group of $\fg$. 
	These $\fh$ agree with the flavor symmetry of the maximal twisted puncture of class S theory of type $\tilde \fg$ where the twist is of order $o$.
	}
	\label{tab:tildeG}
\end{table}

We can interpret the right hand side of \eqref{eq:MIIBdualSW} as the \Nequals{(2,0)} theory of type $\tilde \fg$ on $S^1$ with the outer automorphism twist, when $\fg=D_k,E_6,E_7$. 
An M$5$-brane probing the singularity in the M-theory side is dualized to a D$3$-brane in the type IIB side as before, and locally the situation around the D$3$ should be the same as in the untwisted case. So in the IR we expect that this D$3$-brane again behaves as a simple puncture. Hence, we get the prediction about the twisted compactification of minimal $(\fg,\fg)$ conformal matter for $G=D_N,E_6,E_7$, that is
\begin{equation}
	\text{minimal $(\fg,\fg)$ conformal matter on $T^2_{w_2}$}
	\to
	T_{\tilde{\fg}}(\underline{\text{full}},\underline{\text{full}},\text{simple}).
	\label{eq:CMtwist}
\end{equation}
As before, $T_{\tilde{\fg}}(A,B,C)$ means the class S theory of type $\tilde \fg$ on a sphere with punctures $A,B$ and $C$, and the underlined punctures are twisted punctures.
\footnote{ For the case with $\fg=E_6, r=3, \tilde{\fg}=D_4$, one of the $\underline{\text{full}}$ puncture has a twist by an order-3 element, while the other $\underline{\text{full}}$ has the inverse twist.} 
Here the symbol ``$\to$" means the theory on the left hand side flows to the right hand side with correctly chosen  holonomies and Coulomb branch parameters.
These class S theories with twisted punctures have been studied in detail in the literature, see e.g.~\cite{Chacaltana:2012zy,Chacaltana:2012ch,Chacaltana:2013oka,Chacaltana:2014nya,Chacaltana:2015bna,Chacaltana:2016shw,Tachikawa:2018rgw}.

Let us check that the flavor symmetry of the resulting theory $T_{\tilde \fg}(\underline{\text{full}},\underline{\text{full}},\text{simple})$ agrees with the field theoretical consideration.
The 6d original theory has the symmetry $\fg\times \fg$.
Let us consider a single factor of $\fg$.
From the field theoretical perspective, as we are introducing a nontrivial Stiefel-Whitney class, we break $\fg$ to some subgroup, which is determined by the particular choice of holonomies on $T^2$  representing a given Stiefel-Whitney class.
Among those subgroups, there are a number of maximal ones.
In Table~\ref{tab:tildeG}, we also tabulated such a maximal subalgebra $\fh\subset \fg$ preserved by the Stiefel-Whitney class, as can be found in Appendix A, Table 6 of \cite{Kac:1999gw}.%
\footnote{The said table contains $\fh= B_{k-1}$ instead of $\fh=C_{k-1}$ for the line $\fg=D_{2k+1}$, $t=4$. This is an example where there can be multiple maximal subalgebras preserved by different holonomies with the same Stiefel-Whitney class. The choice $\fh=C_{k-1}$ has previously appeared in the physics literature in \cite{Witten:2000nv}.}
Therefore the question is whether the twisted maximal puncture $\underline{\text{full}}$ has the flavor symmetry $\fh$.
The twisted maximal puncture of class S theory of type $\tilde \fg$, twisted by an outer automorphism of order $o$, has the flavor symmetry which is given by the Langlands dual of the subgroup of $\tilde \fg$ preserved by the corresponding automorphism, whose results are tabulated in Table 1 of \cite{Chacaltana:2012zy}. 
We indeed see that the symmetry $\fh$ preserved by the Stiefel-Whitney class of order $t$ is given by the flavor symmetry of the maximal puncture twisted by an outer automorphism of order $o$, where $o|t$.

\subsubsection{$(D,D)$ conformal matter}
\label{sec:DD}
The discussions we gave above was based on a chain of string dualities.
Let us confirm the outcome via a direct field-theoretical computations.
We start with the case of the $(D_n,D_n)$ conformal matter theory.
The analysis we just carried out above predicts the following:
\begin{itemize}
\item when $n=2n'$ is even, we will have the class S theory 
\begin{equation}
T_{D_{n'+1}}(\underline{\text{full}},\underline{\text{full}},\text{simple})
\label{Deven}
\end{equation}
The Coulomb branch parameters have dimensions $6$, $8$, $10$, \ldots, $n+2$.
\item when $n=2n'+1$ is odd, we will have the class S theory
\begin{equation}
T_{A_{2n'-2}}(\underline{\text{full}},\underline{\text{full}},\text{simple}).
\label{Dodd}
\end{equation}
The Coulomb branch parameters have dimensions $3$, $5$, $7$, \ldots, $n-1$.
\end{itemize}

Let us now analyze the compactification field-theoretically.
As already recalled above, on a generic point of its one-dimensional tensor branch, the low-energy theory is a $\fusp(2n-8)$ gauge theory with $2n$ fundamental hypers, with $\fso(4n)$ flavor symmetry.
We compactify the system on $T^2$ with a nontrivial $\bZ_2$ Stiefel-Whitney class for both the gauge and flavor symmetries, using the decomposition $\fsu(2)\times \fso(n-4)\subset \fusp(2n-8)$ and $\fsu(2)\times \fusp(2n) \subset \fso(4n)$.
We then have a 4d theory with a Coulomb branch parameter $u$, for whose generic value we have an infrared-free $\fso(n-4)$ gauge theory with $n$ hypermultiplets in the vector representation;
the flavor symmetry is $\fusp(2n)$.

The singularity structure of the space of $u$ and the central charges can be determined as described in Sec.~\ref{sec:rank1central}.
We easily compute $k_\text{generic}=n-4$, $c_\text{generic}=(2n^2-13n+22)/12$, $a_\text{generic}=(7n^2-53n+110)/48$.
Using $I=1$, $t=2$ and $d=-(n-3)$, we find \begin{equation}
k_\text{SCFT}=n+2,\quad 
c_\text{SCFT}=\frac{2n^2+5n-26}{12},\quad
a_\text{SCFT}=\frac{7n^2+19n-106}{48}.
\end{equation} The central charges $(a,c)$ can be also expressed using the effective number of vector multiplets and hypermultiplets as follows:\begin{equation}
(n_v)_\text{SCFT}=\frac12(n+6)(n-3),\qquad (n_h)_\text{SCFT}=(n+4)(n-2).
\end{equation}

When $n=2n'$ is even, these central charges are indeed those of the class S theory \eqref{Deven}, as can be computed from the results of \cite{Chacaltana:2012zy,Chacaltana:2013oka}.
When $n=2n'+1$ is odd, these central charges are \emph{not} those of the class S theory \eqref{Dodd} derived using string duality.
This is due to the following.

On the one hand, in the field theoretical analysis just performed, we used the $\bZ_2$ Stiefel-Whitney class of $D_{2n}$ which preserves the $\fusp(2n)$ subalgebra.
On the other hand, in the duality chain analyzed in the previous subsection, only the $D_n\times D_n$  subalgebra of $D_{2n}$ is visible.
Now the crucial fact is that the largest $\fusp$ algebra preserved by the $\bZ_2$ Stiefel-Whitney class of $\SO(2n)$ is  $\fusp(n)$ when $n$ is even, but $\fusp(n-3)$ when $n$ is odd; this fact is explained e.g.~in Appendix A of \cite{Kac:1999gw}.
This nicely matches the conclusion of our duality chain in the previous subsection:
we found two twisted full punctures of class S theory of type $D_{n'+1}$ or of type $A_{2n'-2}$, depending on whether $n=2n'$ is even or $n=2n'+1$ is odd,
where one twisted full puncture has the symmetry $\fusp(2n')=\fusp(n)$ in the former case and $\fusp(2n'-2)=\fusp(n-3)$ in the latter case.
The flavor symmetries of two twisted full punctures are known to combine, and give $\fusp(4n')=\fusp(2n)$ when $n$ is even and $\fusp(4n'-4)=\fusp(2n-6)$ when $n$ is odd.

Our interpretation when $n=2n'+1$ is odd is then as follows: the class S theory \eqref{Dodd} which we identified by a chain of string dualities
is the theory with $\fusp(2n)$ symmetry we discussed above
deformed by a mass term  preserving $\fusp(2n-6)$ symmetry.
We note that this class S theory \eqref{Dodd}, $T_{A_{2n'-2}}(\underline{\text{full}},\underline{\text{full}},\text{simple})$, was studied recently in \cite{Tachikawa:2018rgw}: it is also known as the $R_{2,2n'-2}$ theory, has $\fusp(4n'-4)\times \fu(1)=\fusp(n-6)\times \fu(1)$ flavor symmetry, and the following central charges:
\begin{equation}
k^{\fusp(n-6)}_\text{SCFT}=2n',\quad 
c_\text{SCFT}=\frac{8n'^2-6n'-1}{12},\quad
a_\text{SCFT}=\frac{14n'^2-9n'-4}{24}.
\label{xxx}
\end{equation} The effective number of vector multiplets and hypermultiplets are known to be \begin{equation}
(n_v)_\text{SCFT}=(2n'+1)(n'-1)\quad,
(n_h)_\text{SCFT}=(2n'-1)^2.
\end{equation}

Let us reproduce these numbers from the compactification.
We first need to understand the structure of the $u$-plane.
For this, we note that the case $n=5$, $n'=2$ was already treated in Sec.~\ref{sec:USp4}.
In that case, the flavor symmetry was broken from $\fusp(10)$ to $\fusp(4)\times \fu(1)\times \fsu(3)$,
and there are a $IV^*$ singularity carrying the $\fusp(4)\times \fu(1)$ symmetry,
an $I_3$ carrying $\fsu(3)$ symmetry, and an additional $I_1$ on the $u$-plane.

In the general case, then, the original flavor symmetry $\fusp(2n)$ is broken to $\fusp(2n-6)\times \fu(1)\times \fsu(3)$,
and we again have singularities of type $IV^*$,  $I_3$, $I_1$ respectively.
The $IV^*$ singularity carries the $\fusp(2n-6)\times \fu(1)$ theory, and $I_3$ should simply be the point where three hypers become massless.
On a generic point on the $u$-plane, we have an infrared-free $\fso(n-4)$ gauge theory with $n-3$ hypers in the vector representation.
We then have $k_\text{generic}=n-4$, $c_\text{generic}=(n^2-8n+17)/6$ and $a_\text{generic}=(7n^2-59n+134)/48$.

The analysis we performed which led to \eqref{rec-k}, \eqref{rec-c} and \eqref{rec-nv} can readily be repeated, and we obtain
\begin{align}
k_\text{SCFT}-k_\text{generic}&= \frac{6I}t,\\
c_\text{SCFT}-c_\text{generic}&= \frac{3(1-d)}{2t}-1,\\
(2a-c)_\text{SCFT}-(2a-c)_\text{generic}&= -\frac{3d}{2t}-\frac12.
\end{align}
Using  $I=1$, $t=2$ and $d=-(n-3)$, we indeed reproduce the central charges in \eqref{xxx}.

Let us end this section by providing a class S construction of the $T^2$ compactification of the $(D_n,D_n)$ conformal matter for odd $n=2n'+1$ with the Stiefel-Whitney twist, preserving the maximal $\fusp(2n)$ flavor symmetry.
To do this, we start from the $T^2$ compactification with the Stiefel-Whitney twist of the $(D_{n+1},D_{n+1})$ theory, which is $T_{D_{n'+2}}(\underline{\text{full}},\underline{\text{full}},\text{simple})$ as we already saw above.
We simply partially close one twisted full puncture $\underline{\text{full}}=\underline{[1^{2n'+2}]}$ to the next-to-full one $\underline{[2,1^{2n'}]}$.
The central charges of the resulting theory $T_{D_{n'+2}}(\underline{\text{full}},\underline{[2,1^{2n'}]},\text{simple})$ can be computed as in \cite{Chacaltana:2013oka}, and the values are consistent with its identification with one half hypermultiplet in the fundamental of $\fusp(2n'+2)$ together with interacting SCFT with the symmetry $\fusp(2n) =\fusp(4n'+2) \supset \fusp(2n'+2)\times \fusp(2n')$.
The case $n=5$, $n'=2$ appears as the entry 14 of the big table of Sec.~3.2.3 of \cite{Chacaltana:2013oka}.

\subsubsection{$(E,E)$ conformal matters}
\label{sec:EE}
Here we would like to check the proposals for $(E,E)$ conformal matters by matching the conformal anomalies.
Those theories have a multi-dimensional tensor branch.
For such cases, we need to inductively apply the analysis of subsection~\ref{sec:rank1central} as it was done in \cite{Ohmori:2015pua} for the untwisted case.
Although we could repeat the analysis of section \ref{sec:rank1central} for individual cases, 
it is more convenient to derive the relations between the increases of the 4d central charges $k_\text{SCFT}-k_\text{generic}$, $c_\text{SCFT}-c_\text{generic}$, $a_\text{SCFT}-a_\text{generic}$ and the 6d Green-Schwarz contribution for the 6d anomaly polynomial.\footnote{In \cite{Ohmori:2015pua}, the linear relation between $a$, $c$ and $k$ and the total anomaly polynomial of the 6d theory was derived. For the twisted case, however, we do not expect such a simple relation to exist.}
Here, we assume that at each step of the induction the Coulomb branch geometry (as long as the metric for $u$ is concerned) is that of the $T^2$ compactified E-string theory. Therefore, the analysis here does not apply to the cases considered in subsection~\ref{sec:hol}.

Let us denote the GS contribution for the 6d anomaly polynomial for a 6d theory $T$ by
\begin{equation}
	A_\text{GS}^T=\sum_{i,j}\frac12\Omega_{ij}I^i_\text{GS}I^j_\text{GS} \supset \alpha_\text{GS}^T p_1(T)^2 + \beta_\text{GS}^T p_1(T)c_2(R) + \kappa_\text{GS}^T p_1(T) \Tr F^2.
\end{equation}
Here $I^i_\text{GS}$ is the GS term for the $i$-th tensor multiplet, and $\Omega_{ij}$ is the corresponding kinetic matrix.\footnote{$\Omega_{ij}$ is the inverse of the negative of the intersection matrix of curves in the F-theory model.}
Suppose that the $l$-th node has $\Omega^{ll}=1$, then we define the 6d theory $\widehat{T}$ as the (possibly a product of) IR SCFT(s) given by activating the tensor vev for the $l$-th node. Then, we have the relation of the GS contributions:
\begin{equation}
	A_\text{GS}^T= A_\text{GS}^{\widehat{T}} + \frac12 \hat{I}^2,
\end{equation}
where $\hat{I}$ is the GS coupling for the $l$-th node when all the other tensor vevs are deactivated. $\hat{I}$ takes the form of
\begin{equation}
	\hat{I}_\text{GS} = d\, c_2 (R) + \frac{1}{4} p_1 (T) + \frac{1}{4} \Tr(F^2_F) - \frac{1}{4} \Tr(F^2_G).\label{GS2}
\end{equation}
This is the same as \eqref{GS}, but this time $d$ is not just the dual Coxeter number of the gauge algebra on the $l$-th node. See the reference~\cite{Ohmori:2014kda} for the detail.
We get the relations between the coefficients:
\begin{equation}
	\delta\alpha_\text{GS} =1/32,\quad
	\delta\beta_\text{GS} =d/4,\quad
	\delta\kappa_\text{GS} =1/16,
	\label{abgk}
\end{equation}
where $\delta\alpha = \alpha^T_\text{GS}-\alpha_\text{GS}^{\widehat{T}}$, $\delta\beta = \beta^T_\text{GS}-\beta_\text{GS}^{\widehat{T}}$ and $\delta\kappa = \kappa^T_\text{GS}-\kappa_\text{GS}^{\widehat{T}}$.

Let $a^T, c^T, k^T$ denote the 4d conformal anomalies of the twisted compactified theory of $T$, and $\delta a = a^T-a^{\widehat{T}}$, $\delta c = c^T-c^{\widehat{T}}$ and $\delta k = k^T-k^{\widehat{T}}$. Then, by the same analysis as in subsection~\ref{sec:rank1central}, we have
\begin{align}
	\delta k_\text{SCFT}-\delta k_\text{generic} &= \frac{12}t ,\\
	\delta c_\text{SCFT}-\delta c_\text{generic} &= \frac{3(1-d)}t -1,\\
	\delta (2a-c)_\text{SCFT}-\delta (2a-c)_\text{generic} &=\frac{-3d}{t}-\frac12,
\end{align}
where $t$ is the order of the Stiefel-Whitney twist.
Then, by induction, we get:
\begin{align}
	\label{eq:afrom6d}
	a_\text{SCFT}-a_\text{generic} &=(\frac{3}{2t}-\frac34)32\alpha_\text{GS} -\frac{12}t\beta_\text{GS},\\
	\label{eq:cfrom6d}
	c_\text{SCFT}-c_\text{generic} &=(\frac{3}{t}-1)32\alpha_\text{GS} -\frac{12}t\beta_\text{GS},\\
	\label{eq:kfrom6d}
	k_\text{SCFT}-k_\text{generic} &=\frac{192}{t}\kappa_\text{GS}.
\end{align}

As said above, the analysis here depends on the assumption that the Coulomb branch geometry at each step of the induction is that of $T^2$ compactified E-string theory without holonomy. This assumption does not hold when the tensor branch effective field theory of the 6d theory involves a node with $\Omega^{ll}=1$ without a gauge group paired to the node. Nevertheless, the formula derived above applies to such cases, due to the reason explained below.

For example, when the 6d theory is the minimal $(E_6,E_6)$ conformal matter, the theory contains two of such nodes. Such a node is a tensor vev deformation of a copy of the E-string theory. The Stiefel-Whitney twist to the $\frac{E_6\times E_6}{\mathbb{Z}_3}$ symmetry formally induces the Stiefel-Whitney twist to the $\frac{E_6\times SU(3)}{\mathbb{Z}_3}$ subgroup of the $E_8$ global symmetry of the E-string theory. 
However, this Stiefel-Whitney twist to the $\frac{E_6\times SU(3)}{\mathbb{Z}_3}$ subgroup can be conjugated into the Cartan elements inside $E_8$, since $E_8$ is simply-connected. Therefore, at the step of shrinking this node, we should get a theory which is a mass deformation of the Minahan-Nemeschansky $E_8$ theory.

The particular mass deformation we get should be able to be computed using the analysis in \cite{Ganor:1996pc, Eguchi:2002fc, Eguchi:2002nx}, but can be more easily guessed as follows. In the 4d, the $\mathbb{Z}_3$ center of the $E_6$ (or equivalently that of $\SU(3)$) remains as a (part of) global symmetry. Moreover, this acts on the Coulomb branch parameter $u$ by $u\to \omega u$ with $\omega^3=1$ as explained Appendix~\ref{app:A}; the argument there is for gauge symmetry and here we use the same argument for the global symmetry. Therefore, the Coulomb branch geometry should have $\mathbb{Z}_3$ isometry. Then a natural guess of the geometry is the 3-fold cover of the geometry of the $T^2$ compactified E-string without holonomy, branching at the $II^*$ singularity. The 3-fold branching make the $II^*$ singularity into the $I_0*$ singularity, and therefore we get the $\mathfrak{su}(2)$ gauge theory with four flavors.
The formulas \eqref{eq:afrom6d}, \eqref{eq:cfrom6d} and \eqref{eq:kfrom6d} automatically applies with $t=3$.

More generally, when we do the compactification with Stiefel-Whitney class twist in $\mathbb{Z}_t$, from an E-string node we get the $t$-fold cover of $II^*$ singularity. For $t=2,3,4,5,6$, the corresponding 4d theory is the Minahan-Nemeschansky $E_6$ theory, the $\mathfrak{su}(2)$ gauge theory with four flavors, the Argyres-Douglas $A_2$ theory, the Argyres-Douglas $A_0$ theory, and a free vector (no singularity), respectively. These $t$'s are the only cases we will see in this paper.
For all of these cases the formulas \eqref{eq:afrom6d}, \eqref{eq:cfrom6d}, \eqref{eq:kfrom6d} hold as stated. Therefore, we can apply those formulas for the 6d theories including E-string nodes.

\paragraph{$(E_6,E_6)$ conformal matter:}
The minimal $(E_6,E_6)$ conformal matter admits a $\mathbb{Z}_3$ valued Stiefel-Whitney class $w_2$ for its global symmetry bundle, and the $T^2$ compactification with the nontrivial $w_2$ is expected to flow to 
\begin{equation}
	T_{D_4}(\underline{\text{full}}, \underline{\text{full}}, \text{simple}),
\end{equation}
where the $\underline{\text{full}}$ means the $\mathbb{Z}_3$ twisted full puncture, with $G_2$ global symmetry.

The anomaly polynomial for the minimal $(E_6,E_6)$ conformal matter is
\begin{equation}
	A \supset \frac{553}{5760} p_1(T)^2 - \frac{89}{48}p_1(T) c_2(R) +\frac18 p_1(T) \Tr F^2_L+ \frac18 p_1(T) \Tr F_R^2,
\end{equation}
where $F_L$ and $F_R$ denotes the field strength for the left and right $E_6$ background, respectively.
Subtracting the one-loop contribution from the tensor branch effective action, which consists of three tensor multiplets and one $\fsu(3)$ gauge multiplet, we get the GS contribution:
\begin{equation}
	A_\text{GS} \supset \frac{3}{32} p_1(T)^2 - \frac{7}{4}p_1(T) c_2(R) +\frac18 p_1(T) \Tr F^2_L+ \frac18 p_1(T) \Tr F_R^2.
\end{equation}
Substituting this to the formulas \eqref{eq:afrom6d}, \eqref{eq:cfrom6d} and \eqref{eq:kfrom6d} with $t=3$, we have
\begin{equation}
	a_\text{SCFT}-a_\text{generic} = \frac{25}4 , \quad
	c_\text{SCFT}-c_\text{generic} = 7 , \quad
	k_\text{SCFT}-k_\text{generic} = 8,
\end{equation}
where $k$ stands for the flavor central charge for both left and right $G_2$ symmetries in 4d.

On the tensor branch, the $\mathbb{Z}_3$ twists break the entire $\SU(3)$ gauge group, therefore on a generic point of the tensor branch the theory flows to three 4d vector multiplets each of which comes from the tensor multiplets. Therefore, the central charges on a generic point is
\begin{equation}
	a_\text{generic} = \frac{5}{8}, \quad
	c_\text{generic} = \frac12, \quad
	k_\text{generic} = 0.
\end{equation}
Therefore, the conformal anomalies at the SCFT point are
\begin{equation}
	a_\text{SCFT} = \frac{55}{8}, \quad
	c_\text{SCFT} = \frac{15}{2}, \quad
	k_\text{SCFT} = 8,
\end{equation}
which are consistent with the data of the fixture \#18 in \cite{Chacaltana:2016shw}.

\paragraph{$(E_7,E_7)$ conformal matter:}
The minimal $(E_7,E_7)$ conformal matter admits a $\mathbb{Z}_2$ valued Stiefel-Whitney class $w_2$ for its global symmetry bundle, and the $T^2$ compactification with the nontrivial $w_2$ is expected to flow to 
\begin{equation}
	T_{E_6}(\underline{\text{full}}, \underline{\text{full}}, \text{simple}),
\end{equation}
where the $\underline{\text{full}}$ means the $\mathbb{Z}_2$ twisted full puncture, with $F_4$ global symmetry.

The anomaly polynomical for the minimal $(E_7,E_7)$ conformal matter is
\begin{equation}
	A \supset \frac{469}{2280} p_1(T)^2 - \frac{125}{24}p_1(T) c_2(R) +\frac3{16} p_1(T) \Tr F^2_L+ \frac3{16} p_1(T) \Tr F_R^2,
\end{equation}
where $F_L$ and $F_R$ denotes the field strength for the left and right $E_6$ background, respectively.
The effective field theory on the tensor branch consist of five tensor multiples, $\fsu(2)\times \fso(7)\times \fsu(2)$ gauge maultiplets, and hypers in the $\frac12 (\mathbf{2},\mathbf{8},\mathbf{1})\oplus \frac12 (\mathbf{1},\mathbf{8},\mathbf{2})$ representation.
Subtracting the one-loop contribution from the tensor branch effective action, we get the GS contribution:
\begin{equation}
	A_\text{GS} \supset \frac{5}{32} p_1(T)^2 - \frac{19}{4}p_1(T) c_2(R) +\frac3{16} p_1(T) \Tr F^2_L+ \frac3{16} p_1(T) \Tr F_R^2.
\end{equation}
Substituting this to the formulas \eqref{eq:afrom6d}, \eqref{eq:cfrom6d} and \eqref{eq:kfrom6d} with $t=2$, we have
\begin{equation}
	a_\text{SCFT}-a_\text{generic} = \frac{57}2 , \quad
	c_\text{SCFT}-c_\text{generic} = 31 , \quad
	k_\text{SCFT}-k_\text{generic} = 18.
\end{equation}
where $k$ stands for the flavor central charge for both left and right $G_2$ symmetries in 4d.

On the tensor branch, the $\mathbb{Z}_2$ twists break the entire $\fsu(2)\times \fso(7) \times \fsu(2)$ gauge algebra into $\fso(4)\subset \fso(7)$, and hypers in one Dirac spinor representation of $\fso(4)$ which remains. 
This comes about as there are states charged in the $(\mathbf{56}_{E_7},\mathbf{2}_{\fsu(2)})$ for each $\fsu(2)$ gauge algebra and the $E_7$ global symmetry coming from the rank $1$ E-string theory adjacent to it. Thus, for consistency, we also need to introduce a Stiefel-Whitney class for both $\fsu(2)$ gauge algebras. Then, due to the presence of the half-hypers in the $ (\mathbf{2},\mathbf{8},\mathbf{1})\oplus (\mathbf{1},\mathbf{8},\mathbf{2})$, we must also incorporate a Stiefel-Whitney class for the $\fso(7)$ gauge algebra. There are different ways to choose the holonomies in this case, but a convenient one is just to embed the Stiefel-Whitney class for $\fsu(2)\simeq\fso(3)$ inside $\fso(7)$ using  $\SO(3)\times \SO(4) \subset \SO(7)$. This breaks the $\fso(3)$ part, but leaves the $\fso(4)$ part unbroken. Furthermore, under its $\fso(3)\times \fso(4)$ subalgebra, the spinor representation of $\fso(7)$ decomposes to $\mathbf{8}\rightarrow (\mathbf{2}_{\fso(3)},\mathbf{2}_{\fso(4)}\oplus \mathbf{2'}_{\fso(4)})$. Therefore, from similar reasons as in the previous cases, one Dirac spinor representation of $\fso(4)$ should remain. 

 From this we see that the central charges on a generic point are
\begin{equation}
	a_\text{generic} = \frac{59}{24}, \quad
	c_\text{generic} = \frac{13}{6}, \quad
	k_\text{generic} = 0.
\end{equation}
Therefore, the conformal anomalies at the SCFT point are
\begin{equation}
	a_\text{SCFT} = \frac{743}{24}, \quad
	c_\text{SCFT} = \frac{199}{6}, \quad
	k_\text{SCFT} = 18.
\end{equation}
The conformal central charges of $T_{E_6}(\underline{\text{full}}, \underline{\text{full}}, \text{simple})$ can be computed using the data in \cite{Chacaltana:2015bna}, and it matches with the above result.

\paragraph{Examples of nilpotent Higgsing} 
We can generalize the previous two examples by doing the nilpotent vev Higgsings compatible with the Stiefel-Whitney twist. 
Here, as an example, we give the nilpotent vev to the one of the $E_6$ symmetry of the $(E_6,E_6)$ conformal matter so that the resulting theory have $\frac{E_6\times SU(6)}{\mathbb{Z}_3}$ symmetry. In 6d, the resulting theory is studied in \cite{Heckman:2016ssk}, and it is the $\mathfrak{su}(3)$ gauge theory coupled with a copy of the E-string theory and 6 fundamental hypermultiplets.

The 4d theory obtained by $T^2$ compactification with the Stiefel-Whitney twist should be a nilpotent Higgsing of $T_{D_4}(\underline{\text{full}},\underline{\text{full}},\text{simple})$, which is the 4d theory corresponding to the $(E_6,E_6)$ conformal matter. The symmetry should be $G_2\times \SU(2)$. A candidate theory found in \cite{Chacaltana:2016shw} as \#19, which is $T_{D_4}(\underline{\text{full}},\underline{A_1},\text{simple})$, where $\underline{A_1}$ is a twisted puncture. Indeed, One can check that the conformal and flavor central charges matches.

We can further Higgs the remaining $E_6$ into $SU(6)$. This result in the $\mathfrak{su}(3)+12F$ theory in 6d, which was already analyzed in subsection~\ref{sec:2.2}.
In the context here we expect that the corresponding 4d theory is $T_{D_4}(\underline{A_1},\underline{A_1},\text{simple})$, which is indeed identified with the 4d rank 1 $\mathfrak{su}(4)$ theory in \cite{Chacaltana:2016shw}.

From $(E_7,E_7)$ conformal matter, we expect that we can get many twisted fixtures of the form $T_{E_6}(\underline{P_1},\underline{P_2},\text{simple})$ with some twisted punctures $\underline{P_1}, \underline{P_2}$. Although it would be interesting to work out the precise correspondence between the Higgsing in 6d and the twisted punctures, we remain it to be investigated in a future work.

\subsection{Other $6d$ SCFT lifts of $5d$ brane webs}
\label{sec:quivers}

Another class of theories we can generalize to is that of $6d$ SCFTs that can be engineered in string theory by a collection of D$6$-branes and NS$5$-branes in the presence of an O$8^-$-plane and D$8$-branes. This type of SCFTs can also be engineered in other methods in string theory, notably, as theories living on M$5$-branes probing a $\bC^2/\bZ_k$ singularity and as F-theory compactifications on a base containing a linear chain of curves with self intersections $-1$, $-2$,  $-2$, \ldots, $-2$, decorated with $\fsu$ or $\fusp$ type gauge algebras. 
Another convenient description is as UV completions of $6d$ gauge theories or semi-gauge theories, where by the latter we mean gauge theories that may contain strongly interacting parts, similarly to the $E$ type conformal matter studied in the previous subsection. This class includes all the rank-one cases, studied in the previous section, as well as the $D$ type conformal matter SCFTs studied in the previous subsection. The type of quiver gauge theories that we can get in this manner were studied in \cite{DelZotto:2014hpa,Heckman:2015bfa,Heckman:2016ssk,Mekareeya:2017jgc}. These include a long quiver of $\fsu$ gauge algebras, residing on the $-2$ curves in the F-theory description, ending with a gauge group that lives on the $-1$ curve. The latter can be either of type $\fusp$, $\fsu$ with antisymmetric matter, or $\fsu(6)$ with a half-hyper in the $\bold{20}$.

One reason to concentrate on these examples is that the $5d$ description can be straightforwardly generalized to these cases. This is due to the fact that, when compactified on a circle, these theories flow to $5d$ IR gauge theories that can be engineered via brane webs\cite{Zafrir:2015rga,Ohmori:2015tka,Hayashi:2015zka}. This implies that we can play the same games discussed in section~\ref{subsec:rank1-5d} also for this case, particularly, study mass deformations and dualities. The latter in particular suggests that this construction leads to SCFTs in this case, which makes them especially interesting. As discussed in the previous section, we expect the resulting theories to be the same as twisted compactifications of $5d$ SCFTs. The latter can be lifted to string theory via a twisted compactification of brane web configurations, where the twist involves a discrete element of $\SL(2,\bZ)$. The latter are known to be either of type $\bZ_2$, $\bZ_3$, $\bZ_4$ or $\bZ_6$, and thus, these are the Stiefel-Whitney twists we expect to have in the $6d$ construction. The first three cases are then generalization of the cases in section 2 while the $\bZ_6$ case has no rank-one example.     

Next we are going to consider the possible theories, their expected global symmetry and for the cases we expect a possible Stiefel-Whitney compactification, we further discuss some of its expected properties. We shall present the theories through their low-energy tensor branch description, from which one can then convert to any of the other descriptions. We shall here only discuss the cases where we allow fundamental hypers only at the end of the quiver; more general cases can then be generated via Higgsing. This choice is convenient as it is easier to analyze in $6d$, and is also natural from the $5d$ viewpoint, as will hopefully become apparent in the few subsections which follow.

\subsubsection{$\bZ_2$}
\label{sec:3.4.1}
\begin{figure}
\center
\includegraphics[width=0.75\textwidth]{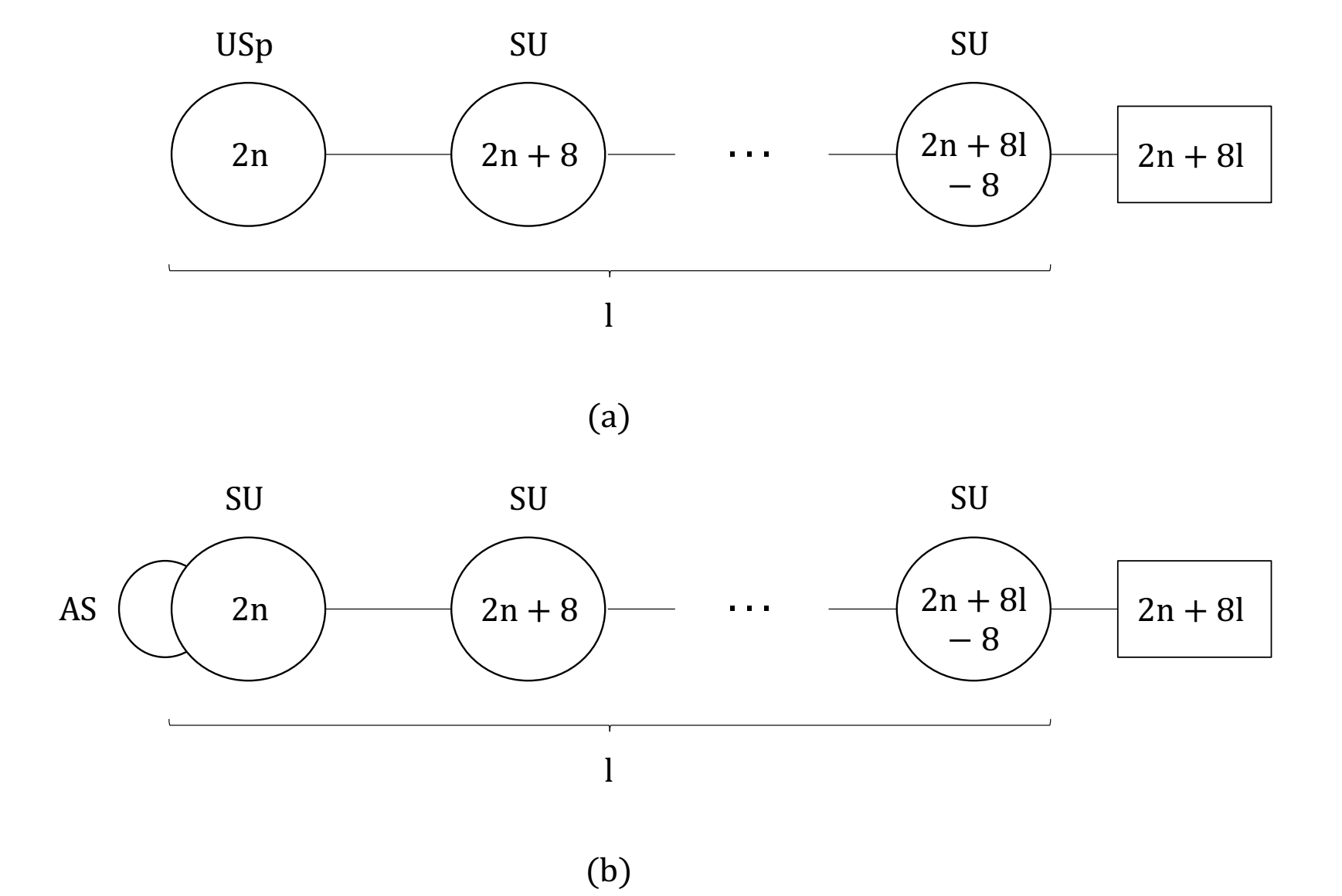} 
\caption{Quivers of $6d$ gauge theories which are the low-energy tensor branch descriptions of certain $6d$ SCFTs.}
\label{6dQuiversZ2}
\end{figure}

Consider the cases where the $6d$ SCFT has a low-energy gauge theory description as the quivers in figure \ref{6dQuiversZ2}. First consider the case in figure \ref{6dQuiversZ2} (a). Here the edge group is of type $\fusp(2n)$, where we include also the $n=0$ case in which the $\fusp$ factor is replaced with the rank-one E-string theory connected to the $\fsu(8)$ via gauging of $\fsu(8)\subset E_8$. For later convenience, we shall also define $N=n+2l$.

This low-energy gauge theory in fact describes two different $6d$ SCFTs\cite{Mekareeya:2017jgc}, where the distinction is given by the non-trivial $\bZ_2$ valued $\theta$ angle of $\fusp(2n)$\footnote{In the $n=0$ case, this is expressed as the choice of the embedding of $\fsu(8)$ in $E_8$, which comes in two distinct variants.}. While these do not affect the perturbative matter spectrum of the theory, it modifies the non-pertrbative spectrum that exists in the SCFT. To see this consider taking the infinite coupling limit of $\fusp(2n)$. In that limit we can replace this part by an $\fsu(2n+8)$ gauging of the $(D_{n+4}, D_{n+4})$ minimal conformal matter SCFT. The latter contains a state in the chiral spinor representation of its $\fso(4n+16)$ global symmetry. As the latter is gauged by $\fsu(2n+8)$, the spinor needs to be decomposed to $\fsu(2n+8)$ representations. There are, however, two inequivalent ways to do this, one where it decomposes to all the even-rank antisymmetric representations of $\fsu(2n+8)$, and one where it decomposes to all the odd-rank ones. The $\theta$ angle exactly captures this difference. We shall generally refer to the case with the even rank decomposition as the $\theta=0$ case, and the odd rank case as the $\theta=\pi$ case.

The global symmetry of the SCFT here for generic $n$ is $\fu(1)\times \fsu(2n+8l)$, where the $\fu(1)$ is the anomaly free combination of all the bifundamental $\fu(1)$ rotations. For the special case of $n=0$ and $\theta=0$, this is further enhanced to $\fsu(2)\times \fsu(2n+8l)$, which will play an important role when we discuss the $\bZ_4$ case.  

Next, consider the case in figure \ref{6dQuiversZ2} (b). Here the edge gauge algebra is of type $\fsu(2n)$, where at this level $n$ is allowed to be half-integer. For later convenience, we shall also define $N=n+2l-1$ in the integer $n$ case. Here the global symmetry of the SCFT for generic $n$ is $\fu(1)\times \fsu(2n+8l)$, where the $\fu(1)$ is the anomaly free combination of all the antisymmetric and bifundamental $\fu(1)$ rotations. For the special case of $n=2$, this is further enhanced to $\fsu(2)\times \fsu(2n+8l)$, which will play an important role when we discuss the $\bZ_4$ case. The case of $n\leq 1$ reduces to the previous case, while the $2n=3$ case will not be of any importance to us here.

So far the discussion of the global symmetry was at the Lie algebra level, and we can next ask what is the global symmetry at the group level. For that we consider what gauge invariant states we expect, and if there is some element in the center of the $\SU$ group that acts trivially on them. In what follows we shall ignore the $\fu(1)$ factor of the global symmetry and in particular will not consider possible modding out involving both the $\fu(1)$ and part of the center of the $\SU$ group.  

Lets first consider the tail of the quiver which is shared by both types of $6d$ SCFTs. The first gauge invariants that we can consider are just the baryons for the edge $\fsu(2n+8l-8)$ gauge algebra. By inspection, we see that these are in the rank $8$ antisymmetric representation of $\fsu(2n+8l)$, or its conjugate. This will be invariant under a $\bZ_2$, $\bZ_4$ or $\bZ_8$ subgroup of $\SU(2n+8l)$. We first note that if $n$ is half-integer then such a group does not exist. Thus, for the theories of \ref{6dQuiversZ2} (b) with $n$ half-integer, the group is globally $\SU(2n+8l)$, up to possible cancellations with the $\fu(1)$. We next restrict to the $n$ integer case.

We can continue and consider forming an $\fsu(2n+8l-16)\times \fsu(2n+8l)$ bifundamental from the $\fsu(2n+8l-16)\times \fsu(2n+8l-8)$ and $\fsu(2n+8l-8)\times \fsu(2n+8l)$ bifundamentals. We can now form gauge invariants through baryons of the $\fsu(2n+8l-16)$ gauge algebra. These transform in the rank $16$ antisymmetric representation of $\fsu(2n+8l)$, or its conjugate, and are consistent with the previous potential modding.

We can continue with this along the quiver, where each step the rank of the resulting antisymmetric representation of $\fsu(2n+8l)$ increases by $8$. These then are all consistent with the potential modding. The only point where something non-trivial arises is when we reach the end of the quiver. For the theories in \ref{6dQuiversZ2} (a), we now have an effective $\fusp(2n)\times \fsu(2n+8l)$ bifundamental, and we need to limit to gauge invariants of $\fusp(2n)$. However, these can be quadratic giving the rank $2$ antisymmetric representation of $\fsu(2n+8l)$. Similarly for the theory in \ref{6dQuiversZ2} (b), we get an $\fsu(2n)\times \fsu(2n+8l)$ bifundamental, and while restricting to $\fsu(2n)$ baryons does not give any new restriction, we can build a gauge invariant also using the antisymmetric. These can be used to form a gauge invariant state in the rank $2$ antisymmetric representation of $\fsu(2n+8l)$.

Therefore, we see that the maximal modding we can do is $\bZ_2$, and indeed $\SU(2n+8l)$, for the theories we are considering, always has the required center. So far we considered only the perturbative states. We next want to say a few words about the non-perturbative states. Particularly, we have mentioned one such states, the one associated with the $\fusp(2n)$ algebra at strong coupling. These lead to gauge variant states in all the even or odd rank antisymmetric representations of $\fsu(2n+8)$ for $\theta=0$ or $\theta=\pi$. We can next use these states to create gauge invariants of $\fsu(2n+8l)$. For $\theta=0$, we can only create rank even antisymmetric representations which are consistent with the $\bZ_2$ modding. However, for $\theta=\pi$, we can create rank odd antisymmetric representations which are inconsistent with the $\bZ_2$ modding.

So, to wrap up the discussion so far, for the class of $6d$ SCFTs in figure \ref{6dQuiversZ2}, the global symmetry is consistent with $\SU(2n+8l)/\bZ_2$ for the cases of $\theta=0$ for the theories in (a) and $n$ integer for the theories in (b). In the other cases it must be $\SU(2n+8l)$. Again we note that this is up to possible modding with the additional $\fu(1)$. 

Next, we shall concentrate on the cases consistent with $\SU(2n+8l)/\bZ_2$. For these cases, we can consider torus compactifications with a non-trivial Stiefel-Whitney class of $\SU(2n+8l)/\bZ_2$, that is we turn on two holonomies commuting up to the $\bZ_2$ element. This can be engineered by using the $\fsu(2)\times \fsu(n+4l)$ subalgebra of $\fsu(2n+8l)$, and embedding the holonomies in the $\fsu(2)$ part, similarly to what was done in the previous sections. This breaks the $\fsu(2)$, and so preserves at most the $\fsu(n+4l)$ subgroup of the global symmetry.  

This is expected to lead to a $4d$ theory and generalizes the construction we did previously for the $(D, D)$ conformal matter to a more general family of theories. Next, we shall rely on the methods used in the previous section to study some of their properties.

First, we wish to consider the effect of the compactification on the gauge groups. Similarly to the previous cases, consistency requires also that there be almost commuting holonomies for all the gauge groups. For the $\fsu$ type gauge algebra, these can be handled in a similar matter as the global symmetry, and they are expected to break part of the gauge symmetry. For the $\fusp(2n)$ algebra, we can use the $\fsu(2)\times \fso(n)$ subgroup of $\fusp(2n)$, and again embed the holonomies in the $\fsu(2)$ part which leaves only the $\fso(n)$ part unbroken.
This then determines the gauge algebra remaining at a generic point on the Coulomb branch spanned by the operators descending from the tensor multiplets. Taking into account all these gauge groups and the tensor multiplets, we find the resulting theory to be of rank $2(l-1)(n+2l)+\left\lfloor{\frac{n}{2}}\right\rfloor + 1$ for the case in figure \ref{6dQuiversZ2} (a), and $2(l-1)(n+2l)+n$ for the case in figure \ref{6dQuiversZ2} (b).  

\begin{figure}
\center
\includegraphics[width=0.75\textwidth]{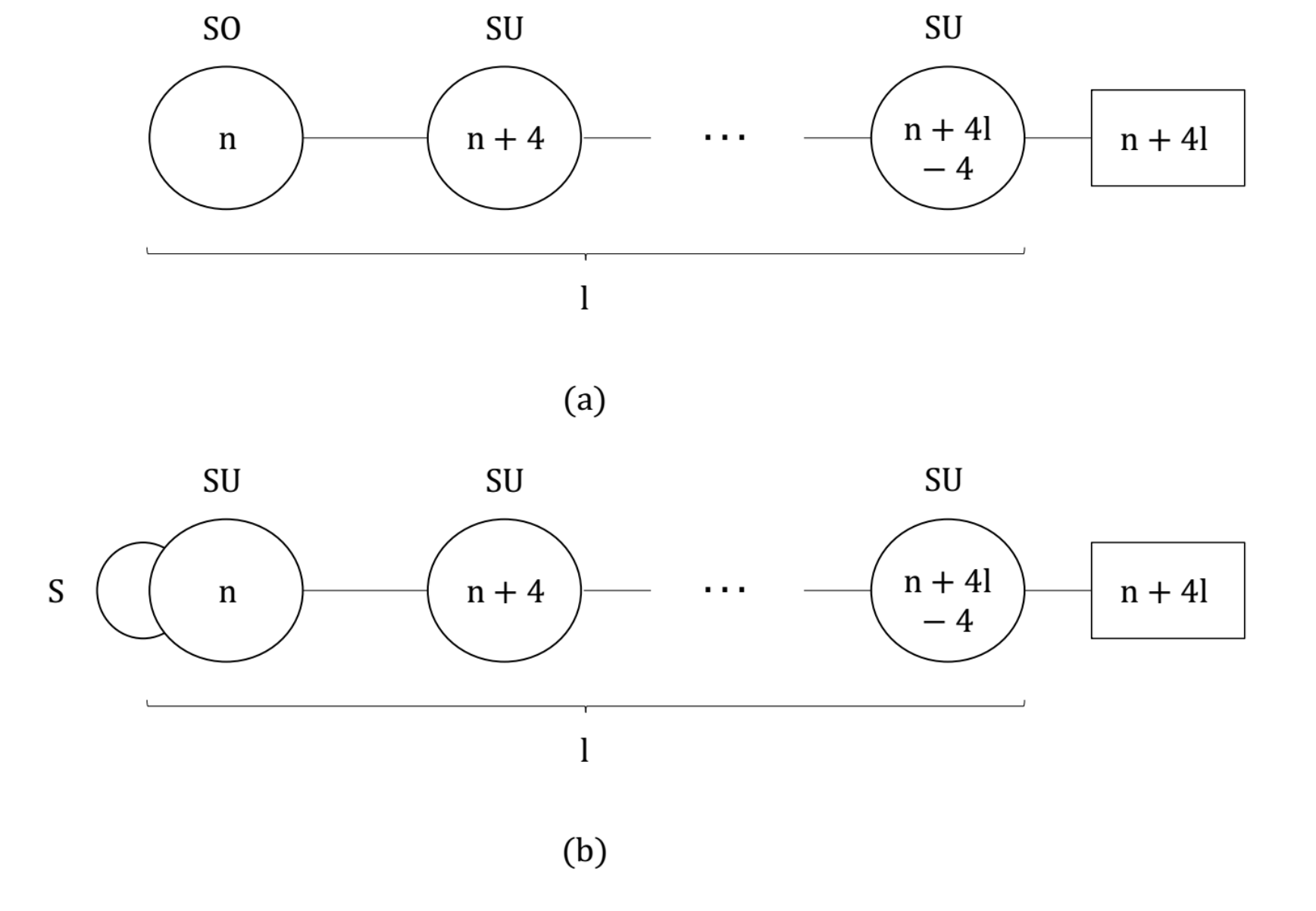} 
\caption{Quivers of the $4d$ gauge theory that is expected to exist, on a generic point on the Coulomb branch spanned by the operators descending from the tensor multiplets, for theories resulting from the Stiefel-Whitney compactification of the theories in figure \ref{6dQuiversZ2}.}
\label{4dQuiversZ2}
\end{figure}

We can next determine the matter content on a generic point on the Coulomb branch spanned by the operators descending from the tensor multiplets. Having already determined the gauge groups, we now need to consider the effect on the hypermultiplets. For the bifundamentals, by essentially the same reasonings as in the previous sections, we conclude that each should contribute a single bifundamental between the reduced groups. This just leaves the antisymmetric. Under the $\fsu(2)\times \fsu(n)$ subgroup of $\fsu(2n)$, it decomposes to the $(\bold{3},\bold{AS}) \oplus (\bold{1},\bold{S})$. As the first term is projected out, we see that the antisymmetric hyper contributes a single symmetric hyper for the resulting $\fsu(n)$ theories. The resulting theories are summarized in figure \ref{4dQuiversZ2}.     

\subsubsection{$\bZ_3$}

\begin{figure}
\center
\includegraphics[width=0.75\textwidth]{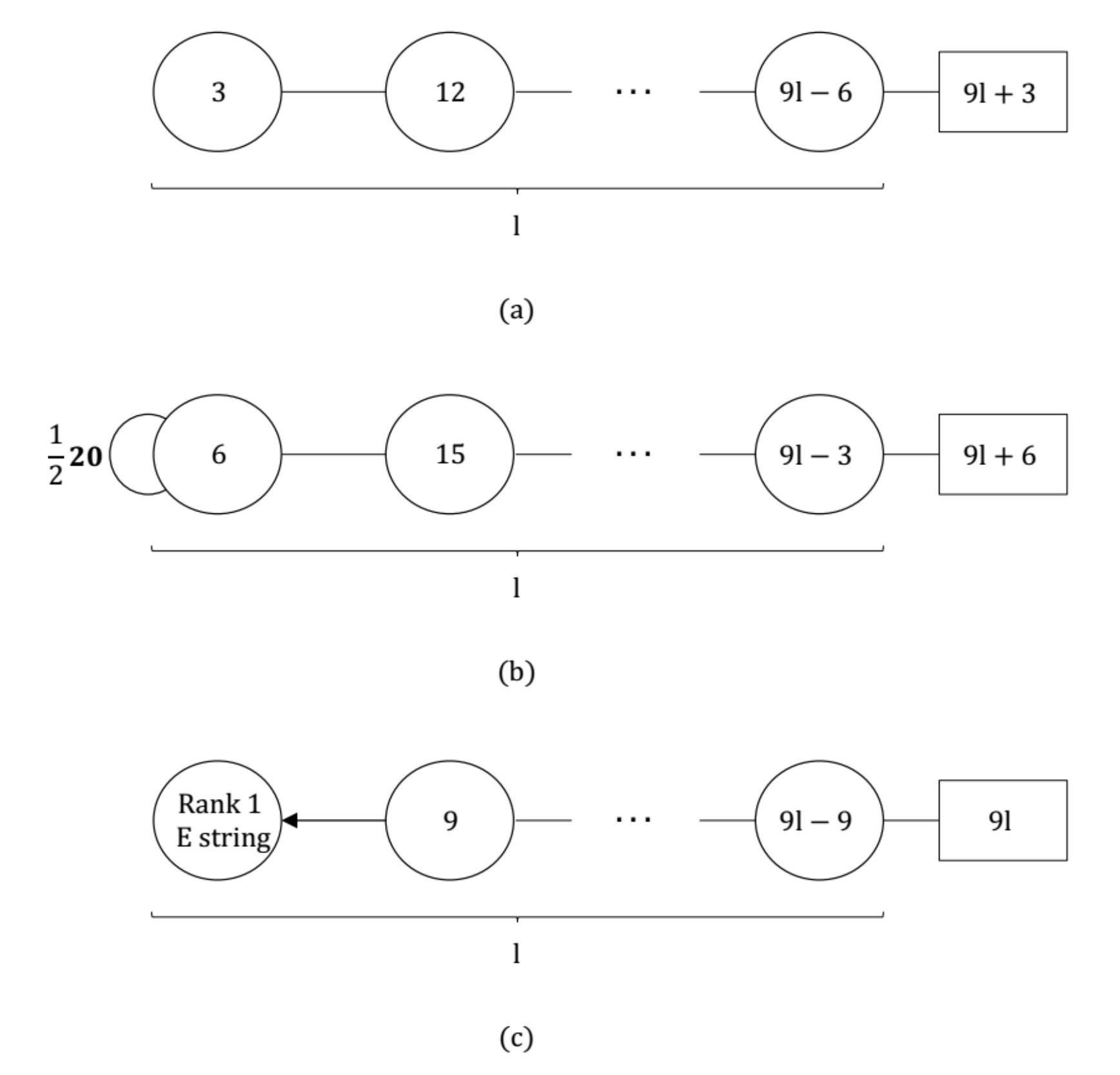} 
\caption{Quivers of $6d$ gauge theories which are the low-energy tensor branch descriptions of certain $6d$ SCFTs. Here all gauge algebras are of type $\fsu$.}
\label{6dQuiversZ3}
\end{figure}

Next, we consider the cases where the $6d$ SCFT has a low-energy gauge theory description as the quivers in figure \ref{6dQuiversZ3}. We start by consider the case in figure \ref{6dQuiversZ3} (a). In the infinite coupling limit,  we expect a $6d$ SCFT with $\fsu(9l+3)$ flavor symmetry algebra, where all the bifundamental $\fu(1)$ groups are anomalous. We can again inquire as to what is the global symmetry at the group level, which we can attempt to answer by considering the spectrum. 

The analysis now mimics the one in the previous section, where we essentially consider baryons made from a progressively large number of bifundamentals, which we shall sometimes refer to as generalized baryons. By inspection, it is clear that the baryons of the $\fsu(9l-6)$ group are in the rank $9$ antisymmetric representation of $\fsu(9l+3)$ or its conjugate. Likewise, those of the adjacent group, are in the rank $18$ antisymmetric representation of $\fsu(9l+3)$ or its conjugate, and so forth. This terminates when we get to the $\fsu(3)$ gauge algebra where baryons of it are in the are in the rank $3$ antisymmetric representation of $\fsu(9l+3)$ or its conjugate. Therefore, by considering the perturbative spectrum, we see that the global symmetry is consistent with being $\SU(9l+3)/\bZ_3$. We reserve a more details study of both the perturbative and the non-perturbative operators to future work.

Next, consider the theory in figure \ref{6dQuiversZ3} (b). In the infinite coupling limit,  we expect a $6d$ SCFT with $\fsu(9l+6)$ symmetry algebra, where again all the bifundamental $\fu(1)$ groups are anomalous. Like the previous cases, We can inquire as to what is the global symmetry at the group level, which we can attempt to answer by considering the spectrum. 

Through a similar analysis as in the previous section, we see that the generalized baryons contribute states that are in the antisymmetric representation of $\fsu(9l+6)$ of ranks rank $9$, $18$, $27$ and so forth, until we get to the $\fsu(6)$ gauge algebra. The generalized baryons there are in the rank $6$ antisymmetric representation of $\fsu(9l+6)$. Furthermore, we can build a gauge invariant from three generalized bifundamentals and the rank $3$ antisymmetric half-hyper. Such a state is then in the rank $3$ antisymmetric representation of $\fsu(9l+6)$. So again, from considerations of the perturbative spectrum, we see that the global symmetry is consistent with being $\SU(9l+6)/\bZ_3$. Again we defer a more detailed study of the spectrum to future work.

We can do the same analysis also to the theory in figure \ref{6dQuiversZ3} (c). Here,  the expected $6d$ SCFT should have an $\fsu(9l)$ symmetry algebra. The basic gauge invariants made from the bifundamentals are all in the antisymmetric representation, whose rank is divisible by $9$, of $\fsu(9l)$. This leaves the contribution of the rank-one E-string theory, whose basic operator is the $E_8$ conserved current operator. Under the $\fsu(9)$ maximal subalgebra of $E_8$, the adjoint of the latter decomposes to $\bold{248} \rightarrow \bold{80} \oplus \bold{84} \oplus \bar{\bold{84}}$, where the latter two representations are the rank $3$ antisymmetric representation of $\fsu(9)$ and its conjugate. From these two, and the bifundamental, one can build states in the rank $3$ antisymmetric representation of $\fsu(9l)$. This suggests that the global symmetry is actually $\SU(9l)/\bZ_3$. 

If this is the case, then we can consider torus compactifications not allowed by $\SU(3N)$ but allowed by $\SU(3N)/\bZ_3$.
Here for convenience we have defined $N=3l+1$, $3l+2$ or $3l$ for cases (a), (b) and (c), respectively. This generalizes the construction we used to realize the $4d$ rank-one $\fsu(4)$ theory, and we expect it to lead to a wide class of $4d$ theories. Next, we shall explore some of their properties.   

First, we can analyze the expected Coulomb branch dimension, and matter content on generic points on the conformal manifold. Similarly to the previous cases, consistency requires also that there be almost commuting holonomies for all the gauge groups. Here all the symmetry algebras are of type $\fsu$, and this can be handled in a similar fashion as the global symmetry, by using the $\fsu(3)\times \fsu(k)$ subalgebra of $\fsu(3k)$. The almost commuting holonomies  then break the $\fsu(3)$ part. Therefore, at a generic point on the Coulomb branch, associated with the operators descending from the $6d$ tensor multiplets, we expect to have some unbroken gauge symmetry. Taking into account these gauge groups and the tensor multiplets, we find the resulting theory to be of rank $\frac{l(3l-1)}{2}$ for the case in figure \ref{6dQuiversZ3} (a), $\frac{l(3l+1)}{2}$ for the case in figure \ref{6dQuiversZ3} (b), and $1+\frac{3l(l-1)}{2}$ for the case in figure \ref{6dQuiversZ3} (c).  

\begin{figure}
\center
\includegraphics[width=0.75\textwidth]{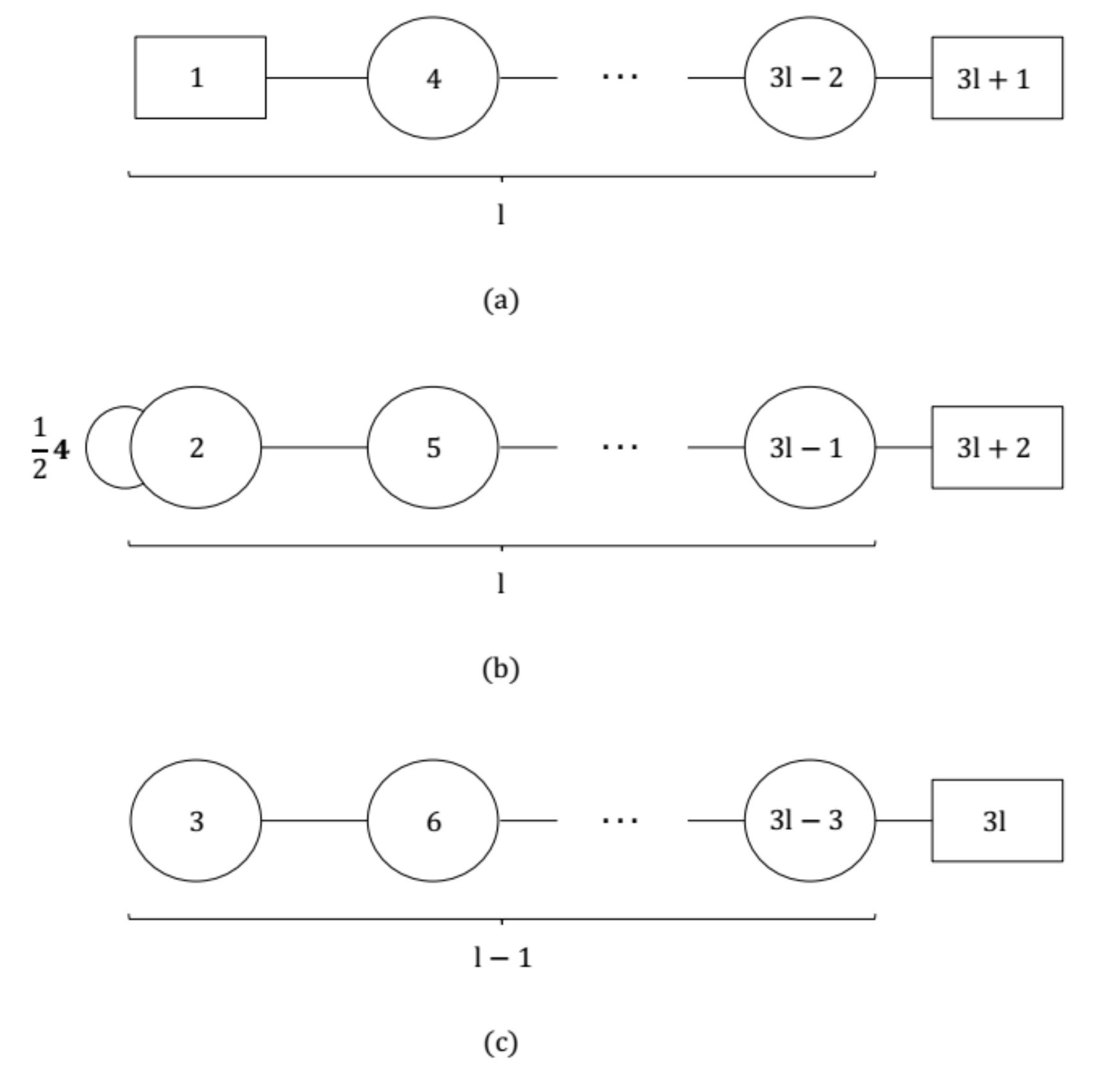} 
\caption{Quivers of the $4d$ gauge theory that is expected to exist, on a generic point on the Coulomb branch spanned by the operators descending from the tensor multiplets, for theories resulting from the Stiefel-Whitney compactificaton of the theories in figure \ref{6dQuiversZ3}.}
\label{4dQuiversZ3}
\end{figure}

We can next determine the matter content on a generic point on the Coulomb branch spanned by the operators descending from the tensor multiplets. Having already determined the gauge groups, we need now consider the effect on the hypermultiplets. For the bifundamentals, as in the previous sections, each one should contribute a single bifundamental between the reduced groups. For the case in figure \ref{6dQuiversZ3} (a), this determines the theory completely. For the case in figure \ref{6dQuiversZ3} (b), we also need to determine what happens to the three index antisymmetric. Under $\fsu(2)\times \fsu(3) \subset \fsu(6)$ we have that $\bold{20}\rightarrow (\bold{2}, \bold{8}) \oplus (\bold{4}, \bold{1})$. Therefore, it provides an half-hyper in the $\bold{4}$ of the gauge $\fsu(2)$. For the case in figure \ref{6dQuiversZ3} (c), we also need to determine what happens to the rank-one E-string theory. However, we do not expect any contribution from the latter when we are at a generic point on its tensor branch. The resulting theories are then summarized in figure \ref{4dQuiversZ3}.

\subsubsection{$\bZ_4$}

\begin{figure}
\center
\includegraphics[width=0.75\textwidth]{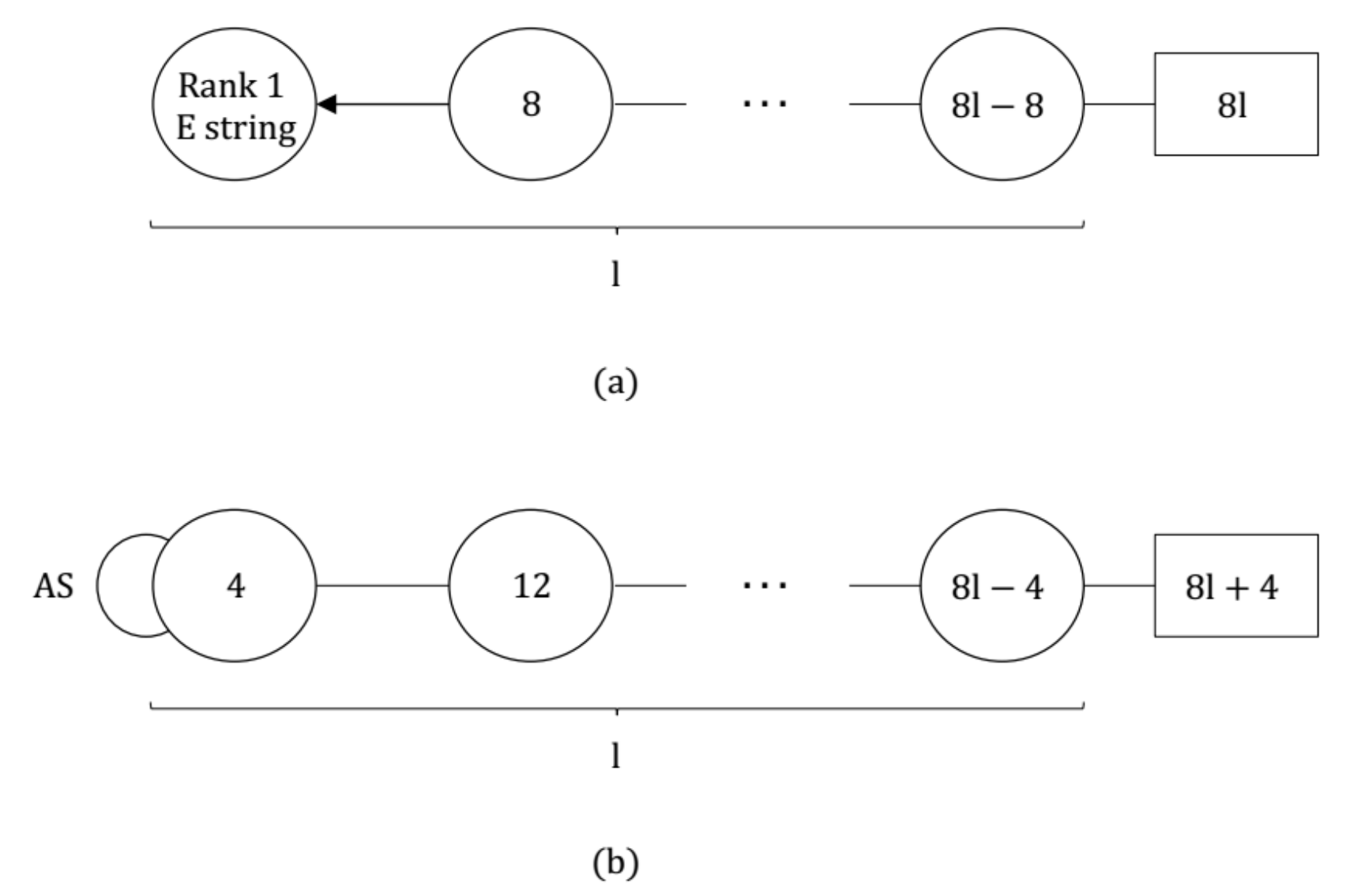} 
\caption{Quivers of $6d$ gauge theories which are the low-energy tensor branch descriptions of certain $6d$ SCFTs. Here all groups are of type $\fsu$.}
\label{6dQuiversZ4}
\end{figure}

Consider the cases where the $6d$ SCFT has a low-energy gauge theory description as the quivers in figure \ref{6dQuiversZ4}. For the case in figure \ref{6dQuiversZ4} (a), we use the embedding of $\fsu(8)$ whose commutant in $E_8$ is $\fsu(2)$, which is the extension of the $\theta=0$ case. For later convenience, we shall define $N=2l$ for figure \ref{6dQuiversZ4} (a), and $N=2l+1$ for figure \ref{6dQuiversZ4} (b). These then are expected to lift to $6d$ SCFTs with an $\fsu(2)\times \fsu(4N)$ global symmetry algebra. We can then inquire as to what is the group structure.

The analysis proceeds as  in our discussion of the $\bZ_2$ cases. However, there is one difference here being the existence of the additional $\fsu(2)$. Particularly, the states uncharged under the $\fsu(2)$ all transform trivially under a $\bZ_4$ subgroup of the center of $\SU(4N)$. Examples of these include the $\fsu(4)$ generalized baryons for figure \ref{6dQuiversZ4} (b), or the generalized baryons formed using the rank $4$ antisymmetric of $\fsu(8)$ that we get from the $E_8$ conserved currents. Furthermore, for the operators that are affected by the generator of this $\bZ_4$, that effect can be canceled by also acting with the center of $\SU(2)$. Thus we conclude that here the global symmetry group is consistent with being $\frac{\SU(2)\times \SU(4N)}{\bZ_4}$, where the $\bZ_4$ generator is given by $(\omega_{\SU(2)}, \omega^{N}_{\SU(4N)})$ for $\omega_G$ the generator for the center of $G$.

We can then again consider torus compactifications consistent with the group, but not with its simply connected version. This generalizes the construction we used to realize the $4d$ rank-one $\fsu(3)$ theory, and we expect it to lead to a wide class of $4d$ theories. Next, we shall explore some of their properties.   

First, we can analyze the expected Coulomb branch dimension, and matter content on generic points on the conformal manifold. Similarly to the previous cases, consistency requires also that there be almost commuting holonomies for all the gauge groups. Here all the algebras are of type $\fsu$, and this can be handled in a similar fashion as the global symmetry, by using the $\fsu(4)\times \fsu(N)$ subgroup of $\fsu(4N)$, where the gauge holonomies break the $\fsu(4)$ part. Counting the contribution from the gauge groups and the tensor multiplets, we find the resulting theories to be of rank $1+l(l-1)$ for the case in figure \ref{6dQuiversZ4} (a), and $l^2$ for the case in figure \ref{6dQuiversZ4} (b).  

\begin{figure}
\center
\includegraphics[width=0.75\textwidth]{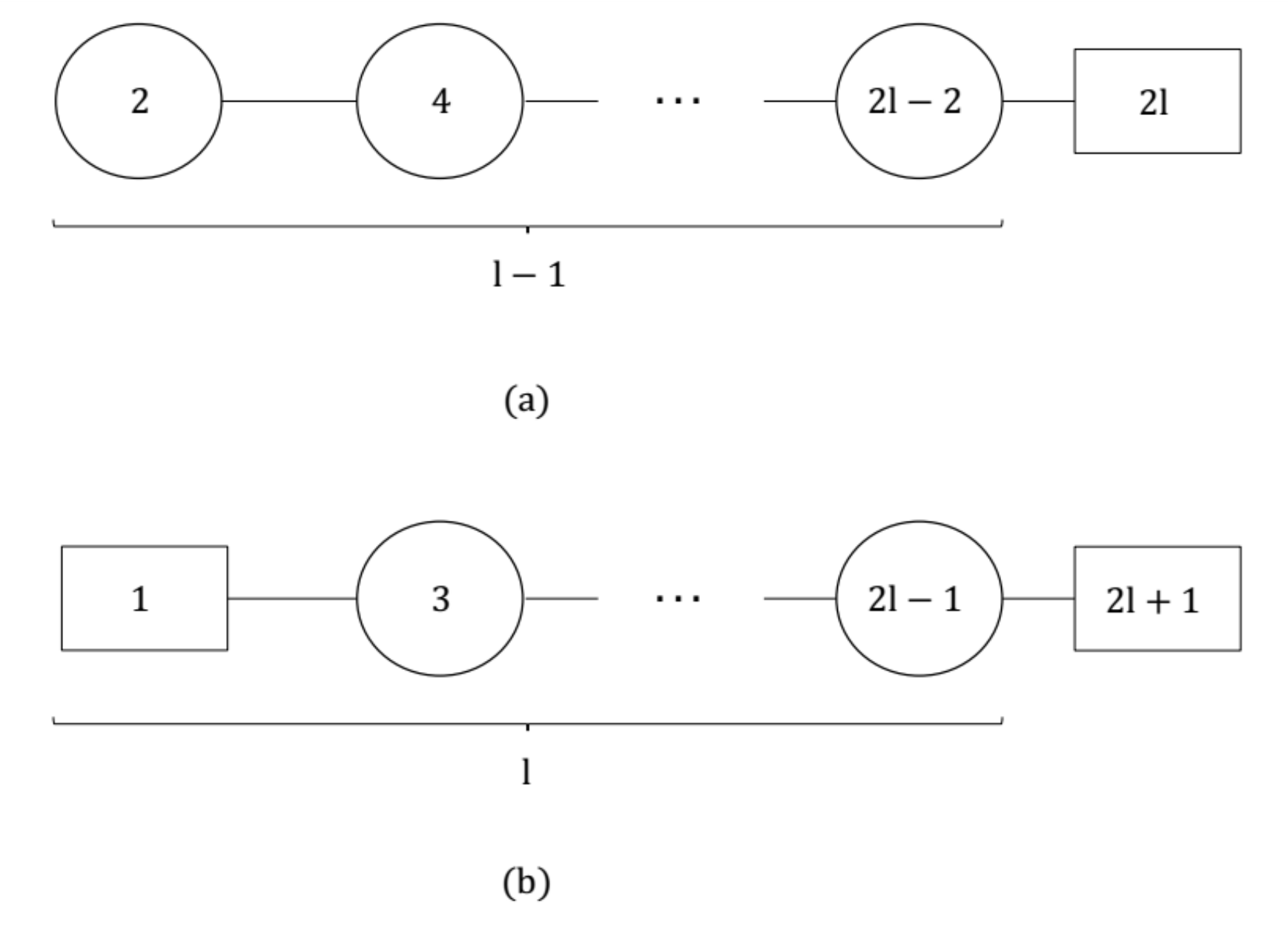} 
\caption{Quivers of the $4d$ gauge theory that is expected to exist, on a generic point on the Coulomb branch spanned by the operators descending from the tensor multiplets, for theories resulting from the Stiefel-Whitney compactificaton of the theories in figure \ref{6dQuiversZ4}.}
\label{4dQuiversZ4}
\end{figure}

We can next determine the matter content on a generic point on the Coulomb branch spanned by the operators descending from the tensor multiplets. Having already determined the gauge groups, we need now consider the effect on the hypermultiplets. For the bifundamentals, as in the previous sections, each one should contribute a single bifundamental between the reduced groups. We again do not expect any matter from the rank-one E-string theory, and the antisymmetric was determined previously to not contribute in this limit. Thus we expect the theories shown in figure \ref{4dQuiversZ4}.     

\subsubsection{$\bZ_6$}

\begin{figure}
\center
\includegraphics[width=0.75\textwidth]{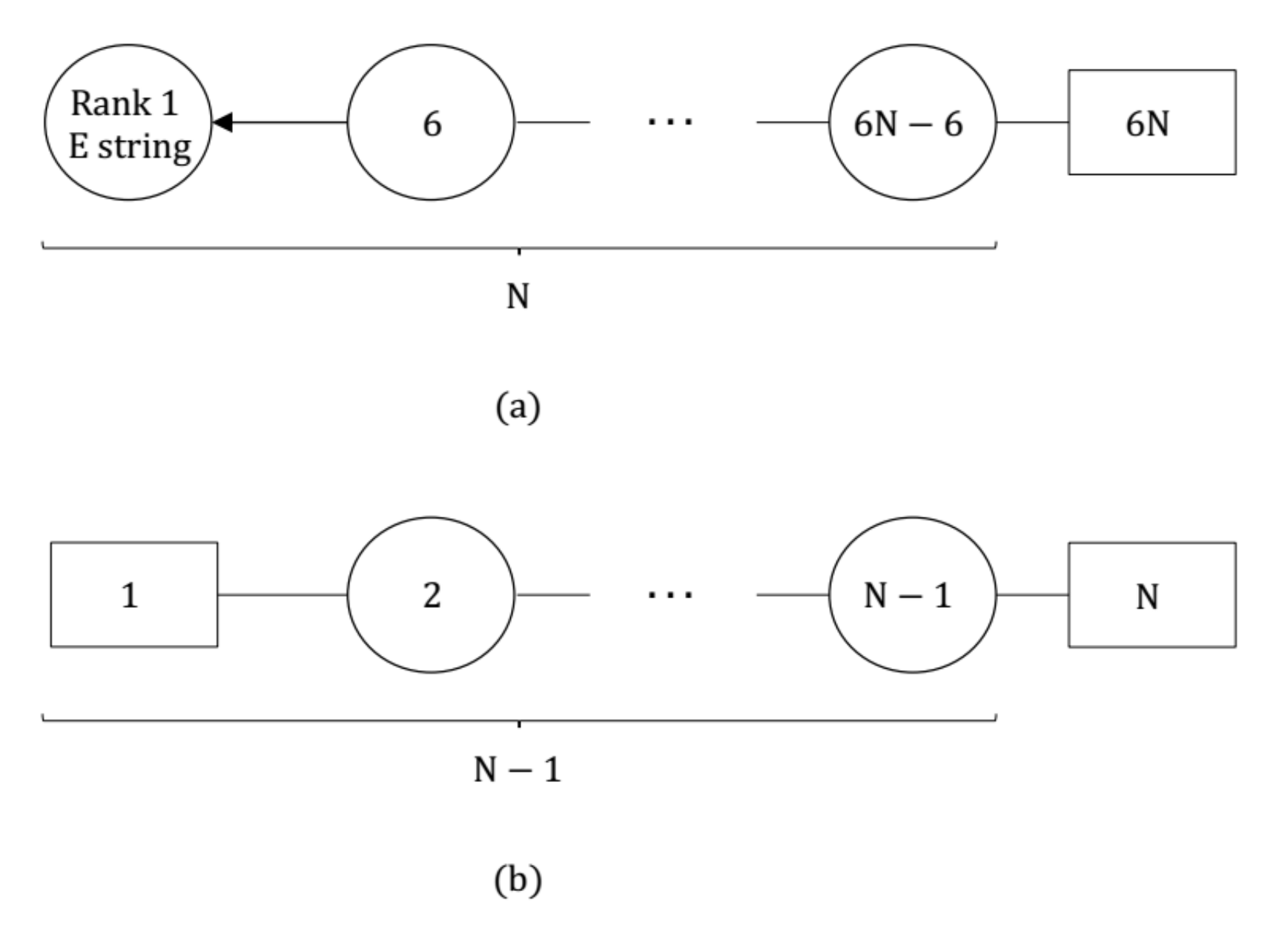} 
\caption{(a) A $6d$ quiver describing a $6d$ SCFT on a generic point on the tensor branch. (b) The Quiver of the $4d$ gauge theory that is expected to exist, on a generic point on the Coulomb branch spanned by the operators descending from the tensor multiplets, for the theory resulting from the SW compactification of the theory in (a).}
\label{QuiversZ6}
\end{figure}

Consider the case where the $6d$ SCFT has a low-energy gauge theory description as the quiver in figure \ref{QuiversZ6} (a). It is expected to lift to a $6d$ SCFT with $\fsu(2)\times \fsu(3) \times \fsu(6N)$ global symmetry algebra. We can then inquire as to what is the group structure. By similar methods to how we tackle the previous cases, we conclude that the perturbative matter transform trivially under a $\bZ_6$ subgroup of the center of $\SU(6N)$.

This leaves the $\fsu(6)$ charged contributions of the $E_8$ conserved currents. Under the $\fsu(2)\times \fsu(3) \times \fsu(6)$ subalgebra of $E_8$, the adjoint of $E_8$ decomposes as $\bold{248}\rightarrow (\bold{3}, \bold{1}, \bold{1}) \oplus (\bold{1}, \bold{8}, \bold{1}) \oplus (\bold{1}, \bold{1}, \bold{35}) \oplus (\bold{2}, \bold{1}, \bold{20}) \oplus (\bold{3}, \bold{1}, \bold{15}) \oplus (\bar{\bold{3}}, \bold{1}, \bar{\bold{15}}) \oplus (\bold{3}, \bold{2}, \bar{\bold{6}}) \oplus (\bar{\bold{3}}, \bold{2}, \bold{6})$. By inspection one can see that all gauge invariants transform trivially under a $\bZ_6$ of the center $\SU(6N)$ properly combined with actions from the centers of $\SU(2)$ and $\SU(3)$.

 This suggests that the global symmetry group here is  $\frac{\SU(2)\times \SU(3) \times \SU(6)}{\bZ_6}$, where the $\bZ_6$ generator is given by $(\omega_{\SU(2)}, \omega_{\SU(3)}, \omega^{N}_{\SU(6N)})$. We can then again consider torus compactifications consistent with the group, but not with its simply connected version. This class of theories does not have a rank-one representative and so was not previously discussed. Nevertheless, we expect it to lead to a wide class of $4d$ theories, and next, we shall explore some of their properties.

Again the properties we are interested in are the Coulomb branch dimension, and the matter content on a generic point on the subspace of the Coulomb branch coming from the $6d$ tensor branch. As the analysis is the same as the ones preformed previously, we shall just state the end result here. We find that the $4d$ theory should have a $1+\frac{N(N-1)}{2}$ dimensional Coulomb branch, and that the matter content on a generic point on the subspace of the Coulomb branch coming from the $6d$ tensor branch is the one illustrated in figure \ref{QuiversZ6} (b).

\subsubsection{Calculating the $4d$ central charges}

We can use our previous algorithm, particularly \eqref{eq:afrom6d}, \eqref{eq:cfrom6d} and \eqref{eq:kfrom6d}, to compute the central charges also for this class of theories. For simplicity, we concentrate on the $\bZ_2$ and $\bZ_3$ cases, as these can be compared against values expected from the $5d$ picture, which we discuss in detail in the next subsection.  

For this we need the value of the Green-Schwarz term, or at least the terms in it relevant for these equations. For this class of theories, it is known that the anomaly polynomial contains\cite{Zafrir:2015rga}:
\bea
I_8 \supset && (n_h - n_v + 29n_t) \frac{7p_1 (T)^2 - 4 p_2 (T)}{5760} - (n_v - n_t + 12 \sum^{n_G}_i \sum^i_{j=1} h_{G_j}) \frac{C_2 (R) p_1 (T)}{48} \nonumber \\ && + (6 +d_{\rho}) \frac{\Tr (F^2_\text{global}) p_1 (T)}{96},
\eea 
where we have used $n_t$, $n_v$, and $n_h$ for the number of tensor, vector and hpermultiplets respectively, $n_G$ for the number of gauge groups, and $h_{G_i}$ for the dual Coxeter number of the group $i$, where the group on the $-1$ curve has $i=1$\footnote{The formula still holds when the $-1$ curve is empty, but then we must take $h_{G_1}=1$.}. The last term is for the non-abelian flavor symmetry that exists at the end of the quiver, where $d_{\rho}$ is the dimension of the gauge group at the end of the quiver.

We can get the contribution from the Green-Schwarz term, by removing the contribution from the tensor, vector and hpermultiplets. We then find:
\be
A^T_\text{GS} \supset \frac{n_t}{32} p_1 (T)^2 - \sum^{n_G}_i \sum^i_{j=1} h_{G_j} \frac{C_2 (R) p_1 (T)}{4} + \frac{\Tr (F^2)_\text{global} p_1 (T)}{16}.
\ee   
We can next evaluate the central charges for each case. In what follows, we find it more convenient to work with, instead of $a$ and $c$, their linear combination $d_H = 24(c-a)$ and $n^{4d}_v = 4(2a-c)$.

We consider the $6d$ SCFTs with tensor branch description as shown in figure \ref{6dQuiversZ2}. From this we see that $n_t = l$ and $\sum^{n_G}_i \sum^i_{j=1} h_{G_j} = \frac{l}{3} (3 l n + 4l^2 -1)$ for (a) and $\sum^{n_G}_i \sum^i_{j=1} h_{G_j} = \frac{l}{3} (l+1)(3 n + 4l -4)$ for (b).

When reduced to $4d$ on a torus with a $\bZ_2$ SW class, we expect a $4d$ theory having the quiver theories in figure \ref{4dQuiversZ2} as a low-energy description on a generic point on the subspace of the $4d$ Coulomb branch that descends from the $6d$ tensor branch. From this we see that:
\be
k_\text{generic} = 2 (n+4l-4) ,
\ee  
for both cases, and
\begin{align}
d_{H}{}_\text{generic} &= \frac{n^2 + n(8l+1) + 16l(l-1)}{2} -1 ,\\
n^{4d}_{v}{}_\text{generic} &= \frac{(3(2l-1)n^2+ 3 (8l^2-8l-1)n+32l^3-48l^2+16l+6)}{6},
\end{align}
for the case in (a), and 
\begin{align}
d_{H}{}_\text{generic} &= \frac{n^2 + n(8l+1) + 16l(l-1)}{2} ,\\
n^{4d}_v{}_\text{generic} &= \frac{l(3n^2+ 12 (l-1)n+16l^2-24l+8)}{3}
\end{align}
for the case in (b).

From this using equation \ref{eq:kfrom6d}, and $t=2$, $I=2$, we find that:
\be
k_\text{SCFT} = 2 (n+4l+2) = 2(N+k+2),
\ee
for both cases. Using equations \ref{eq:afrom6d} and \ref{eq:cfrom6d}, together with $t=2$, we further find:
\begin{align}
d_{H}{}_\text{SCFT} &= \frac{n^2 + n(8l+1) + 8l(2l+1)}{2} - 1\\
& = \frac{(N+k)(N+k+1)}{2}+k-1,\\
n^{4d}_{v}{}_\text{SCFT} &= \frac{(2l-1)(3n^2+ 3 n(10l+1)+40l^2-4l-6)}{6} \\
&= \frac{(k-1)(3N^2+ 3 N(3k+1)-2k^2-5k-6)}{6},
\end{align}
for the case in (a), and
\begin{align}
d_{H}{}_\text{SCFT} &= \frac{n^2 + n(8l+1) + 8l(2l+1)}{2} \\
&= \frac{(N+k)(N+k+1)}{2}+k-1,\\
n^{4d}_{v}{}_\text{SCFT} &= \frac{l(3n^2+ 6 n(5l+1)+40l^2-24l-22)}{6} \\
&= \frac{(k-1)(3N^2+ 3 N(3k+1)-2k^2-5k-6)}{6},
\end{align}
for the case in (b). In the second part we have used the parameters $N$ and $k$ introduced previously for later convenience. As we shall show in the next subsection, these match the values computed using the $5d$ constructions. 

We next move on to the $\bZ_3$ case. Now we consider the $6d$ SCFTs with tensor branch description as shown in figure \ref{6dQuiversZ3}. From this we see that $n_t = l$ and $\sum^{n_G}_i \sum^i_{j=1} h_{G_j} = \frac{3}{2} l^2(l+1)$, \space $\frac{3}{2} l(l+1)^2$, \space $\frac{1}{2} l(3l^2-1)$ for (a), (b) and (c) respectively.

When reduced to $4d$ on a torus with a $\bZ_3$ SW class, we expect a $4d$ theory having the quiver theories in figure \ref{4dQuiversZ3} as a low-energy description on a generic point on the subspace of the $4d$ Coulomb branch that descends from the $6d$ tensor branch. From this we see that:
\be
k_\text{generic} = 2 (N-3) ,
\ee  
for all cases, and
\begin{align}
d_{H}{}_\text{generic} &= \frac{(N-1)(N-2)}{2} ,\\
n^{4d}_{v}{}_\text{generic} &= \frac{(N-1)(2N^2-7N+2)}{18},
\end{align}
for the case in (a), 
\begin{align}
d_{H}{}_\text{generic} &= \frac{N^2 - 3N +6}{2} ,\\
n^{4d}_v{}_\text{generic} &= \frac{(N-2)(2N^2 - 5N - 1)}{18}
\end{align}
for the case in (b), and
\begin{align}
d_{H}{}_\text{generic} &= \frac{N^2 - 3N -2}{2} ,\\
n^{4d}_v{}_\text{generic} &= \frac{2N^3-9N^2+9N+18}{18}
\end{align}
for the case in (c). Here we have used the parameter $N$ that was defined when these $6d$ SCFTs were first introduced.

From this, using equations \ref{eq:kfrom6d}, \ref{eq:afrom6d} and \ref{eq:cfrom6d}, together with $t=3$, $I=3$, we find that:
\begin{align}
k_\text{SCFT} &= 2(N+3), \\
d_{H}{}_\text{SCFT} &= \frac{(N-1)(N+2)}{2},\\
n^{4d}_{v}{}_\text{SCFT} &=  \frac{2N^3 - 3 N^2 - 5N + 6}{6}.
\end{align}

 The values of $k_\text{SCFT}$ and $d_{H}{}_\text{SCFT}$ matches those computed using the $5d$ realization in \cite{Zafrir:2016wkk}. The form of $n^{4d}_{v}{}_\text{SCFT}$ can be argued from the $5d$ realization using a physically reasonable conjecture. This is explained in the end of appendix \ref{app:B}. The result matches the one computed here.    

\subsubsection{Relation to $5d$ compactifications}
\label{bwc}

Like in the previous section, we can relate these SW compactifications to twisted compactifications of $5d$ SCFTs. The idea is again to first reduce the $6d$ SCFT along a circle with an holonomy which we will choose to be the diagonal one in the pair of almost commuting holonomies. This should lead to some $5d$ theory, which can be determined from the results of \cite{Zafrir:2015rga,Ohmori:2015tka,Hayashi:2015zka}. 

The resulting $5d$ theories can be represented via brane webs, and in many cases, the action of the second holonomy can be upgraded to a geometric action. Specifically, we consider the cases where the modded discrete group is $\bZ_2$, $\bZ_3$ , $\bZ_4$ and $\bZ_6$. In those cases we generally have an $\fsu(lN)$ global symmetry, for discrete group $\bZ_l$, with the diagonal holonomy breaking, up to $\fu(1)$ factors, $\fsu(lN)\rightarrow \fsu(N)^l$, and the second holonomy then acting on these as a cyclic shift. 

\begin{figure}
\center
\includegraphics[width=1\textwidth]{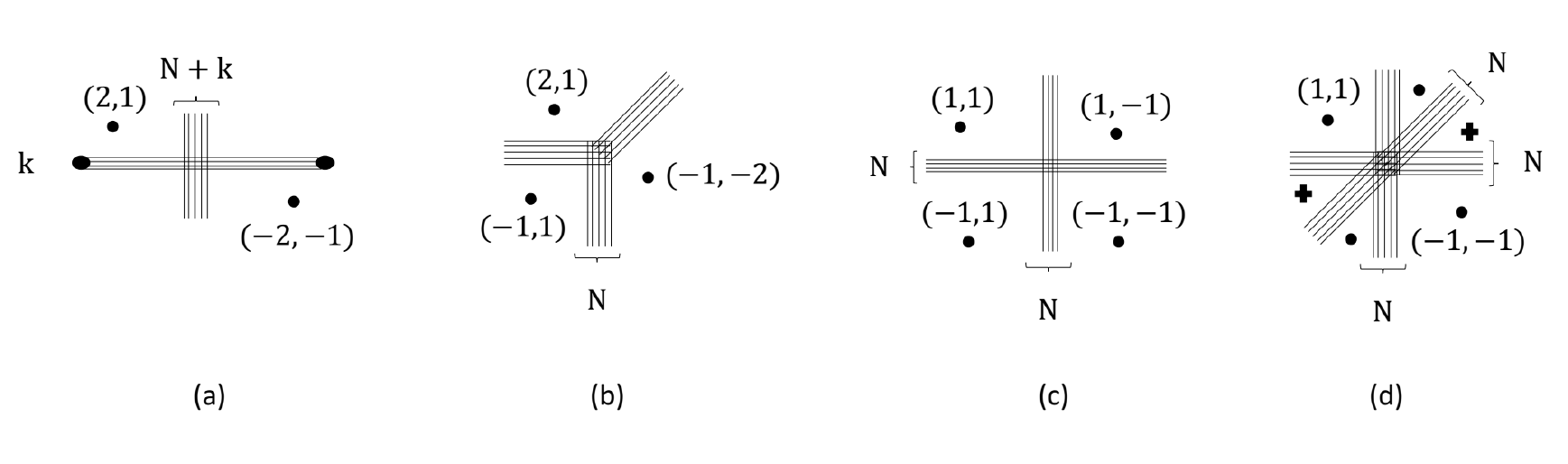} 
\caption{The brane webs describing the effective $5d$ theory that results from the compactification of the $6d$ SCFTs introduced in this subsection. Specifically, (a) is for the $\bZ_2$ cases in figure \ref{6dQuiversZ2}, where $k=2l$ for (a) and $k=2l+1$ for (b), and $N$ is as defined in that section. (b) is for the $\bZ_3$ cases in figure \ref{6dQuiversZ3}, where $N$ is as defined in that section. (c) and (d) are for the $\bZ_4$ and $\bZ_6$ cases, respectively, where again $N$ is as defined there. Here we use black dots for $7$-branes, where we have written their $(p,q)$-charges next to them. If no charges are written, then it is a D$7$-brane. We also use black $+$ for $(0,1)$ $7$-branes. The configurations are drawn such that they show the discrete symmetry permuting the flavor symmetry, here manifested by a rotation in the plane of the web accompanied with the appropriate $\SL(2,\bZ)$ transformation.}
\label{MoreWebs}
\end{figure}

This structure then has a manifestation on the $5d$ representation. Specifically, the web description of the $5d$ theory can be arranged so as to make this structure manifest. This has been done, for the theories considered here, in figure \ref{MoreWebs}. The webs drawn there exhibit the $\fsu(N)^l$ global symmetry, and also a discrete symmetry implementing the cyclic permutation.

To get the  compactified theory with Stiefel-Whitney twists, we also need to take the zero radius limit, and compactify the $5d$ theory to $4d$ with the second holonomy. The zero radius limit should be implemented on the web by some mass deformation, preserving the $\fsu(N)^l$ symmetry and the discrete action. Here we have implicitly assumed that the second holonomy is present, and hence the need to preserve the discrete symmetry. The natural choice then is the deformation associated with sending the free $7$-branes in figure \ref{MoreWebs} to infinity. This leads to $5d$ SCFTs that can be represented by the brane webs in figure \ref{MoreWebs}, but with the free $7$-branes removed.

Finally, we need to consider the $4d$ compactification. Due to the action of the second holonomy, this should be equal to the compactification of the $5d$ SCFT with a twist under the associated discrete symmetry element. This is the relation that we alluded to in the beginning of this subsection. This relation becomes more useful when the discrete symmetry has a geometric manifestation on the brane web. This is true for the cases we consider here, where the discrete symmetry can be lifted to the entire brane web configuration. When lifted, the discrete symmetry is generally manifested by a combination of an $\SL(2,\bZ)$ transformation and a rotation in the plane spanned by the web.

The advantage in this case is that one can then use the web description to uncover various properties of the resulting $4d$ theories. This was studied extensively in \cite{Zafrir:2016wkk}, for the $\bZ_2$ and $\bZ_3$ cases, where the extension for the $\bZ_4$ and $\bZ_6$ cases is straightforward. As covered in \cite{Zafrir:2016wkk}, the notable properties that can uncovered using the web description are the dimension of the Higgs branch, mass deformations and dualities. Next we broadly review some of the results that can be uncovered in this way, referring to \cite{Zafrir:2016wkk} for the details. 

We start by considering dualities. Here the idea is similar to the one we discussed in section 2.2, and so we will not repeat it here. Generally we can uncover many dualities from the $5d$ brane webs. These are usually of the form where gauging a theory in this class, for some $\bZ_k$, and a class S theory is dual to a different gauging of a different class S theory and a theory in this class, for some $\bZ_{k'}$. As these dualities usually relate a theory to another with $k=k'$, it is difficult to use this to study the complete class of theories.

\begin{figure}
\center
\includegraphics[width=1\textwidth]{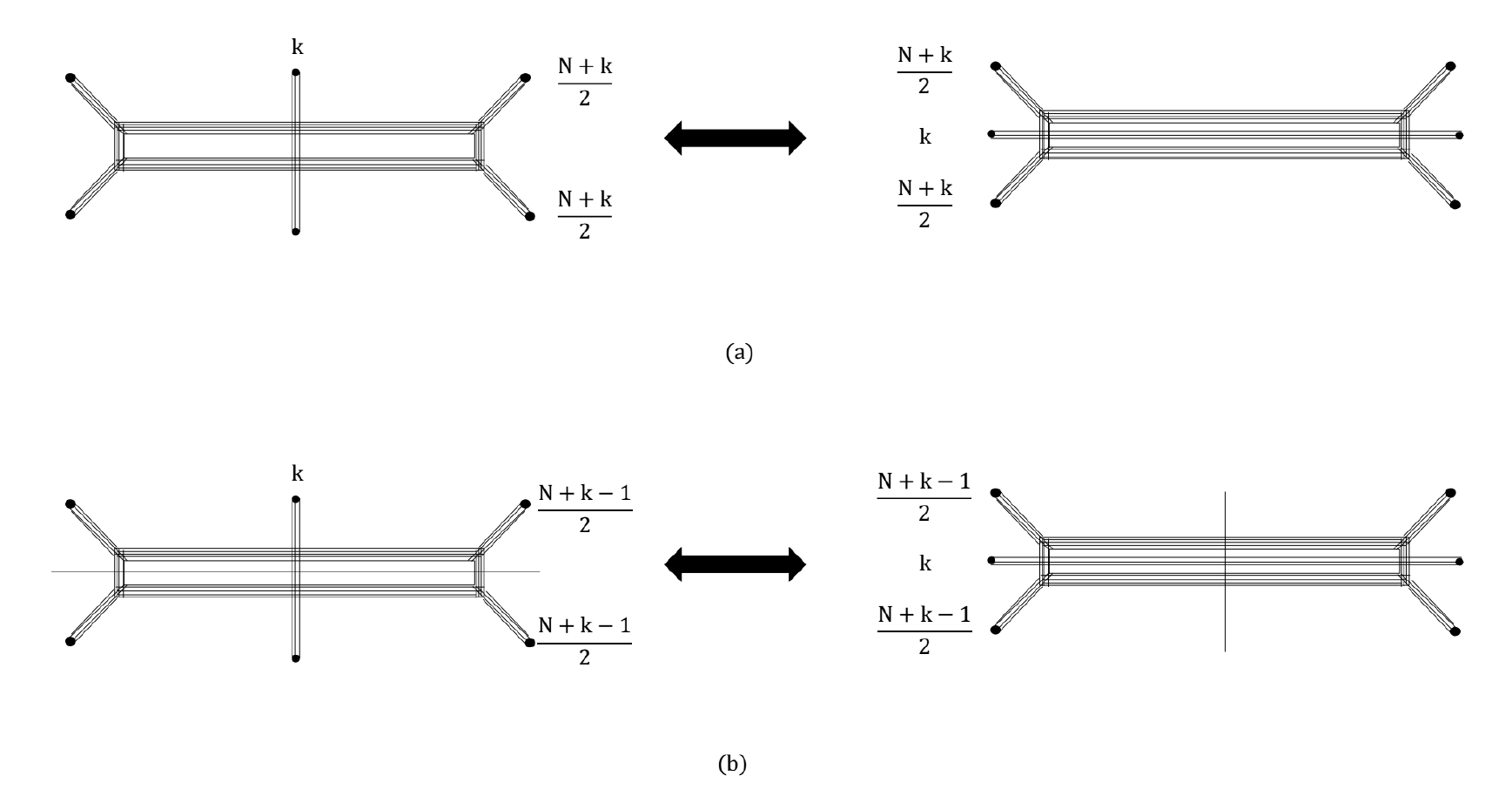} 
\caption{Two dualities expected from the $5d$ construction. (a) is for the case when $N+k$ is even and (b) is when it is odd. The figure on the left reduces to a $4d$ gauging of the SCFT expected from the Stiefel-Whitney compactification of the $6d$ SCFTs discussed in section~\ref{sec:3.4.1}. The figure on the right is the same configuration, after an S-duality and continuation past infinite coupling. It describes a gauging of a $4d$ SCFT that can be described in class S. This manipulation should reduce to a duality in $4d$.}
\label{Duality}
\end{figure}

An exceptional case is the $\bZ_2$ case, where there is a duality relating gauging of a class S theory to a gauging of the $\bZ_2$ twisted $5d$ SCFT. The brane manipulations leading to this duality are shown in figure \ref{Duality}. From them we get the following duality. In the case of figure \ref{Duality} (a), which applies when $N+k$ is even, one side describes a $\fusp(N+k)$ gauging of the $4d$ SCFT we get from the SW compactification in this case. This is dual to an $\fso(N+2k)$ gauging of an SCFT describable in class S as a compactification of an $A_{N+2k-3}$ $(2,0)$ theory on a sphere with the three puncture: $[1^{N+2k-2}], [(\frac{N+k-2}{2})^2,k], [\frac{N+k-2}{2},\frac{N+3k-4}{2}]$.

In the case of figure \ref{Duality} (b), which applies when $N+k$ is odd, one side describes an $\fsu(N+k)+1AS$ gauging of the $4d$ SCFT we get from the SW compactification in this case. This is dual to an $\fsu(N+2k-1)+1S$ gauging of an SCFT describable in class S as a compactification of an $A_{N+2k-4}$ $(2,0)$ theory on a sphere with the three puncture: $[1^{N+2k-3}], [(\frac{N+k-3}{2})^2,k], [\frac{N+k-1}{2},\frac{N+3k-5}{2}]$.

From this duality we can note two things. First the class S side is indeed a conformal theory, which is a non-trivial test that the resulting $4d$ theory is an SCFT. Second we can use this to uncover the central charges of the resulting $4d$ theory. Comparing against the results computed in the previous subsection, we find they indeed match.

We can next consider many other dualities, now relating a theory to another in the same class, and in all cases, using the central charges uncovered from the previous duality, the dualities relate conformal theories and the central charges match. When the resulting theories can be identified with class S theories, like in the $D$ type conformal matter, these dualities can sometimes be mapped to class S dualities. Furthermore, we can sometime combine multiple dualities so as to arrive at dualities only between Lagrangian theories or class S theories, which can then be checked by other methods. The resulting consistency of the web of dualities strongly suggests that the $4d$ theories resulting from these compactifications are SCFTs obeying these duality relations.    

For the remaining cases we in general do not have a frame with only known theories. In the $\bZ_3$ case, we can identify some of the theories with class S theories, particularly for $\bZ_3$ twisted $D_4$, but in most of the other cases these theories appear to be new. We expect these to also be $4d$ SCFTs, and the resulting webs of dualities appear consistent with that, but the evidence here is weaker than in the $\bZ_2$ case.

Another property we can infer from the web is the behavior under mass deformation. As mentioned in section \ref{subsec:rank1-5d}, these are both ones leading to other SCFTs, as well as those leading to $5d$ gauge theories. The former appears to reduce to flows between $4d$ SCFTs, at least in the cases we checked so far. The latter, however, reduce to flows sending the $4d$ theory to an IR free one.

Generically, the mass deformations sending the $5d$ SCFT to another $5d$ SCFT, consistent with the discrete symmetry, are visible from the brane web, so this flow chart should be derivable using this method. Deformations to $5d$ gauge theories, however, can be hard to see for $\bZ_k$  greater than $\bZ_2$.

\subsection{$\bZ_5$}
\label{sec:Z5}

We have previously mentioned that this construction can be related to twisted $5d$ compactifications. The latter in turn can be represented via a string theory construction, where the twisted discrete symmetry is manifested geometrically, usually involving a subgroup of $\SL(2,\bZ)$. The latter limits the possible discrete groups to the list studied above.

\begin{figure}
\center
\includegraphics[width=0.75\textwidth]{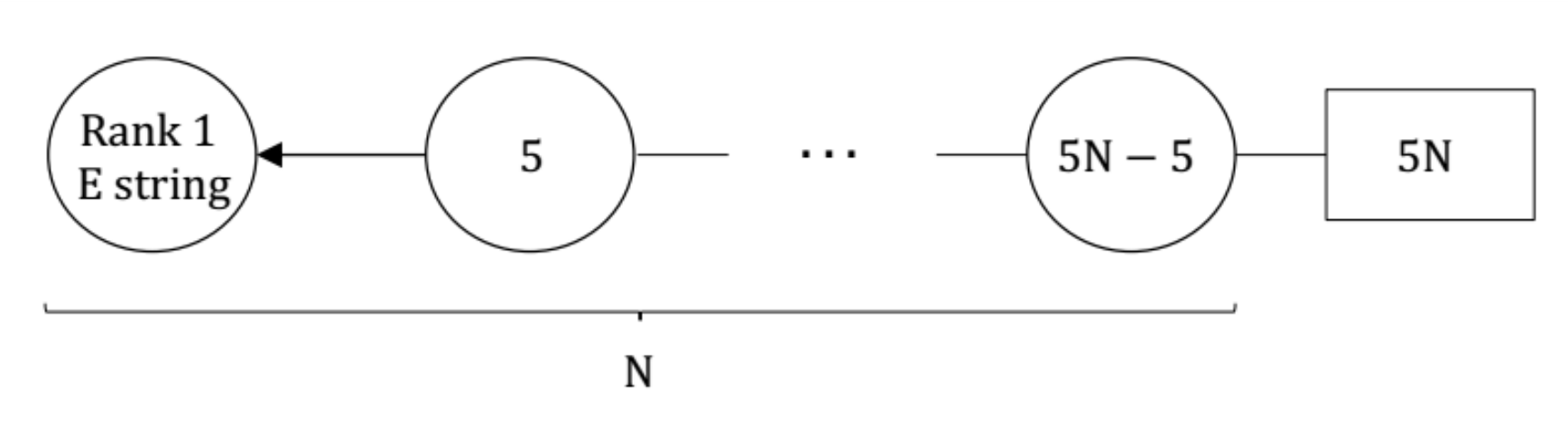} 
\caption{A $6d$ quiver describing a $6d$ SCFT on a generic point on the tensor branch.}
\label{QuiversZ5}
\end{figure}

Nevertheless, we can construct $6d$ SCFTs that appear to support Stiefel-Whitney compactification with other discrete groups. Consider the case where the $6d$ SCFT has a low-energy gauge theory description as the quiver in figure \ref{QuiversZ5}. It is expected to lift to a $6d$ SCFT with $\fsu(5) \times \fsu(5N)$ global symmetry algebra. As should now be clear from our previous similar analysis, all gauge invariants built from both the perturbative matter transform trivially under a $\bZ_5$ subgroup of $\SU(5N)$. 

We can also check the $\fsu(5)$ charged contributions of the $E_8$ conserved currents. Under the $\fsu(5) \times \fsu(5)$ subgroup of $E_8$, the adjoint of $E_8$ decomposes as $\bold{248}\rightarrow (\bold{24}, \bold{1}) \oplus (\bold{1}, \bold{24}) \oplus (\bold{10}, \bold{5}) \oplus (\bar{\bold{10}}, \bar{\bold{5}}) \oplus (\bold{5}, \bar{\bold{10}}) \oplus (\bar{\bold{5}}, \bold{10})$. By inspection one can see that all gauge invariants transform trivially under a $\bZ_5$ in the center $\SU(5N)$ properly combined with the center of $\SU(5)$.   

Like in the previous cases we can also use the $4d$ theories resulting from direct torus reduction of these $6d$ SCFTs to study the spectrum. We have indeed performed this for the $N=2$ case, where we again find it consistent with the global symmetry being $\frac{\SU(5)\times \SU(10)}{\bZ_5}$. Therefore, it seems that these theories have a non-simply connected global symmetry and, thus, we can consider their compactification with a non-trivial Stiefel-Whitney class.

We can next look at some specific cases. For $N=1$, we have just the rank $1$ E-string theory whose global symmetry group is simply connected so we expect the construction to reduce just to a deformation of the rank $1$ $4d$ $E_8$ theory. As noted in section \ref{sec:conformal-matter}, we expect this deformation to lead to the Argyres-Douglas $A_0$ theory. 

The next case is $N=2$. Here we expect to get a rank $2$ theory with an $SU(2)$ global symmetry, possessing a single free hyper in the doublet representation of this $SU(2)$ on generic points on the Coulomb branch. We can use this and the formulas \ref{eq:kfrom6d}, \ref{eq:afrom6d} and \ref{eq:cfrom6d}, to compute the central charges. Using $t=5$ and $I=5$, we find that:
\begin{align}
k_\text{SCFT} &= 14, \\
c_\text{SCFT} &= \frac{39}{10},\\
a_\text{SCFT} &= \frac{19}{5}.
\end{align}

From the last two we can further compute: $n^{4d}_{v}{}_\text{SCFT} =  \frac{74}{5}$. Interestingly applying the procedure outlined in appendix \ref{app:B} to this case gives Coulomb branch operator dimensions of $\frac{12}{5}$ and $6$. This is indeed consistent with the calculated value of $n_v$. 

To our knowledge, there is no known theory with these Coulomb branch dimensions, making this a somewhat interesting example. Unfortunately the lack of a $5d$ construction reduces our ability to study this theory in more detail. Nevertheless, the construction seems to be well defined and the computed central charges seem consistent with known bounds.\footnote{%
In addition, the Coulomb branch dimensions are consistent with the constraint in Argyres-Martone \cite{Argyres:2018urp}.
This pair $(\frac{12}{5}, 6)$ of the Coulomb branch dimensions do not directly appear in the v4 of Caorsi-Cecotti \cite{Caorsi:2018zsq}, but it can be made compatible by dropping an unwritten assumption in it. 
The authors thank Professor S.~Cecotti for helpful discussions on this point.
} 
Furthermore, we can formulate a reasonable proposal for the structure of the Coulomb branch. We suppose that the Coulomb branch is described by two variables $z_1$ and $z_2$, both of dimension $\frac{6}{5}$. However we enforce the following identifications: 
\begin{equation}
I:\quad z_1 \leftrightarrow z_2,\qquad\qquad
II:\quad (z_1, z_2) \rightarrow (e^{\frac{2\pi i}{5}} z_1, e^{-\frac{2\pi i}{5}} z_2).
\label{eq:ident}
\end{equation}
The gauge invariant combinations are then: $z^5_1 + z^5_2$ and $z_1 z_2$, giving a Coulomb branch spanned by operators of dimensions $6$ and $\frac{12}{5}$, respectively. 
This is similar to the structure of the Coulomb branch of the rank $N$ E-string theories, but here $z_i$ have dimension $\frac{6}{5}$, and the modded-out discrete group is different. Regarding the former point, giving that for $N=1$ we expect to get the Argyres-Douglas $A_0$ theory, whose Coulomb branch spanning operator has dimension $\frac{6}{5}$, this does not seem unreasonable.   

Overall, we seem to find this construction consistent and gives seemingly interesting theories also for these $\bZ_5$ cases. It might be interesting to better understand the $5d$ picture in this case, though we shall not pursue it here.

\section*{Acknowledgments}
KO is supported in part by the National Science Foundation grant PHY-1606531 and also  by the Paul Dirac fund.
YT is partially supported  by JSPS KAKENHI Grant-in-Aid (Wakate-A), No.17H04837 
and JSPS KAKENHI Grant-in-Aid (Kiban-S), No.16H06335,
and also by WPI Initiative, MEXT, Japan at IPMU, the University of Tokyo.
GZ is supported in part by World Premier International Research Center Initiative (WPI), MEXT, Japan.

\appendix

\section{The residual gauge symmetry and the quotient on $u$}
\label{app:A}
Here we would like to discuss in more detail the reason why the physical Coulomb branch parameter $u_\text{phys}$
is the power $u_\text{naive}^t$ of the naive Coulomb branch parameter $u_\text{naive}$ coming from the 6d tensor multiplet by the order $t$ of the Stiefel-Whitney twist.
We are going to argue that
\begin{itemize}
	\item When compactified with nontrivial Stiefel-Whitney twist, the tensor branch effective theory gives a $\mathbb{Z}_t$ discrete gauge symmetry in 4d, and
	\item this $\mathbb{Z}_t$ acts on the naive Coulomb branch parameter $u_\text{naive}$ by $u_\text{naive} \mapsto \omega_t u_\text{naive}$, where $\omega_t$ is a (primitive) $t$th root of unity,
\end{itemize}
and therefore $u^t$ is the gauge invariant physical coordinate.

Let us consider the concrete case of $\mathbb{Z}_3$ twist of the 6d $\fsu(3)$ gauge theory of 12 flavors, explained in Section~\ref{sec:2.2}.
As explained in that section, we have the holonomies \eqref{eq:SU12holonomy} along the directions of the compactifying torus. Note that there are matter fields in the bifundamental representation of gauge $\fsu(3)$ and the flavor $\fsu(12)$ algebra, which forces
\begin{equation}
	w_2(\mathcal{P}_\text{gauge}) = w_2(\mathcal{P}_\text{global}),
	\label{eq:w2Pgauge}
\end{equation}
where $\mathcal{P}_\text{gauge}$ and $\mathcal{P}_\text{global}$ stands for the gauge and global symmetry bundles. 
Therefore, the gauge bundles that should be summed up in the path integral is actually $\SU(3)/\bZ_3$ bundles with non-trivial $w_2$ along the compactifying $T^2$, and trivial $w_2$ along the rest of the dimensions.
The gauge holonomies along $T^2$ can be taken to be \eqref{eq:SU3holonomy}.
Due to these holonomies, the gauge symmetry is broken down to the commutator of $P,Q$ in $\SU(3)$, which is the center $\mathbb{Z}_3$ of $\SU(3)$.
Therefore, in 4d, there is the $\mathbb{Z}_3$ gauge symmetry remains.

In general, $\mathbb{Z}_t$-valued Stiefel-Whitney twist on the global symmetry induces a Stiefel-Whitney twist on the gauge symmetry $G$ via \eqref{eq:w2Pgauge}, 
which is realized by some holonomies $A,B$ satisfying $Q^{-1}P^{-1}QP = g_t$ where $g_t$ is an element of the center $Z(G)$ with $g_t^t=1$ generating $\mathbb{Z}_t \subset Z(G)$. Since  $g_t$ is in the center, it remains in 4d as a residual gauge symmetry, possibly embedded inside a larger group.

The next step is to study the action of this $\mathbb{Z}_t$ symmetry on the naive Coulomb branch parameter $u_\text{naive}$ defined by
\begin{equation}
	u_\text{naive} = \exp\left(a + 2\pi i \int_{T^2} \sB\right),
	\label{eq:naiveu}
\end{equation}
where $a$ and $\sB$ are the scalar and the self-dual 2-form fields in the 6d tensor multiplet coupled to the $G$ vector fields.
This field $u$ is invariant under the 1-form gauge transformation of $\sB\to \sB+\mathrm{d}\Lambda$, since $\sB$ is quantized so that the cohomology of $\mathrm{d}\Lambda$ is integer valued.
Without the Stiefel-Whitney twist, this ensures that $u_\text{naive}$ is gauge invariant, but with SW twist we need to carefully consider the variation of $u_\text{naive}$ under the residual discrete symmetry,
since the tensor field get transformed by the gauge transformation as, roughly,
\begin{equation}
	\delta_{\lambda} \sB \propto \Tr \lambda F.
	\label{eq:btransf}
\end{equation}
where $F$ is the gauge field strength. 

To compute the exact amount of  the gauge variation, it is convenient to state \eqref{eq:btransf} in a more invariant way.
The precise meaning of \eqref{eq:btransf} is that the modification of the Bianchi identity for the 3-form field strength $\sH$ is given by 
\begin{equation}
	\mathrm{d}\sH = c_2(F),
\end{equation}
where $c_2(F)$ is the second Chern number of the gauge bundle $\mathcal{P}_{gauge}$.
The local solution to the Bianchi identity is 
\begin{equation}
	\sH = \mathrm{d}\sB + \CS(A),
	\label{eq:hsol}
\end{equation}
where $\CS$ stands for the local expression of the Chern-Simons functional depending of the $\SU(3)$ connection $A$.
The gauge variation of $\sB$ can be determined by demanding $\sH$ to be invariant.

Consider $T^3=T^2\times S^1$, and put the holonomies $P,Q$ along the first two directions. 
Take the coordinate $x$ on the third direction with $x\sim x+1$, and set the configuration of $\sB$ satisfying $\sB(x=1) = g\cdot \sB(0)$ where $g\cdot$ is the action of the center element $g$ of $\SU(3)$. 
The gauge transformation by $g$ does not cause problems on the gauge field, since it is invariant under $g$.
The value of $\int_{T^3}\sH $ is
\begin{equation}
	\begin{split}
		\int_{T^3}\sH &= \int_{x=0}^{x=1}\int_{T^2}\mathrm{d} \sB\\
			    &= \int_{T^2}(g\cdot \sB - \sB).
	\end{split}
\end{equation}
There is no contribution form the second term of \eqref{eq:hsol} since $x$ component of the $\SU(3)$ connection is zero in $0<x<1$.
On the other hand, we can perform the gauge transformation with a gauge parameter $g(x)\in \SU(3)$ with $g(0) =\mathds{1}$ and $g(1)=g^{-1}$.
After this gauge transformation, $\int_{T^3}\mathrm{d}(g(x)\cdot \sB)$ vanishes since $g(x)\cdot \sB$ is a well-defined 2-form on $T^3$.
Therefore, from the gauge invariance of $\sH$, we have
\begin{equation}
	\begin{split}
	\sH &= \int_{T^2}g\cdot \sB - \int_{T^2}\sB\\
	&= \int_{T^3} \CS(A),
	\end{split}
\end{equation}
where the $A$ is the connection on $T^3$ with holonomies $P,Q,g^{-1}$.
Note that while $a$ is not flat as an $\SU(3)$ connection, its projection onto $\SU(3)/\bZ_3$ is flat.
$\CS$ invariants of such connections on $T^3$ were studied in \cite{Borel:1999bx}, as
\begin{equation}
	\CS(A) = \omega_t,
\end{equation}
for a (primitive) $t$th root of unity $\omega_t$.
This concludes that the naive $u_\text{naive}$ defined by \eqref{eq:naiveu} is acted nontrivially by $\mathbb{Z}_t$ and therefore $u_\text{phys} = u^t_\text{naive}$ is the physical coordinate.

\section{Dimensions of $4d$ Coulomb branch operators}
\label{app:B}
In a previous paper \cite{Ohmori:2015pua},
 a method was devised to compute $a$, $c$ and $k$ of the $4d$ \Nequals2 SCFTs
 obtained by $T^2$ compactifications of very Higgsable 6d SCFTs without Stiefel-Whitney twist.
This method was extended in Sec.~\ref{sec:rank1central}  and Sec.~\ref{sec:EE} of this paper to the case with Stiefel-Whitney twist.
The computation  involved a recursive construction: at each step, we shrink a $-1$ curve, and we determined how $a$, $c$ and $k$ change.
By carefully studying this procedure, we can make an informed guess
providing us with not only $n_v=4(2a-c)$ but also individual scaling dimensions of Coulomb branch operators.

This section contains three subsections: 
In Sec.~\ref{B:without}, we motivate our prescription and discuss those examples whose Coulomb branch dimensions can be compared to the ones available in the existing literature,
in the case without the Stiefel-Whitney twist
In Sec.~\ref{B:with}, we extend the discussions to the case with the Stiefel-Whitney twist.
In Sec.~\ref{B:classS}, we give a justification for our prescription when the resulting 4d theory can be straightforwardly analyzed by the class S methods.

\subsection{Without Stiefel-Whitney twist}
\label{B:without}
\subsubsection{The idea and simple  examples}
\label{B:simple}
Consider a single step, where a tensor multiplet from a $-1$ curve is coupled to a vector multiplet of gauge group $G$ gauging a $6d$ SCFT obtained from previous steps.
Let us say that the 6d SCFT in the last step produces $4d$ Coulomb branch operators $v'_1$, \ldots, $v'_n$.
Furthermore, the gauge invariants of the vector multiplet of the group $G$ give rise to operators $v_1$, \ldots, $v_r$.
The tensor branch would produce another Coulomb branch operator $u$.
At the origin $u=0$ we have the new 6d SCFT.

A simple guess for the Coulomb branch operators at the origin is that they are given by \begin{equation}
U_0=u;\qquad U_i=u^{c_i} v_i; \qquad U'_i=u^{c'_i} v'_i.
\end{equation} for some positive integers $c_i$. 
We always have $\dim(u)=6$.
Since we know $n_v$ before and after we shrink the $-1$ curve, we know $\sum c_i+ \sum c'_i$: it is given by \begin{equation}
\sum c_i+ \sum c'_i = -d -1
\end{equation}
where $d$ is as always the coefficient of the Green-Schwarz coupling to $c_2(R)$, as in \eqref{GS}.

Let us first consider the case where we just have a single $-1$ curve with a gauge group $G$ and hypers in some representation.
We have $-d=h^\vee(G)$ in this case. So we need a set of integers satisfying \begin{equation}
\sum_{i=1}^{\rank G} c_i= h^\vee -1.
\end{equation}
A good candidate for such $c_i$ is given by the comarks of the Dynkin diagram; this is one of the basic properties of the dual Coxeter number.

Even after this guess, we need to decide the permutation when we combine $u^{c_i}$ and  $v_i$. 
Indeed, if the set $\{u^{c_i} v_i\}$ gives the required $n_v$, the set $\{u^{c_i} v_{\sigma(i)}\}$ for any permutation $\sigma$ also does the same.
Comparing with the examples for which the Coulomb branch dimensions are known in the existing literature,
we found that the ordering so that $c_i \le c _j$ and $\dim(v_i)\le \dim(v_j)$ when $i\le j$ does the job for almost all the cases, except when $G=SO(2n)$ with $n\ge 6$.
On this last case, see the end of Sec.~\ref{B:classS}.

With this preparation, it is not so difficult to come up with a rule in the general case. 
Suppose we are given a very Higgsable theory. 
We first study the sequence of blow-downs to shrink all curves.
Then we assign a number $n_i$ to each node using the following algorithm:
\begin{itemize}
\item One first assigns arrows so that each arrow goes from a curve $X$ which is shrunk
to all curves $Y$ which intersects $X$.
\item One assigns $n_i=1$ to the last curve to be shrunk.
\item If a curve $X$ is connected by one arrow to another curve $Y$, $n_X=n_Y+1$.
\item If a curve $X$ is connected by two arrows to curves $Y_1$ and $Y_2$, $n_X=n_{Y_1}+n_{Y_2}+1$.
\end{itemize}
An example in the case of the $(E_8,E_8)$ conformal matter is given below:
\begin{equation}
\includegraphics[width=.5\textwidth,valign=m]{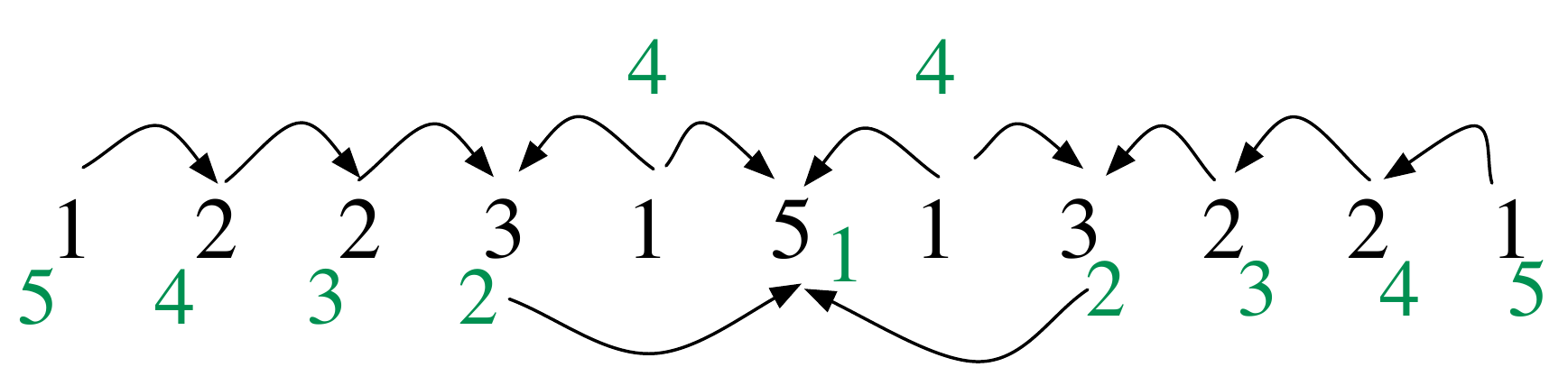}.
\end{equation}

Now, the Coulomb branch dimensions of the $4d$ theory is given by the following rule.
For each curve, \begin{itemize}
\item let the gauge group be $G$,
\item let the degree of its invariants be $d_1,\ldots, d_r$,
\item let the comarks be $c_1,\ldots, c_r$, and
\item let the number assigned by the algorithm above be $n$.
\end{itemize}
Then this curve contributes $1+\rank G$ Coulomb branch operators to the resulting 4d SCFT. 
They have scaling dimensions \begin{equation}
6n,\quad
6n c_1+d_1,\quad
6n c_2+d_2,\quad
\ldots,\quad
6n c_r+d_r.
\end{equation}
Let us now provide simple examples.

\paragraph{$(E_8,E_8)$ conformal matter:} 
Applying this rule to the $(E_8,E_8)$ conformal matter, we find the following scaling dimensions:
\begin{equation}
\begin{array}{rrrrrrrrrrrrrr}
30\\
24\\
18& 18+2 \\
12&12+2&2\cdot 12+6 \\
24\\
6&6+2&2\cdot 6+6& 2\cdot 6+8& 3\cdot 6+12\\
24\\
12& 12+2& 2\cdot 12+6\\
18&18+2\\
24\\
30\\
\end{array}
\end{equation}
which matches the result of the $E_8$ class S theory $T_{E_8}(\text{full},\text{full},\text{simple})$ given in \cite{Chcaltana:2018zag}.\footnote{%
See in particular  the entry \url{https://golem.ph.utexas.edu/class-S/E8/fixtures/select?puncture1\%5Bid\%5D=1&puncture2\%5Bid\%5D=1&puncture3\%5Bid\%5D=69}
in their database.}

\paragraph{$(E_7,E_7)$ conformal matter:} 
We can do the same exercise with the $(E_7,E_7)$ conformal matter. 
The diagram is now 
\begin{equation}
\includegraphics[width=.15\textwidth,valign=m]{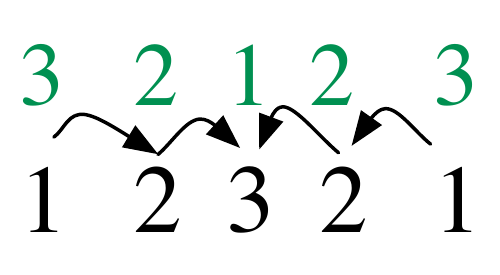}
\end{equation}
which results in the Coulomb branch dimensions \begin{equation}
\begin{array}{rrrrrrrrr}
18\\
12& 12+2\\
6&  6+2& 6+4&  2\cdot 6+6\\
12& 12+2\\
18
\end{array}
\end{equation}
which matches the $E_7$ class S theory  $T_{E_7}(\text{full},\text{full},\text{simple})$  given in \cite{Chacaltana:2017boe}.\footnote{%
See in particular the entry \url{https://golem.ph.utexas.edu/class-S/E7/fixtures/select?puncture1\%5Bid\%5D=1&puncture2\%5Bid\%5D=1&puncture3\%5Bid\%5D=44}
in their database.}

\subsubsection{More examples}
\label{sec:more}

Here we illustrate the previously mentioned rule with various examples. All these examples are drawn from $6d$ SCFTs that can be engineered as theories on M$5$-branes probing an M$9$-plane on a $C^2/\Gamma_{A_k}$ singularity, or theories related to them by flows. These provide a large class of families that lead to $4d$ SCFTs when compactified on a torus, and thus are ideal test cases for the conjecture.

Before we move to actual examples, we discuss how to present the results. These are a series of Coulomb branch dimensions, that is the number $d_{\Delta}$ of Coulomb branch operators of dimension $\Delta$. These are quite unwieldy to work with. Instead it is convenient to draw inspiration from index computations and define the function
\be
C(\nu) = \sum d_{\Delta} \nu^{\Delta}, 
\ee  
where the sum is over all Coulomb branch operators. In general, this expression can be summed into a meromorphic function, which is easier to manipulate than a list of operator dimensions. The list can be recovered from $C(\nu)$ by a Taylor expansion around $\nu=0$. We also note that:
\be
d_C = \lim_{\nu\to 1} C(\nu),\qquad 
n_v = \lim_{\nu\to 1} (2\partial_{\nu}C(\nu)-C(\nu)) .
\ee

 We begin with a few general comments on our chosen class of theories. The compactifications of these theories to $4d$ on a torus without flux were analyzed in \cite{Mekareeya:2017jgc}, and it was found that these lead to $4d$ SCFTs describable in $A$ type class S theory. We shall next consider several examples from this class. All theories in this class can be realized as F-theory compactifications on the curves $-1, -2, -2, -2, \ldots, -2$ where on the $-2$ we always have groups of type $SU$ and on the $-1$ curve we can have groups of type $SU$ or type $USp$. Therefore, this case has the following special features. First the comarks are all $1$ here. Second we can blow-down the cures along the chain one by one starting from the $-1$ curve. Thus we see that we need to assign $n=1$ to the rightmost $-2$ curve, $n=2$ to its neighbor and so fort, until the $-1$ curve, to which we assign $n=l$, where there are $l-1$ ($-2$)-curves.

Next we consider specific examples. We start with the theories in figure \ref{6dQuiversZ2}. These are $6d$ gauge theory descriptions of the $6d$ SCFTs in question on the tensor branch. Using our algorithm, we predict that:
\be
C(\nu) = \frac{\nu^6 (\nu^{k-1} - 1)(\nu^{2N+k-2} + \nu^{2N+k-3} + \nu^{2N+k-4} - \nu^{2k-2} - \nu^{k-1} -1)}{(\nu-1)^2(\nu+1)(\nu^2+\nu+1)} \label{CBZ2D} ,
\ee
where $k=2l, N=n+2l$ for the case in figure \ref{6dQuiversZ2} (a) and $k=2l+1, N=n+2l-1$ for the case in figure \ref{6dQuiversZ2} (b).

These theories are known to reduce to class S theories of type $A_{2N+2k-3}$ associated to a three punctured sphere with punctures $[1^{2N+2k-2}]$, $[(N+k-1)^2]$ and $[N+k-2,N,k]$. It is a straightforward exercise in class S theory to show that the spectrum of Coulomb branch operators expected from this theory matches (\ref{CBZ2D}).

As our second example, consider the theories in figure \ref{6dQuiversZ3}. These are $6d$ gauge theory descriptions of the $6d$ SCFTs in question on the tensor branch. Using our algorithm, we predict that:
\be
C(\nu) = \frac{\nu^6 (\nu^{N-1} - 1)(\nu^{N-2} - 1)(\nu^{N-1}+\nu^{N-2}+1)}{(\nu-1)^2(\nu+1)(\nu^2+\nu+1)} \label{CBZ3D} ,
\ee
where $N=3l+1$ for the case in figure \ref{6dQuiversZ2} (a), $N=3l+2$ for the case in figure \ref{6dQuiversZ2} (b) and $N=3l$ for the case in figure \ref{6dQuiversZ2} (c).

These theories are known to reduce to class S theories of type $A_{3N-4}$ associated to a three punctured sphere with punctures $[1^{3N-3}]$, $[(N-1)^3]$ and $[2N-3,N]$. It is a straightforward exercise in class S theory to show that the spectrum of Coulomb branch operators expected from this theory matches (\ref{CBZ3D}).

\subsection{With Stiefel-Whitney twists}
\label{B:with}
The above method can be extended also to cases with a non-trivial Stiefel-Whitney class on the torus. Next we shall present the algorithm followed by a series of examples. The rule that we shall now present has been tested in many known example, though in this case we do not have as large an example pool as the previous case. Therefore, some modification might be necessary in cases that do not fit nicely with the examples considered here.

\subsubsection{The idea and simple examples}
\label{B:simpleT}

In what follows we consider torus compactifications with non-trivial Stiefel-Whitney class valued in $\bZ_t$. The initial part of the rule is unchanged. We again first study the sequence of blow-downs to shrink all curves, and assign a number $n_i$ to each node using the previous algorithm. Next, for each curve, we define the following: \begin{itemize}
\item let the part of the gauge group unbroken by the Stiefel-Whitney class be $G_\text{ub}$,
\item let the degree of its invariants be $d_1,\ldots, d_r$, and
\item let the number assigned by the algorithm above be $n$.
\end{itemize}
Then this curve contributes $1+\rank G_\text{ub}$ Coulomb branch operators to the resulting 4d SCFT. There are essentially two slightly different cases. One is when the group $G_\text{ub}$ is of type $\SO(2l)$ for integer $l$, and the other is when $G_\text{ub}$ is any other group. Note that here we view the case of an empty $-1$ curve as $\SO(2l)$ for $l=0$, and so this case belongs to the first case. 

We shall first consider the second case. Here the scaling dimensions of the $1+\rank G_\text{ub}$ Coulomb branch operators contributed by the curve are conjectured to be:  \begin{equation}
6n,\quad
6n +d_1,\quad
6n +d_2,\quad
\ldots,\quad
6n +d_r.
\end{equation}  

In the first case, and calling the dimension of the Coulomb branch operator associated with the Pfaffian of the $\SO$ group as $d_r$, we now have: \begin{equation}
6n,\quad
6n +d_1,\quad
6n +d_2,\quad
\ldots,\quad
6n +d_{k-1},\quad
\frac{6n}{t} + d_r.
\end{equation}

Applying this rule to the compactification of the $(E_6,E_6)$ conformal matter with $\bZ_3$ SW class, we find the following scaling dimensions:
\begin{equation}
\begin{array}{r}
12/3\\
6\\
12/3\\
\end{array}
\end{equation}
which matches the result for the expected $\bZ_3$ twisted $D_4$ class S theory  $T_{D_4}(\underline{\text{full}},\underline{\text{full}},\text{simple})$ \cite{Chacaltana:2016shw}.

We can do the same exercise with the $(E_7,E_7)$ conformal matter with $\bZ_2$ SW class. We now find: \begin{equation}
\begin{array}{rrr}
18/2\\
12\\
6&  6+2&  6/2 + 2\\
12\\
18/2
\end{array}
\end{equation}
which matches the result for the expected $\bZ_2$ twisted $E_6$ class S theory $T_{E_6}(\underline{\text{full}},\underline{\text{full}},\text{simple})$\cite{Chacaltana:2015bna}.

\subsubsection{More examples}

Here we shall consider additional examples. Consider first the cases in figure \ref{6dQuiversZ2}. These $6d$ SCFTs can be compactified with $\bZ_2$ SW class, and we can apply our formula to predict the dimension of Coulomb branch operators. As discussed in section \ref{bwc}, using the $5d$ brane web construction, one can also uncover dualities obeyed by these theories from which the dimension of Coulomb branch operators can also be found, allowing us to compare the results. We shall next perform this for selected cases.

We shall start with the case in figure \ref{6dQuiversZ2} (a) for $l=1$. Here the $6d$ gauge algebra is $\fusp(2n)$, and in $4d$ we expect an $\fso(n)$ gauge algebra on a generic point on the Coulomb branch descending from the $6d$ tensor branch. Applying our prescription, we expect to have exactly one Coulomb branch operator for each of dimensions, $6, 8, 10, ... , n+3, n+5$ for $n$ odd, and $6, 8, 10, ... , n+2, n+4 ; \frac{n}{2}+3$ for $n$ even. That is:
\be
C(\nu)_{n\text{ odd}} = \frac{\nu^6(\nu^{n+1}-1)}{\nu^2-1} , 
\qquad 
 C(\nu)_{n\text{ even}} = \frac{\nu^6(\nu^{n}-1)}{\nu^2-1} + \nu^{\frac{n}{2}+3}.
\ee

From the $5d$ picture, we saw that this theory obeys the following duality. For $n$ even, we can double-gauge the $\fusp(2n+8)$ global symmetry with a $\fusp(n+4)$. The gauging is then conformal and, from the duality in section \ref{bwc} applied to this case, is dual to a double-gauging by $\fso(n+6)$ of a class S SCFT with $\fso(2n+12)$ global symmetry. For $n$ odd, we can gauge the $\fusp(2n+8)$ global symmetry with an $\fsu(n+4)$ with an antisymmetric hyper. The gauging is then conformal and dual to a gauging by $\fsu(n+5)$ with a symmetric hyper of the same class S SCFT, but with $\fso(2n+10)$ global symmetry. Using the duality, and the class S technology in \cite{Chacaltana:2010ks}, we can compute the dimension of the Coulomb branch operators of the $4d$ SCFT resulting from the SW compactification. We find that it matches the results expected from our prescription. 

We next consider the case in figure \ref{6dQuiversZ2} (b) for $l=1$. Here the $6d$ gauge algebra is $\fsu(2n)$, and in $4d$ we expect an $\fsu(n)$ gauge algebra on a generic point on the Coulomb branch descending from the $6d$ tensor branch. Applying our prescription, we expect to have exactly one Coulomb branch operator for each of dimensions, $6, 8, 9, 10, ... , n+4, n+5, n+6$.
That is:
\be
C(\nu)_{n\text{ odd}} = \nu^6 + \frac{\nu^8(\nu^{n-1}-1)}{\nu-1} .
\ee

From the $5d$ picture, we see that this theory obeys the following duality. For $n$ even, we can gauge the $\fsu(n+4)$ global symmetry with a $\fusp(n+4)$. The gauging is then conformal and dual to a gauging by $\fso(n+7)$ of a class S theory with $\fsu(n+7)$ global symmetry. For $n$ odd, we can gauge the $\fsu(n+4)$ global symmetry with an $\fsu(n+4)$ with an antisymmetric hyper. The gauging is then conformal and dual to a gauging by $\fsu(n+6)$ with a symmetric hyper of the same class S theory, but with $\fsu(n+6)$ global symmetry. We again find that the matching of the Coulomb branch operators expected from the duality is consistent with our prescription.

We next consider the cases with $l=2$. For the SCFT in figure \ref{6dQuiversZ2} (a), we expect $\fso(n)\times \fsu(n+4)$ gauge groups in $4d$. Applying our prescription, we expect to have following spectrum of Coulomb branch operators: $6, 8, 9, 10, ... , n+8, n+9, n+10; 12, 14, 16, ... , n+9, n+11$ for $n$ odd, and $6, 8, 9, 10, ... , n+8, n+9, n+10; 12, 14, 16, ... , n+8, n+10; \frac{n}{2}+6$ for $n$ even. That is:
\begin{align}
C(\nu)_{n\text{ odd}} &= \frac{\nu^6(\nu^{n+7}+\nu^{n+6}+\nu^{n+5}-\nu^{6}-\nu^{3}-1)}{\nu^2-1} , \\
 C(\nu)_{n\text{ even}} &= \nu^6(\frac{(2\nu^{n+6}+\nu^{n+5}-\nu^{6}-\nu^{3}-1)}{\nu^2-1} + \nu^{\frac{n}{2}}).
\end{align}

From the $5d$ picture, we see that this theory obeys the following duality. For $n$ even, we can gauge the $\fsu(n+8)$ global symmetry with a $\fusp(n+8)$. The gauging is then conformal and dual to a gauging by $\fso(n+12)$ of a class S theory with $\fsu(n+12)$ global symmetry.  
For $n$ odd, we can gauge the $\fsu(n+8)$ global symmetry with an $\fsu(n+8)$ with an antisymmetric hyper. The gauging is then conformal and dual to a gauging by $\fsu(n+11)$ with a symmetric hyper of the same class S theory, but with $\fsu(n+11)$ global symmetry. We again find that the matching of the Coulomb branch operators expected from the duality is consistent with our prescription. 

We can preform the same analysis for the case in figure \ref{6dQuiversZ2} (b). In $4d$ we expect $\fsu(n)\times \fsu(n+4)$ gauge groups. Applying our prescription, we expect to have the following spectrum of Coulomb branch operators: $6, 8, 9, 10, ... , n+8, n+9, n+10; 12, 14, 15, 16, ... , n+10, n+11, n+12$.
That is:
\be
C(\nu)_{n\text{ odd}} = \frac{\nu^6(\nu^2+1)(\nu^{n+5}-\nu^6+\nu^5-\nu^3+\nu-1)}{\nu-1} .
\ee

From the $5d$ picture, we see that this theory obeys the following duality. For $n$ even, we can gauge the $\fsu(n+8)$ global symmetry with a $\fusp(n+8)$. The gauging is then conformal and dual to a gauging by $\fso(n+13)$ of a class S SCFT with $\fsu(n+13)$ global symmetry. For $n$ odd, we can gauge the $\fsu(n+8)$ global symmetry with an $\fsu(n+8)$ with an antisymmetric hyper. The gauging is then conformal and dual to a gauging by $\fsu(n+12)$ with a symmetric hyper of the same family of SCFTs, but with $\fsu(n+12)$ global symmetry. We again see that the matching of Coulomb branch operators is consistent with our prescription.

Finally we want to consider a few $\bZ_3$ examples. Here we consider the theories in figure \ref{6dQuiversZ3}. The gauge groups expected in $4d$ on a generic point in the subspace of the Coulomb branch descending from the $6d$ tensor branch are shown in figure \ref{4dQuiversZ3}. Using this and our prescription, we can predict the spectrum of Coulomb branch operators for these theories. Here we do not know of any generic dualities relating these to known theories and so we have to test this in a different manner. Here we shall combine them and discuss them uniformly using the $N$ defined in section~\ref{sec:more} when we considered direct compactifications of these theories.

For low $N$, these theories can be identified with known theories. Specifically, $N=1$ should be empty, $N=2$ should be $4$ free half-hypers, $N=3$ should be $\fsu(2)+4F$ and $N=4$ the rank $1$ $\fsu(4)$ SCFT, as we saw in Sec.~\ref{sec:rank1}. This can be motivated using the $5d$ realization as twisted compactifications of $5d$ SCFTs, see \cite{Zafrir:2016wkk}.
 On general principles, we expect the effective number of vectors $n_v = 4(2a-c)$ to be some cubic polynomial in $N$. Given that, and the known values at the four points $N=1,2,3,4$, we can completely determine the polynomial to be:
\be
n_v = \frac{1}{3} N^3 - \frac{1}{2} N^2 - \frac{5}{6} N + 1 . \label{nvZ3}
\ee
We can now match this against the expectation from our prescription. Performing the calculation we indeed find that we recover (\ref{nvZ3}). As an example, let us do the next cases, $N=5,6$. For $N=5$, we expect two Coulomb branch operators of dimension $6$ and $8$. This gives $11+15 = 26 = \frac{125}{3} - \frac{25}{2} - \frac{25}{6} + 1$. For $N=6$, we expect four Coulomb branch operators of dimension $4, 6, 8$ and $9$. This gives $7+11+15+17 = 50 = \frac{216}{3} - \frac{36}{2} - \frac{30}{6} + 1$.

\subsection{Partial justification}
\label{B:classS}

Let us give an independent argument to compute the Coulomb branch dimensions
using the class S technology.
At this moment, this only applies to the linear chain $-1$, $-2$, $-2$, \ldots $-2$ with E-string on the leftmost $-1$.

\paragraph{First example.}
As an example, consider $-1$, $-2$, $-2$, $-2$ with $\varnothing$, $SU(1)$, $SU(2)$, $SU(3)$ on top. 
On a generic point on the four-dimensional tensor branch, the 4d gauge theory is given by a class S-theory of type $SU(4)$, with  four simple punctures at $z_{1,2,3,4}$, and the full puncture at $z=\infty$. 
The Seiberg-Witten curve in Gaiotto's form is given by \cite{Gaiotto:2009we} \begin{equation}
\lambda^4 + \Phi_2(z) \lambda^2 + \Phi_3(z) \lambda + \Phi_4(z) =0, \qquad \lambda=xdz/z
\end{equation} where \begin{align}
\Phi_2(z)&=\frac{u_2^{(0)} + u_2^{(1)}z }{(z-z_1)(z-z_2)(z-z_3)(z-z_4)} (dz)^2,\\
\Phi_3(z)&=\frac{u_3^{(0)} }{(z-z_1)(z-z_2)(z-z_3)(z-z_4)} (dz)^3,\\
\Phi_4(z)&= 0.
\end{align}
In the standard class S argument, we assign dimension $k$ to $\Phi_k$, and dimension 0 to $z$.
Then we see that $\Phi_2(z)$ contains two dimension-2 operators 
and $\Phi_3(z)$ contains one dimension-3 operator, accounting for the Coulomb branch of $SU(2)\times SU(3)$.

Now, we know that when a single $z_i$ goes to zero, we have the Minahan-Nemeschansky $E_8$ theory.\footnote{This suggests that there should be a puncture of a hitherto-unknown type at $z=0$, which somehow generates the $E_8$ theory.
It would be interesting to understand the nature of this puncture.}
Therefore, we  assign dimension $6$ to $z_i$.
We also need to take the invariants under the action $z_i \leftrightarrow z_j$.
This gives rise to operators of dimension $6$, $12$, $18$ and $24$.
Furthermore, we would like to keep $\Phi_k$ to have dimension $k$,
but now the variable $z$ has dimension $6$.
This makes \begin{equation}
\Delta(u_2^{(0)})=12+2,\quad
\Delta(u_2^{(1)})=6+2,\quad
\Delta(u_3^{(0)})=6+3.
\end{equation}

This reproduces the outcome of our prescription given in Sec.~\ref{B:simple}.
Furthermore, this analysis can be readily generalized to any linear $SU$ chain on $-1$, $-2$, \ldots, $-2$, as long as $-1$ has the E-string on it.

\paragraph{Second example.}
Next, let us move on to the case with the Stiefel-Whitney twist.
Consider again $-1$, $-2$, $-2$, $-2$, but with $\varnothing$, $SU(5)$, $SU(10)$, $SU(15)$ on top, and introduce $\mathbb{Z}_5$ Stiefel-Whitney twist.
The four-dimensional theory on the generic point on the Coulomb branch is still given by the Gaiotto curve above, but the $z_i$'s there correspond to the `naive tensor moduli'
and we need to introduce $w_i=(z_i)^{1/5}$, whose dimension is $6/5$,
as discussed in the main text.

Generalizing the identifications we proposed in \eqref{eq:ident}, 
we propose that we perform the identifications under 
\begin{equation}
I:\quad w_i\leftrightarrow w_j;\qquad
(w_i,w_j)\mapsto (\gamma w_i,\gamma^{-1} w_j)
\end{equation} where $\gamma=e^{2\pi i/5}$.
The invariant combinations are the symmetric polynomials of $z_i$, 
together with the generalized Pfaffian $w_1w_2w_3w_4$ satisfying
\begin{equation}
(w_1 w_2 w_3 w_4)^5 = z_1 z_2 z_3 z_4.
\end{equation}

The end result is that we replace an operator of dimension $24$ is replaced by an operator of dimension $24/5$,
and reproduces our prescription in Sec.~\ref{B:simpleT}.
It is clear that this result can be again extended to arbitrary linear chain of $SU$ groups which  allows a suitable $\mathbb{Z}_{2,3,4,5,6}$ Stiefel-Whitney twist.

\paragraph{Third example.}

Finally, let us discuss two non-$SU$ cases. 
Consider $-1$, $-2$ with $\varnothing$, $E_7$.
The gauge theory on the generic point on the tensor branch is $E_7$ with three $\mathbf{56}$s, and has a class S description of type $E_7$,
with two simple punctures ($E_7(a1)$) and two punctures of another type ($A_3+A_2+A_1$),
see Sec.~4.1 of \cite{Chacaltana:2017boe}.
We put two punctures of type $A_3+A_2+A_1$ at $z=0$ and $z=\infty$,
and two simple punctures at $z_1$, $z_2$.
There is an exchange symmetry between $z_{1,2}$.

The differential $\Phi_k$ has the form \begin{equation}
\Phi_k(z)=\frac{u_k (dz)^k}{z^{k-c_k}(z-z_1)^{c_k} (z-z_2)^{c_k} }
\end{equation}
where $c_k$ is the pole structure of the degree-$k$ differential at the simple puncture.
The fact that the class S theory is a gauge theory fixes the pole structure at the other puncture to be complementarily given by $k-c_k$.
As can be seen in the table in Sec.~2.3 of \cite{Chacaltana:2017boe},
the pole structure $c_k$ of the simple puncture is $\{1,2,2,2,3,3,4\}$ and equals the comarks.

Now, we note that in our setting $z_{1,2}$ are not parameters but the vevs of vector multiplets,
and that we have the Minahan-Nemeschansky $E_8$ theory when $z_1\to 0$.
We again assign $\Delta(z_{1,2})=6$, and take the symmetric combination.
We also would like to keep $\Phi_k$ to have dimension $k$.  
Therefore,  we are forced to assign \begin{equation}
\Delta(u_k)=k+c_k\Delta(z) = k+ 6c_k.
\end{equation}
This reproduces the prescription we gave in Sec.~\ref{B:simple}.

Similarly, we can check that the pole structure of the simple puncture equals the comark 
also for $E_6$  \cite{Chacaltana:2014jba}, 
of the type $SU(N)$  \cite{Gaiotto:2009we}, 
and also of the type $SO(2N)$ \cite{Chacaltana:2011ze}.
The pole structure of the simple puncture of $E_8$ does not  equal to comarks, according to \cite{Chcaltana:2018zag}.
But this does not concern for us, since $E_8$ gauge group never appears in our construction.

We need a longer comment for $SO(2N)$.
In this case,  the pole structure of the simple puncture of $SO(2N)$ is indeed $\{1,1,2,2,\ldots;1\}$, and the comark $1$ is always associated to $u_2$, $u_4$ and the Pfaffian.
This suggests that we should assign three $c_i=1$'s to $\tr \phi^2$, $\tr\phi^4$, and $\mathop{\mathrm{Pfaff}}\phi$ and $N-3$ $c_i=2$'s to other operators,
as already announced in Sec.~\ref{B:simple}.
This rule deviates from a simpler rule of sorting both the comarks and the $G$ invariants in the increasing order only when $N\ge 6$. It will be interesting if this prediction can be compared against actual computations.

\bibliographystyle{ytphys}
\baselineskip=.9\baselineskip
\let\bbb\bibitem\def\bibitem{\itemsep1pt\bbb}
\bibliography{swref}

\end{document}